\documentclass[10pt,a4paper,superscriptaddress,aps,prd,nofootinbib,notitlepage,aas_macros,longbibliography]{revtex4-1}
\usepackage{lmodern}

\usepackage[T1]{fontenc}
\usepackage[utf8]{inputenc}
\usepackage{color}
\usepackage{amsmath}
\usepackage{amssymb}
\usepackage{graphicx}
\usepackage{esint}
\usepackage[normalem]{ulem}
\usepackage[stable]{footmisc}
\usepackage{natbib}
\usepackage[unicode=true,
 bookmarks=true,bookmarksnumbered=false,bookmarksopen=false,
 breaklinks=false,pdfborder={0 0 1},backref=false,colorlinks=true]
 {hyperref}
\hypersetup{pdftitle={Measuring gravity at cosmological scales},
 pdfauthor={Luca Amendola, Dario Bettoni,Ana Marta Pinho, Santiago Casas},
 pdfkeywords={dark energy, modified gravity, large-scale structure},
 citecolor=blue,filecolor=blue,linkcolor=blue,urlcolor=blue}

\makeatletter

\pdfpageheight\paperheight
\pdfpagewidth\paperwidth

\definecolor{somegreen}{cmyk}{0,0.49,0.98,0.09}
\definecolor{red}{rgb}{1,0,0}
\definecolor{magenta}{cmyk}{0,1,0,0}
\definecolor{violet}{cmyk}{0,1,0,0}
\definecolor{darkgreen}{rgb}{0,0.65,0.05}
\definecolor{antiquefuchsia}{rgb}{0.33, 0.1, 0.89}

\usepackage{aas_macros}

\makeatother

\begin{document}

\title{Measuring gravity at cosmological scales}

\author{Luca Amendola}
\affiliation{ITP, Ruprecht-Karls-Universität Heidelberg, Philosophenweg 16, 69120
Heidelberg, Germany}
\author{Dario Bettoni}
\affiliation{ITP, Ruprecht-Karls-Universität Heidelberg, Philosophenweg 16, 69120
Heidelberg, Germany}
\affiliation{Departamento de Física Fundamental, Universidad de Salamanca, E 37008 Salamanca, Spain}
\author{Ana Marta Pinho}
\affiliation{ITP, Ruprecht-Karls-Universität Heidelberg, Philosophenweg 16, 69120
	Heidelberg, Germany}
\author{Santiago Casas}
\affiliation{AIM, CEA, CNRS, Université Paris-Saclay, Université Paris Diderot, Sorbonne Paris Cité, F-91191 Gif-sur-Yvette, France}

\begin{abstract}
This paper is a pedagogical introduction to models of gravity and how to constrain them through cosmological observations. We focus on the Horndeski scalar-tensor theory and on the quantities that can be measured with a minimum of assumptions.

Alternatives or extensions of General Relativity have been proposed
ever since its early years. Because of Lovelock theorem, modifying
gravity in four dimensions typically means adding new degrees of
freedom. The simplest way is to include a scalar field coupled to
the curvature tensor terms. The most general way of doing so without incurring in the Ostrogradski instability is the Horndeski Lagrangian and its extensions. Testing
gravity means therefore, in its simplest term, testing the Horndeski
Lagrangian. Since local gravity experiments can always be evaded by
assuming some screening mechanism or that baryons are decoupled, or
even that the effects of modified gravity are visible only at early
times, we need to test gravity with cosmological observations in the
late universe (large-scale structure) and in the early universe (cosmic
microwave background). In this work we review the basic tools to test
gravity at cosmological scales, focusing on model-independent measurements.
\end{abstract}

\date{\today}
\maketitle

\section{Introduction}

Gravity is the force that shapes the overall temporal and spatial structure of the Universe. There is no much need then to explain why it is important to test its validity at all scales and regimes. The huge progress in collecting cosmological data achieved in the last couple of decades makes possible, for the first time, to test gravity and measure its properties at astrophysical and cosmological scales. In order to test a theory one  either has to build a set of alternatives against which to compare the standard model, or to parametrize the deviations from it in some meaningful and general way: both approaches are referred to as ``modified gravity''.

Lovelock's theorem \cite{1972JMP....13..874L} states that Einstein's
gravity is the unique local diffeomorphism invariant theory of a tensor field in 4D with second-order
equations of motion. It is clear then that modifying gravity often implies
adding new degrees of freedom, either scalars, vectors, or tensors. Adding
a mass to the graviton, for instance, requires an additional tensor
field; including more derivatives is also equivalent to adding more propagating
degrees of freedom. 
Other options based
on torsion, non-metricity, or non-locality can also be contemplated, see for instance the review \cite{Clifton:2011jh}.

In this paper we review the main properties of an important class of modified gravity based on a single scalar field, the so-called Horndeski Lagrangian (HL). This model is general enough to display most of the phenomenology of non-Einsteinian gravity: generalized Poisson equation, Yukawa corrections to Newton's potential, presence of anisotropic stress, change in the gravitational wave speed, instabilities, ghosts. Still, the HL is relatively simple in that contains a single  propagating degree of freedom in addition to General Relativity. Using the HL as a paradigm of modified gravity, we focus on its observability at various scales, from the local environment to galaxy clusters, with emphasis on cosmological observations. Although the recent measurement of the gravitational wave speed \cite{LIGO:2017qsa} severely constrain one of the phenomenological time-dependent parameters of the Horndeski model, as we will see, the other parameters  are still mostly unconstrained and open to theoretical and observational investigation.
A major topic of this review is the question of which properties of  gravity can be measured  as model-independently  as possible. 

We will not try to cover exhaustively the field of research in modified gravity; good reviews are already available \cite{Clifton:2011jh,2015LNP...892.....P}. Rather, we wish to discuss pedagogically some aspects or issues that are generic to the quest for traces of modified gravity.

We assume units such that $c=8\pi G_N=M_{\textrm{Planck}}^{-2}=1$ and metric signature $-+++$. An overdot denotes derivation with respect to cosmic time $t$, a prime with respect to $\log a$. A comma will refer to partial derivative, i.e. $\partial_\mu\phi \equiv \phi_{,\mu}$. Also $\Box\phi=g^{\mu\nu}\nabla_\mu\nabla_\nu\phi$ where $\nabla_\mu$ is the covariant derivative. Greek indexes run over space and time coordinates, Latin indexes over space coordinates only.

\section{Beyond Einstein}

The re-discovery of the most general scalar-tensor theory that gives
second order equations of motion, Horndeski action \cite{Horndeski:1974wa}
or Covariant Galileons \cite{Deffayet:2009wt}, and their extensions
\cite{Zumalacarregui:2013pma,Gleyzes:2014qga,Gleyzes:2014dya,Crisostomi:2016czh,Crisostomi:2016tcp,Achour:2016rkg,Langlois:2015cwa}
provides a very general framework for such theories (see \cite{Kobayashi:2019hrl} for a recent review). 
 The HL is defined as the sum of four terms $\mathcal{L}_{2}$ to $\mathcal{L}_{5}$. Defining with  $X=-g_{\mu\nu}\phi^{,\mu}\phi^{,\nu}/2$ the canonical kinetic
term,
the four terms are specified by two non-canonical kinetic functions $K(\phi,X)$ and $G_3(\phi,X)$ and by
two coupling functions $G_{4,5}(\phi,X)$, all of them in principle arbitrary: 
\begin{equation}
S=\int d^{4}x\sqrt{-g}\sum_{i=2}^{5}\mathcal{L}_{i}+S_{m}
\end{equation}
where $S_m$ is the action for matter fields -- dark matter, baryons and radiation -- and
\begin{align}
\begin{aligned}\mathcal{L}_{2}= & K(\phi,X)\,,\\
\mathcal{L}_{3}= & -G_{3}(\phi,X)\Box\phi\,,\\
\mathcal{L}_{4}= & G_{4}(\phi,X)R+G_{4,X}\left[\left(\Box\phi\right)^{2}-\left(\nabla_{\mu}\nabla_{\nu}\phi\right)^{2}\right]\,,\\
\mathcal{L}_{5}= & G_{5}(\phi,X)G_{\mu\nu}\nabla^{\mu}\nabla^{\nu}\phi-\frac{G_{5,X}}{6}\Bigl[\left(\Box\phi\right)^{3}
 -3\left(\Box\phi\right)\left(\nabla_{\mu}\nabla_{\nu}\phi\right)^{2}+2\left(\nabla_{\mu}\nabla_{\nu}\phi\right)^{3}\Bigr]\,.
\end{aligned}
\label{eq:HL}
\end{align}
Note that $G_3$ and $G_5$ must have an $X$ dependence, otherwise they are  total derivatives and could be rewritten -- after integration by parts -- as $K$ and $G_4$ respectively.\footnote{Notice that the number of these functions cannot be reduced by fields redefinitions without going beyond Horndeski action \cite{Bettoni:2013diz,Zumalacarregui:2013pma}.} As usual, each term in the HL has dimension $mass^4$. Often one chooses the scalar field to have dimensions of mass, but this is not necessary. As already mentioned, the Horndeski Lagrangian is the most general Lagrangian for a single
scalar which gives second-order equations of motion for both the scalar
and the metric on an arbitrary background. This is a necessary, but
not sufficient, condition for the absence of instabilities, as we will see later on. The terms $\mathcal{L}_{4},\mathcal{L}_{5}$
couple the field $\phi$ to the Ricci scalar $R$ and the Einstein
tensor $G_{\mu\nu}=R_{\mu\nu}-Rg_{\mu\nu}/2$. As a consequence, $G_{4,5}$ are the gravity-modifying coupling function.
The background equations of motion of the HL are given for completeness in the Appendix, although we do not need them in the following. It is enough to realize that the large freedom offered by the HL allows one to find a background evolution that satisfies all observational constraints.

Let us now briefly discuss some useful limits of the HL.
\begin{itemize}
\item If $G_{4}=1/2$  and $G_5=0$ (it is actually sufficient $G_5=const$)   the HL reduces to the Einstein--Hilbert Lagrangian with a scalar field having a non-canonical kinetic sector given by $\mathcal{L}_{2},\mathcal{L}_{3}$.  The canonical form is obtained for $K=X-V(\phi)$ and $G_3=0$ ($G_3=const$ is sufficient). $\Lambda$CDM is recovered for $K=-2\Lambda$.
\item The ``minimal'' form of modified gravity within the HL is provided by $G_4=G_4(\phi)$ and $G_5=const$: this is then equivalent to a Brans--Dicke scalar-tensor model, again with a non-canonical kinetic sector. 
\item
The original Brans-Dicke model is recovered assuming a  kinetic sector, $K=(\omega_{BD}/\phi) X,G_3=0$, and $G_4(\phi)=\phi/2$.
\item
If the kinetic sector  vanishes, $K_{,X}=G_3=0$, then we reduce ourselves to a $f(R)$ model \cite{Sotiriou:2008rp}, whose Lagrangian is $\mathcal{L}_R=(R+f(R))/2$. In fact, this model  is equivalent to a scalar-tensor theory with $G_4(\phi)=e^{2\phi/\sqrt{6}}/2$ and a potential $K(\phi)=-(Rf_{,R}-f)/2$ where $\phi=\sqrt{6}/2\log (1+f_{,R})$. This relation should then be inverted to get $R=R(\phi)$ and used to replace $R$ with $\phi$ in $K(\phi)$. 

\item If one sets $G_i(\phi,X)=G_i(X)$ then the Lagrangian is invariant under the shift $\phi\rightarrow \phi +c$ with $c=const$. This shift-symmetric version of the HL is connected to the Covariant Galileon when the functional dependence of the $G_i$ is fixed \cite{Deffayet:2009wt} and is able to produce the accelerated expansion without a potential that makes the field slow roll.

\end{itemize}

In general, the equations
of motion for the scalar will couple it to the matter energy density.
The full
set of equations of motion has been studied in several papers, for
instance in \cite{DeFelice2011,DeFelice:2011bh}. Any modification
of the HL, or addition of terms (except the so-called Beyond Horndeski
terms), based on the same scalar field, will introduce higher order
equations of motion and associated instabilities, as a consequence
of the Ostrogradsky theorem \cite{Ostrogradski:1850,Woodard:2015zca}.\footnote{See \cite{Chen:2012au} for a discussion on how to exorcise Ostrogradski ghosts in non-degenerate theories.} Of course one can in
principle add several scalar fields, but on grounds of simplicity
this is rather unnatural. Notice that we do not demand that the $\phi$
drives the present-day accelerated expansion. It could be, after all,
that the modification of gravity and the accelerated expansion are
independent phenomena. It would be very interesting, though, to explain the latter in terms of the former.

\section{Decomposition in modes and stability}

Einstein's gravity is carried by a massless spin-2 field, the metric. 
Being represented by a symmetric matrix, a metric in four dimensions has 10 degrees of freedom (DOF). These 
DOF can be collected according to how they behave under spatial rotations, i.e. as scalars, vectors and tensors. There are then four scalars (4 DOF), two divergence-free vectors (4 DOF) and one traceless, divergence-free tensor (2 DOF). However, only the tensor DOF propagate, that is, they are  subject to linearized equations of motion second-order in the time derivatives. The other DOF obey constraint equations, fully determined by the matter content. This should have been expected, since a massless tensor field, as the gravitational field, has only two independent degrees of freedom. 

The two propagating  degrees of freedom are associated to the two polarizations
$+,\times$ of the gravitational waves. To see that there are no other
propagating DOF, one can proceed by linearly expanding the metric
around Minkowski
\begin{equation}
g_{\mu\nu}=\eta_{\mu\nu}+h_{\mu\nu}
\end{equation}
and keeping only scalar terms, i.e. functions that can be obtained
from scalar or from derivatives of scalars. The most general such
metric is then
\begin{align}
ds^{2} & =-(1+2\Psi)dt^{2}+2B_{,i}dx^{i}dt+((1+2\Phi)\delta_{ij}+2E_{,ij})dx^{i}dx^{j}
\end{align}
Inserting this metric into the Einstein-Hilbert Lagrangian without matter and developing
to second order, one finds the second-order action in Minkowski space
\begin{align}
S_{g}=\frac{1}{2}\int d^{3}xdt[8B_{,i}^{,i}\dot{\Phi}+4\Phi_{,i}\Psi_{,i}-4\dot{\Phi} \dot{E}_{,i}^{,i}+2\Phi_{,i}^{2}-6\dot{\Phi}^{2}]
\end{align}
 The linearly perturbed equations of motion can be obtained then by the Euler-Lagrange
equations with respect to $\Phi,\Psi,E,B$, but here we need only
identify the degrees of freedom. When one varies the action with respect  to $B$, one gets the
constraint $\dot\Phi=0$ which then shows that $\Phi$ is not
a propagating DOF. The same is true for $\Psi$, since there are no time derivatives for it. As we know, in fact, the potentials $\Phi,\Psi$ are
determined by the matter distribution through two constraints, the Poisson equations, which do not involve time derivatives.
So there are no scalar propagating DOF in Einstein
gravity without matter.

The same holds for the vector degrees of freedom. If one
considers instead the tensor DOF in $h_{\mu\nu}$
\begin{equation}
ds^{2}=-dt^{2}+(\delta_{ij}+h_{ij})dx^{i}dx^{j}
\end{equation}
where, after imposing the traceless, divergence-less conditions,  and considering
a wave propagating in direction $x_{3}$,
\begin{equation}
\bar{h}_{ij}=h_{ij}-\frac{1}{2}\eta_{ij}h=\left(\begin{array}{ccc}
h_{+} & h_{\times} & 0\\
h_{\times} & -h_{+} & 0\\
0 & 0 & 0
\end{array}\right)
\end{equation}
 one finds that the two modes $h_{\alpha}=\{h_{+},h_{\times}\}$ obey
 in vacuum the same gravitational wave equation, $\Box h_{\alpha}=0$, analogous to electromagnetic waves. GWs propagate therefore with speed $c_T$ equal to unity.
 
The same procedure can be applied to the HL. One finds then in  absence of matter fields \cite{DeFelice:2011bh,Bellini2014}
\begin{equation}
S=\int d^{3}xdt\{Q_{S}[\dot{\varphi}^2-\frac{c_{s}^{2}}{a^{2}}(\partial_{i}\varphi)^{2}]+\sum_{\alpha=1}^{2}Q_{T}[\dot{h_{\alpha}}^{2}-\frac{c_{T}^{2}}{a^{2}}(\partial_{i}h_{\alpha})^{2}]\}\label{eq:hlsec}
\end{equation}
where $\varphi$ is the scalar mode perturbation and $h_{\alpha}$
the two tensor modes, and $c_S,c_T$ their speed of propagation, respectively. As expected, HL has now three propagating DOF,
plus those belonging to the matter sector.

The four coefficients $Q_{S},c_{S},Q_{T},c_{T}$ depend on the HL functions. Their expression will be given in Sec. \ref{sec:yukawa}.
From the classical point of view, stability
is guaranteed when $Q_{x},c_{x}^{2}$ (with $x=S,T$) have the same
sign. In this case in fact the equations of motion are well-behaved
wave equations with speed $c_{x}$, whose amplitude is constant
(or decaying in an expanding space), rather than growing exponentially
as it would happen for $c_{x}^{2}<0$ (gradient instability). For
the quantum stability, however, one must also require $Q_{x}>0$ (or
more exactly, the same sign of the kinetic energy of matter particles, assumed by convention
to be positive), since otherwise the Hamiltonian is unbounded from
below, which means particles can decay into lower and lower energy
states, without limit, generating so-called ghosts. Therefore, for
the overall stability of the theory one requires $Q_{x},c_{x}^{2}>0$.

\section{The quasi-static approximation}\label{sec:qsa}

In what follows, we put ourselves in Fourier space. That is, we replace
every perturbation variable $X(\vec{x},t)$ with a plane wave parametrized by the
comoving wavevector $\vec{k}$, $X(\vec{x},t)=X_{k}(t)e^{i\vec{k}\cdot\vec{x}}$.
Since we deal only with linearized equations, this simply means replacing
every perturbation variable or their time-derivative with its corresponding
Fourier coefficient $X_{k}$ or its time derivative $\dot{X}_{k}$,
and every space derivative $\partial_{i}^{(n)}$of order $n$ with
$(ik_{i})^{n}X_{k}$. We drop from now on the $k$ subscripts. We then assume that the so-called \emph{quasi-static
approximation }(QSA) is valid for the evolution of perturbations.
This implies that we are observing scales well inside the cosmological
horizon, $\hat{k}\equiv k_{\text{}}/(aH)\gg1$, where $k$ is the
comoving wavenumber, and also inside the Jeans length of the scalar,
$c_{\text{S}}\hat{k}\gg1$, such that the terms containing $k$ (i.e.,
the spatial derivatives) dominate over the time-derivative terms.
For the scalar field, this means we  neglect its wavelike nature, and convert its Klein-Gordon differential equation into a Poisson-like constraint equation.
If $c_{S}\approx1$, the scales at which the QSA is valid correspond
to all sub-horizon scales, which are also the observed scales in the
recent Universe. For models with $c_S\to 0$, the QSA might be valid only in a narrow range of scales, or even be completely lost in the non-linear regime.

Let us explain in more detail the QSA procedure by using standard
gravity as an example. Let us write down the perturbation equations
for a single pressureless matter fluid in $\Lambda$CDM. 
From now on, we adopt the FLRW perturbed metric in the so-called longitudinal gauge, namely
\begin{align}
ds^{2} & =-(1+2\Psi)dt^{2}+a^2(t)(1+2\Phi)\delta_{ij} dx^{i}dx^{j}
\end{align}
If we use
$N=\log a$ as time variable, so that  $\dot{x}=Hx',$ the coefficients
of the perturbation variables become dimensionless, and we are left
with \cite{2010deto.book.....A}
\begin{align}
\delta' & =-\theta-3\Phi'\\
\theta' & =-\left(2+\frac{H'}{H}\right)\theta+\hat{k}^{2}\Psi\label{eq:euler}\\
\hat{k}^{2}\Psi & =-\frac{3}{2}\Omega_{m}\left(\delta+3\hat{k}^{-2}\theta\right)\label{eq:stan3}\\
\Psi & =-\Phi\label{eq:stan4}
\end{align}
where instead of the matter density $\rho_{m}$, we use $\Omega_m$,
\begin{equation}
\rho_{m}=3H^{2}\Omega_{m}\label{eq:3ho-1-1}
\end{equation}
and where $\delta\equiv \delta\rho_m/\rho_m$, $\theta=ik_{i}v^{i}/aH$ if $v^{i}=adx^{i}/dt$ is the peculiar
velocity, so that $\theta=ik_{i}(x^{i})'$. A glance at these equations
tells us that, as an order of magnitude, $\delta\sim\theta\sim\hat{k}^{2}\Psi\sim\hat{k}^{2}\Phi$. Moreover, we assume
 $X\sim X',X''$ for every perturbation variable $X=\{\delta,\theta,\Psi,\Phi\}$
(unless there is an instability,   see below) and, consequently, $\hat{k}^2 X\gg X',X''$. Therefore for
 $\hat{k}\gg 1$ the equations become
\begin{align}
\delta' & =-\theta\\
\theta' & =-\left(2+\frac{H'}{H}\right)\theta-\frac{3}{2}\Omega_{m}\delta
\end{align}
and one can derive the well-known second-order growth equation with
dimensionless coefficients
\begin{equation}\label{eq:deltapp-qsa}
\delta''+\left(2+\frac{H'}{H}\right)\delta'-\frac{3}{2}\Omega_{m}\delta=0
\end{equation}
The same QSA procedure can be followed for more complicate systems.
When a coupled scalar field is present, its perturbation is of the same order
as the gravitational potentials, $\delta\phi\sim\Psi\sim\Phi$. 

The QSA  says nothing about the background behavior.  Additional conditions might be imposed, for instance that the background scalar field slow rolls so that the kinetic terms, proportional to the derivatives $\phi',\phi''$, are negligible with respect to the potential ones. This is indeed expected in order to produce an accelerated regime not too dissimilar from $\Lambda$CDM but, first, one can have acceleration driven by purely kinetic terms, and second, acceleration can be produced even with a significant fraction of energy in the  kinetic terms. So slow-roll approximation and QSA should be kept well distinguished. However, in some formula below we will explicitly make use of the slow-roll approximation on top of the QSA.

Let
us emphasize  that the QSA applies only for classically stable
systems. Imagine a scalar field obeying a second-order equation in Fourier space
\begin{equation}
\phi''+F\phi'+c_S^2 k^2\phi=S
\end{equation}
where from now on we use the physical wavenumber 
\begin{equation}
k_{\mathrm{phys}}=\frac{k_{\mathrm{com}}}{a}
\end{equation}
instead of the comoving one, and
where $F,S$  are the friction and the source, respectively, depending in general on the background solution and on other coupled fields. If $c_S^2<0$, the solution $\phi$ will increase asymptotically as $e^{|c_S| k\log a}=a^{|c_S|k}$ and in this case $\phi''\sim |c_S^2| k^2\phi$ will not be negligible with respect to $c_S^2 k^2\phi$ as we assumed in the QSA. 

For simplicity, from now on we assume that the space curvature has
been found to be vanishing, so $|\Omega_{k0}|\ll1$. Using Einstein's field equations and a pressureless perfect
fluid for matter, we can derive from the HL two generalized Poisson
equations in Fourier space, one for $\Phi$
and one for $\Psi$: 
\begin{align}
k^{2}\Phi & =\frac{1}{2}Y(k,z)\eta(z,k)\rho_{m}(z)\delta_{m}(z,k)\,\label{eq:poisson-1-1}\\
k^{2}\Psi & =-\frac{1}{2}Y(k,z)\rho_{m}(z)\delta_{m}(z,k)\,\label{eq:poisson-2-1}
\end{align}
(we remind that in our units $8\pi G_N=1$) where $z$ is the redshift,
$k$ the physical wavenumber,  $\delta_{m}$ the matter density contrast
and $\eta$ and $Y$ are two functions of scale and time that parametrize
deviations from standard gravity. In some papers the function $Y$ is also called $\mu$. Comparing with eqs. (\ref{eq:stan3},\ref{eq:stan4}),  
we see that in Einstein's General Relativity they reduce to $\eta=Y=1$. Clearly, the \emph{anisotropic stress}
$\eta$, or gravitational slip, is defined as
\begin{equation}
\eta=-\frac{\Phi}{\Psi}
\end{equation}
(From now on, all the perturbation quantities are meant to be root-mean-squares
of the corresponding random variables, and therefore positive definite;
we can therefore define ratios like $\eta$). A value of $\eta\not=1$ can be generated
in standard General Relativity only by off-diagonal spatial elements of the energy-momentum
tensor. For a perturbed fluid, these elements are quadratic in the velocity, $T_{ij}\sim \rho v_{i}v_{j}$,
and therefore vanish at first order for non-relativistic particles.
Free-streaming relativistic particles can instead induce a deviation from $\eta=1$:
this is the case of  neutrinos. However, they play a substantial
role only during the radiation era and are negligible today \cite{Weinberg:2003ur}. 
Therefore, $\eta\not=1$ in the late universe means that gravity is
modified, unless there is some hitherto unknown abundant form of hot
dark matter.

In the QSA one can show that for HL \cite{DeFelice:2011hq,Amendola:2012ky}
\begin{align}
\eta & =h_{2}\left(\frac{1+k^{2}h_{4}}{1+k^{2}h_{5}}\right)\,,\,Y=h_{1}\left(\frac{1+k^{2}h_{5}}{1+k^{2}h_{3}}\right)\,,\label{eq:etay-1}
\end{align}
for suitably defined functions $h_{1-5}$ of time alone that depend
only on $K,G_{3,4,5}$. Their full form will be given in Sec. \ref{sec:yukawa} along with
another popular parametrization of the HL equations proposed in \cite{Bellini2014}.
In general, the functions $h_{3,4,5}$ are proportional to $\mu^{-2}$,
where $\mu$ is a mass scale. In the simplest cases $\mu$ corresponds to the standard mass $m$, i.e. the second derivative
of the scalar field potential, plus other terms proportional to $\phi'$ or $\phi''$. These kinetic terms are expected to be subdominant if $\phi$ drives acceleration today or, more in general, during an evolution that is not  strongly oscillating, so often we can assume that $h_{3,4,5}$ scale simply as $m^{-2}$. This approximation will  be  adopted in the explicit expression for $f(R)$ and conformal coupling that are given below. 
If the scalar field drives acceleration
one expects $m$ to be very small, of order $H_{0}\approx10^{-33}$eV. In this case, at the observable sub-horizon scales,
$\eta\to h_{2}h_{4}/h_{5}$ and $Y\to h_{1}h_{5}/h_{3}$. If instead
this scale is of the order of the linear scales that can be directly
observed (e.g. 100 Mpc), then one could observationally detect the
$k$-dependence of $Y,\eta$ and find that at sufficiently large scale,
such that $k\ll m$, $\eta\to h_{2},Y\to h_{1}$.

The same form of $Y,\eta$ can be obtained also in other theories not based on scalars that produce second-order equations of motion, namely, in bimetric models \cite{Konnig:2014dna} and in vector models \cite{DeFelice:2016uil}.

It is worth stressing the fact that the time dependence of $Y,\eta$, expressed by the functions $h_{1-5}$, is essentially arbitrary. Given observations at several epochs, one can always design a HL that exactly fits the data, no matter how precise they are. In contrast, the space dependence, which in Fourier space becomes the $k$ dependence, is very simple and fixed. The reason is that the HL equations are by definition second-order, and therefore contain at most factors of $k^2$. The $k$ dependence is therefore potentially a more robust test for the validity of the HL than the time one. A model with two coupled scalar fields would instead generate for $Y,\eta$ a ratio of polynomials of order $k^4$ (see e.g. \cite{2015PhRvD..92b4009V}).
Clearly, one has to remember that all this is valid at linear scales: if the $k$ dependence is important only at non-linear scales, e.g. for $k>1$ Mpc$^{-1}$, then it might be completely lost. 

Another equivalent form that we will employ often is
\begin{align}
Y\eta & =h_1 h_{2}\left(1+\frac{\alpha_s k^2}{m^2+k^2}\right)\,,\,Y=h_1 \left(1+\frac{\alpha_t k^2}{m^2+k^2}\right)\,.\label{eq:etay-alt}
\end{align}
where
\begin{eqnarray}
\alpha_t &\equiv&(h_{5}-h_{3})/h_{3}\nonumber\\
\alpha_s &\equiv &(h_{4}-h_{3})/h_{3}\label{eq:aam}\\
m^{2}&\equiv &1/h_{3}\nonumber
\end{eqnarray}
This form has a simple physical interpretation. $Y\eta$ is the modifier of the $\Phi$-Poisson equation, just as $Y$ is the modifier of the $\Psi$-Poisson equation. The parameters $\alpha_t, \alpha_s$ are the strengths of the fifth-force mediated by the scalar field for $\Psi$ (the metric {\it time-time} perturbed component) and for $\Phi$ (the metric {\it space-space} perturbed component), respectively. Finally, $m$ is the effective mass of the scalar field, and $\lambda\equiv1/m$ its spatial range. This interpretation will be discussed  in the next section.

A particularly simple case is realized with the  $f(R)$ models, where $f(R)$ is the function of the curvature $R$ that is to be added to the Einstein-Hilbert Lagrangian. In this case in fact 
\begin{align}
\eta & =1-\frac{1}{2}\frac{k^2}{(3/4)m_R^2+k^2}\,,\quad
Y=\frac{1}{1+f_{,R}}\left(1+\frac{1}{3}\frac{k^2}{m_R^2+k^2}\right)\,.\label{eq:etay-fr}
\end{align}
The derivative $f_{,R}$ is often negligible at the present epoch, in order to reproduce a viable cosmology. In this case,  $m^2_R=(3f_{,RR})^{-1}$  and, for large $k$, $\eta\to 1/2$  and $Y\to 4/3$, regardless of the specific  $f(R)$ model.

Another simple case is conformal scalar-tensor theory, with $G_3=G_5=0$, $G_4=F(\phi)/2$, and $K=(1-3\alpha_t)F(\phi)X-V(\phi)$, where $\alpha_t=(F_{,\phi}/F)^2/2$. In this form,  the strength of the fifth force is $\alpha_t$. In this case we have
\begin{equation}
\eta  = 1-\frac{2\alpha_t F k^2}{(1+\alpha_t)Fk^2+M^2}\,,\quad
Y =\frac{1}{F}\left(1+\frac{\alpha_t k^2}{k^2+M^2}\right)\,.\label{eq:etay-conf}
\end{equation}
where $M^2=V_{,\phi\phi}$. When $M$ is vanishingly small, $\eta\to 1-2 \alpha_t/(1+\alpha_t)$ and $Y=(1+\alpha_t)/F$. Comparing with eq. (\ref{eq:etay-fr}), we see that for $f(R)$, $\alpha_t=-\alpha_s=1/3$.

\section{Potentials in real space}

In real space, one can derive the modified Newtonian potential for
a radial mass density distribution $\rho(r)$ by inverse Fourier transformation.
Let us start with eq. (\ref{eq:etay-alt}) 
\begin{equation}
Y=h_{1}\left(1+\frac{\alpha_t k^{2}}{m^{2}+k^{2}}\right),
\end{equation}
 For a \emph{non-linear
static structure} (e.g. the Earth or a galaxy) the local density is
much higher than the background average density, so $\delta_{m}(k)=[\rho_{m}(k)-\rho_{m}]/\rho_{m}\approx\rho_{m}(k)/\rho_{m}$
where $\rho_{m}=\langle\rho_{m}(k)\rangle$ is the background
density and
\begin{equation}
\rho_m(k)=\int \rho(r) e^{i\mathbf{k}\mathbf{r}} d^3 r
\end{equation}
is the Fourier transform of $\rho(r)$. The Poisson equation (\ref{eq:poisson-2-1}) becomes
then
\begin{equation}
k^{2}\Psi=-\frac{1}{2}Y(k)\rho_{m}(k)\label{eq:phiy}
\end{equation}
In real space and for a radial configuration, this reads
\begin{equation}
r^{-2}\frac{\partial}{\partial r}\left(r^{2}\frac{\partial}{\partial r}\right)\Psi=\frac{1}{2}\sigma(r)\label{eq:phireq}
\end{equation}
(since we use the physical $k$, now $r$ refers to the physical distance) where
\begin{equation}
\sigma(r)=\frac{V}{(2\pi)^{3}}\int e^{i\mathbf{k}\mathbf{r}}Y(k)\rho_{m}(k)d^{3}k
\end{equation}
is the inverse Fourier transform of $Y(k)\rho_{m}(k)$ and $V$ is an
arbitrary large volume that encompasses the structure. Assuming that $\Psi$ vanishes at infinity, eq. (\ref{eq:phireq})
has the general solution
\begin{align}
\Psi(y) & =-\frac{h_{1}}{4}\int_{0}^{\infty}dr\int_{-1}^{1}dz\left(\frac{1}{|\mathbf{r}-\mathbf{y}|}+\frac{\alpha_t e^{-m|\mathbf{r}-\mathbf{y}|}}{|\mathbf{r}-\mathbf{y}|}\right)\rho(r)r^{2}\\
 & =\Psi_{N}+\Psi_{Y}
\end{align}
where $z=\cos\theta$ and where $\Psi_{N}$ is the standard Newtonian
potential, while $\Psi_{Y}$ is the Yukawa correction proportional
to $\alpha_t$. This can be solved for any given radial density distribution
$\rho(r)$. For $m\to\infty$ (or $\alpha_t \to0)$ we are back to the
Newtonian case. 

Let us focus now on the modified gravity part. This can be analytically
integrated in some simple cases. We write 
\begin{align}
\Psi_{Y}(y) & =-\frac{h_{1}}{4}\alpha_t\int_{0}^{\infty}dr\int_{-1}^{1}dz e^{-m|\mathbf{r}-\mathbf{y}|}{|\mathbf{r}-\mathbf{y}|}\rho(r)r^{2}\\
 & =-\frac{h_{1}}{4}\alpha_t\int_{0}^{\infty}\rho(r)r^{2}dr\int_{-1}^{1}\frac{e^{-m\sqrt{r^{2}+y^{2}-2ryz}}}{\sqrt{r^{2}+y^{2}-2ryz}}dz\\
 & =-\frac{h_{1}}{4}\alpha_t\int_{0}^{\infty}\rho(r)r^{2}drF(y,r)
\end{align}
where $F$ has two parts
\begin{align}
F_{1}(y,r) & =\frac{e^{-m(r-y)}-e^{-m(r+y)}}{mry}\,\,,r>y\\
F_{2}(y,r) & =\frac{e^{m(r-y)}-e^{-m(r+y)}}{mry}\,\,,r<y
\end{align}
For a mass point at the origin, for instance, one has $\rho(r)=M\delta_{D}^{(3)}(r)$
where $\delta_{D}^{(3)}$ is the Dirac delta function in 3D, defined
for any regular function $f(r)$ as $\int d^{3}rf(r)\delta_{D}^{(3)}(r)=4\pi\int f(r)r^{2}dr\delta_{D}^{(3)}=f(0)$,
and therefore
\begin{align}
\Psi_{Y}(y) & =-\frac{h_{1}}{4}\alpha_t\int_{0}^{\infty}dr\int_{-1}^{1}dz\frac{e^{-m|r-y|}}{|r-y|}\rho(r)r^{2}\\
 & =-\frac{h_{1}}{8\pi}\alpha_t M\frac{e^{-m|y|}}{|y|}
\end{align}
i.e. the so-called Yukawa correction. The total potential is then
\begin{equation}
\Psi(r)=-h_{1}\frac{G_{N}M}{r}(1+\alpha_t e^{-mr})
\end{equation}
where we reintroduced for a moment Newton's constant $G_{N}$. As anticipated, $\alpha_t$ gives the strength of the Yukawa interaction and $\lambda\equiv1/m$ its spatial range. The
prefactor $h_{1}$ renormalizes the product $G_N M$, so that only the product $h_{1}G_N M$
is then observable (beside $\alpha_t,m$). Sometimes $h_{1}G_N$ is denoted
$G_{\mathrm{eff}}$ because it can be seen as a renormalization of Newton's
constant.

A typical dark matter halo can be approximated by a Navarro-Frenk-White profile \cite{Navarro:1995iw}  with scale $r_{s}$ and density
parameter $\rho_{0}$,
\begin{equation}
\rho(r)=\frac{\rho_{0}}{\frac{r}{r_{s}}(1+\frac{r}{r_{s}})^{2}}
\end{equation}
In this case  we have  \cite{2017JCAP...07..023P} 
\begin{align}
\Psi_{Y}(y) & =-2\pi h_{1}\alpha_t\int_{0}^{\infty}\rho(r)r^{2}drF(y,r)\\
 & =\frac{2\pi h_{1}\alpha_t\rho_{0}}{y}r_{s}^{3}[e^{-m(r_{s}+y)}(\mathrm{Ei}(mr_{s})-\mathrm{Ei}[m(r_{s}+y)])\\
 & -e^{m(r_{s}+y)}\mathrm{Ei}[-m(r_{s}+y)]+e^{m(r_{s}-y)}\mathrm{Ei}(-mr_{s})]
\end{align}
where $\mathrm{Ei}(x)$ is the ExpIntegral function,\begin{equation}
\text{Ei}(x)=-\int_{-x}^{\infty}\frac{e^{-t}}{t}dt\;.
\end{equation}

Exactly the same procedure can be applied to the second potential $\Phi$,
which obeys another Poisson equation
\begin{equation}
k^{2}\Phi=\frac{1}{2}Y\eta\rho_{m}\delta_{m}\,
\end{equation}
One has now
\begin{equation}
Y\eta=h_{1}h_{2}\left(1+\frac{\alpha_s k^{2}}{m^{2}+k^{2}}\right)\,
\end{equation}
Notice that the mass
$m$ is the same for $\Psi,\Phi$: there is just one boson, not two.
The real-space expression for $\Phi$ for a point-mass  $M$ is then identical to the one
for $\Psi$ with $\alpha_s$ in place of $\alpha_t$ and $h_{1}h_{2}$
in place of $-h_{1}$,
\begin{equation}
\Phi(r)=h_{1}h_2\frac{G_{N}M}{r}(1+\alpha_s e^{-mr}).
\end{equation}
Finally, the so-called lensing potential $\psi(r)=\Psi(r)-\Phi(r)$  is responsible for the gravitational lensing of source images in the linear regime. In this regime, given an elliptical source at distance $r_s$ characterized by semiaxes of angular extent $\theta^s_i={(\theta^s_x,\theta^s_y)}$, the image we see is distorted by intervening matter into a new set of semiaxes $\theta_j=(\delta_{ij}+D_{ij})^{-1}\theta^s_i$ where the distortion matrix is proportional to $\psi(r)$
\begin{equation}
D_{ij}=\int_0^{r_s} dr' (1-\frac{r'}{r_s})r'\psi_{,ij}
\end{equation}
All observations of gravitational lensing lead therefore ultimately to an estimation of $\psi(r)$. What is observed in practice is the power spectrum of ellipticities, i.e. the correlation of ellipticities of galaxies in the sky due to a non-zero $\psi(r)$ along the line of sight (see e.g.  \cite{Dodelson:2003ft}, chap. 10).

From eqs. (\ref{eq:poisson-1-1},\ref{eq:poisson-2-1}) we see then that
\begin{equation}\label{eq:lensing-psi}
k^{2}\psi  =-\frac{1}{2}Y(k,z)(1+\eta(z,k))\rho_{m}(z)\delta_{m}(z,k)\,.
\end{equation}
In our formalism, the lensing potential in real space amounts then to
\begin{equation}\label{gwpot}
\Psi(r)-\Phi(r)=-h_{1}(1+h_2)\frac{G_{N}M}{r}(1+A_L e^{-mr})\,,
\end{equation}
where
\begin{equation}\label{eq:AL}
A_L=\frac{h_2\alpha_s +\alpha_t}{1+h_2} \,.
\end{equation}
Since $h_{2}$ is in general different from
unity, the mass $M^{(\Psi)}\equiv h_{1}M$ one infers at infinity from the $\Psi$
potential (often called {\it dynamical mass}) is different from the mass $M^{(\Phi)}\equiv h_{1}h_{2}M$
one infers from the $\Phi$ potential or the one from the lensing combination
$\Psi-\Phi$, i.e. $M^{(\Psi-\Phi)}=h_{1}(1+h_{2})M/2$ ({\it lensing mass}). These masses
of course coincide in standard gravity. As we will see below, one
can indeed compare observationally the estimations and extract $\eta$
by taking suitable ratios.

\section{The parameters of the Yukawa correction}

\label{sec:yukawa}
In \cite{Bellini2014} it has been shown that the HL perturbation equations can be entirely written in terms of four functions of time only, $\alpha_{K,B,M,T}$, given as
\begin{align}
M_{*}^{2}\equiv & 2\left(G_{4}-2XG_{4X}+XG_{5\phi}-\dot{\phi}HXG_{5X}\right)\label{eq:planckmass}\\
HM^2_{*}\alpha_{\textrm{M}}\equiv & \dot{(M^2_{*})}\label{eq:am}\\
H^{2}M_{*}^{2}\alpha_{\textrm{K}}\equiv & 2X\left(K_{X}+2XK_{XX}-2G_{3\phi}-2XG_{3\phi X}\right)+\label{eq:ak}\\
 & +12\dot{\phi}XH\left(G_{3X}+XG_{3XX}-3G_{4\phi X}-2XG_{4\phi XX}\right)+\nonumber \\
 & +12XH^{2}\left(G_{4X}+8XG_{4XX}+4X^{2}G_{4XXX}\right)-\nonumber \\
 & -12XH^{2}\left(G_{5\phi}+5XG_{5\phi X}+2X^{2}G_{5\phi XX}\right)+\nonumber \\
 & +4\dot{\phi}XH^{3}\left(3G_{5X}+7XG_{5XX}+2X^{2}G_{5XXX}\right)\nonumber \\
HM_{*}^{2}\alpha_{\textrm{B}}\equiv & 2\dot{\phi}\left(XG_{3X}-G_{4\phi}-2XG_{4\phi X}\right)+\label{eq:ab}\\
 & +8XH\left(G_{4X}+2XG_{4XX}-G_{5\phi}-XG_{5\phi X}\right)+\nonumber \\
 & +2\dot{\phi}XH^{2}\left(3G_{5X}+2XG_{5XX}\right)\nonumber \\
M_{*}^{2}\alpha_{\textrm{T}}\equiv & 2X\left(2G_{4X}-2G_{5\phi}-\left(\ddot{\phi}-\dot{\phi}H\right)G_{5X}\right)\label{eq:at}
\end{align}
This parametrization (collectively called $\alpha_i$) is linked to the physical properties of the HL. Briefly, $\alpha_T$ expresses the deviation of the GW speed from $c$, $c_T^2=1+\alpha_T$; $\alpha_K$ is connected to the field kinetic sector; $\alpha_B$ to the mixing ("braiding") of the scalar and gravitational kinetic terms; $M_{\star}$ is the time-dependent effective reduced Planck mass and $\alpha_M$ its running. They are designed so that $\alpha_i=0$ for $\Lambda$CDM. They do not vanish, in general,  for standard gravity with a non-$\Lambda$CDM background expansion, nor for non-standard gravity with a $\Lambda$CDM expansion.
Several observational limits on these parameters in specific models have already been obtained (see e.g. \cite{Kreisch:2017uet}).

It is clear that  cancellations can occur among terms belonging to different $G_i$ sectors. However, one should distinguish between dynamical cancellations, i.e. involving a particular background solution for $\phi(t),H(t)$, and algebraic cancellations, which only depend on some special choice for the functions $G_i(\phi,X)$. The former ones, if they exist at all and are not unstable, can be guaranteed only for some particular set of initial conditions, and might occur only for some period, unless the solution happens to be an attractor. The algebraic cancellations, however,
 are independent of the background evolution and therefore valid at all times. Therefore, usually only the second class is regarded as an interesting one.
 
We can now express the four coefficients introduced in eq. (\ref{eq:hlsec}) that determine the stability of the HL as
\cite{Bellini:2014fua}
\begin{align}
Q_{\text{S}} & =\frac{M_{*}^{2}(2\alpha_K+3\alpha_B^2)}{(2-\alpha_{\textrm{B}})^{2}}\,,\label{eq:qs}\\
c_{\text{S}}^{2} & =\frac{\left(2-\alpha_{\textrm{B}}\right)\alpha_1+2\alpha_2}{2\alpha_K+3\alpha_B^2}\,,\nonumber \\
Q_{\text{T}} & =\frac{M_{*}^{2}}{8}\,,\label{eq:qt}\\
c_{\text{T}}^{2} & =1+\alpha_{\textrm{T}}\,\nonumber 
\end{align}
where  
\begin{eqnarray}
\alpha_{1}  &\equiv &  \alpha_{B}+\left(\alpha_{B}-2\right)\alpha_{T}+2\alpha_{M}\label{eq:an}\\
\alpha_{2}  &\equiv & \alpha_B\xi+\alpha'_B-2\xi-3(1+w_m)\tilde{\Omega}_m
\end{eqnarray}
and where $\xi=H'/H$ and $\tilde{\Omega}_m=\frac{\rho_m}{3M_{\star}^2H^2}=1-\frac{\rho_{HL}}{3M_{\star}^2H^2}$ (with this last relation one can get rid of $\rho_m$ everywhere). Here "matter" represents all the components beside the scalar field, i.e. baryons, dark matter, neutrinos, radiation. The matter equation of state $w_m=\sum_i w_i\Omega_i/\Omega_m$ is then an effective value for all the matter components.
Note that $c_S^2=1$ in the standard minimally coupled scalar field case $K=X-V(\phi), G_3=G_5=0$ and $G_4=1/2$.

The relation between the "observable" parameters $h_{1-5}$ that enter the Yukawa correction and the "physical" parameters $\alpha_{K,B,M,T}$ is 
\begin{eqnarray}\label{eq:h-alpha}
h_{1} & = & \frac{\alpha_{T}+1}{M_{\star}^2},\\
h_{2} & = & \frac{1}{\alpha_{T}+1},\label{eq:h2}\\
h_{3} & = & \frac{1}{2H^{2}\mu^2}\left(
(2-\alpha_B)\alpha_1+ 2\alpha_2\right)\\
h_{4} & = & \frac{1}{H^{2}\mu^{2}}\left(
\alpha_1+\alpha_2
\right)\\
h_{5} & = & \frac{1}{H^{2}\mu^{2}}\left(\frac{\alpha_{M}+1}{\alpha_{T}+1}\alpha_{1}+\alpha_2\right)
\end{eqnarray}
where\footnote{With respect to the mass defined in \cite{Bellini:2014fua}, we have $\mu^2=M^2/H^2$. }
\begin{eqnarray}
\mu^2&\equiv&-3[2\xi^2+\xi'+\xi(3+\alpha_M)]\alpha_B-3\xi\alpha_2
\end{eqnarray}
Two remarks are in order. First, the quantity $\mu^2$  acts as an effective squared mass in the perturbation equation of motion for $\phi$; we need to assume therefore that it is non-negative to avoid instability below some finite value of $k$.
Second,  the expressions for $\alpha_i$ and $h_i$ are completely general and do not assume the QSA. The QSA is needed only when we connect the theory to observations through $Y,\eta$.

Considering now only pressureless matter, from the background equations in Appendix \ref{sec:appendix} we see that,
\begin{equation}
\xi=-\frac{3}{2}-\frac{p_{HL}}{2H^2M^2_{\star}}=-\frac{3}{2}(1+w_{HL}\tilde{\Omega}_{HL})\label{eq:xiom}
\end{equation}
where $w_{HL}=p_{HL}/\rho_{HL}$. 
In a $\Lambda$CDM background, $2\xi+3\Omega_m=0$, and $\mu^2$ simplifies to $\mu^2=-3\xi(\alpha_M\alpha_B+\alpha_2)=-3\xi(\alpha_M\alpha_B+\alpha_B \xi+\alpha_B'+3(\Omega_m-\tilde\Omega_m))$.
Notice that $\alpha_K$ does not appear in the $h_i$-$\alpha_i$ relation: this means that the kinetic parameter $\alpha_K$ is not an observable in the QSA linear regime. In Sec. \ref{sec:obs} we will discuss which combinations of $\alpha_i$ are really model independent (MI) observables in cosmology.

Assuming Einstein--Hilbert action for the gravitational sector and a canonical kinetic term for the scalar field, we have $M_{\star}^{2}=1$ and $\alpha_{B,M,T}=0$, so $\alpha_1=0$ and 
\begin{eqnarray}
\mu^{2}&=&-9\xi\tilde{\Omega}_{HL}(1+w_{HL})\\
\alpha_2&=& 3\tilde{\Omega}_{HL}(1+w_{HL})
\end{eqnarray}
Therefore $h_1=h_2=1$ and
\begin{equation}
h_{3,4,5}=-\frac{1}{3\xi H^2}=\frac{2}{9H^2(1+w_{HL}\tilde{\Omega}_{HL})}
\end{equation}
so that, as per construction, $Y,\eta\to 1$.

It is worth noticing that the stability conditions $Q_T,Q_S,c_S^2>0$ imply $(2-\alpha_B)\alpha_1+2\alpha_2>0$ and therefore $h_3>0$ if one also requires $\mu^2>0$. As we have seen, $\lambda=\sqrt{h_3}$ is the range of the fifth-force interaction, so it makes sense that it is positive definite for stable systems. 
In the standard Brans-Dicke model with a potential $V(\phi)$, for instance, and neglecting several subdominant kinetic terms,  we have 
\begin{equation}
\mu^2=\frac{3\alpha_M m_{\phi}^2 \phi'}{3H^2}
\end{equation}
where $m_\phi^2=V_{,\phi\phi}\,$, and therefore finally
\begin{equation}
 h_3=\frac{3+2\omega }{2\phi m_\phi^2}
\end{equation}
where $\phi=M^2_*$ (notice that in Brans-Dicke $\phi$ has dimensions {\it mass}$^2$ and therefore $m_{\phi}$ is dimensionless), so the fifth-force range is
\begin{equation}
\lambda=m^{-1}=(m_\phi M_*)^{-1}\sqrt{\frac{3+2\omega}{2}}
\end{equation}

Assuming a $\Lambda$CDM expansion and $\alpha_T=0$, the conditions for stability during the matter era simplify to    
$\alpha_K>-3\alpha_B^2/2$ and 
\begin{equation}(2-\alpha_B)(\alpha_B+2\alpha_M+3\Omega_m) +2\alpha_B'-6\tilde\Omega_m >0\label{eq:stab}
\end{equation} 
Generalizing, we have that for a background parametrized by a (possibly time-dependent)  EOS $w_{HL}$ and for matter with an effective $w_m$, one has 
\begin{equation}(2-\alpha_B)(\alpha_B+2\alpha_M)-3[1+w_{HL}+(w_m-w_{HL})\tilde\Omega_m]\alpha_B+2\alpha_B'+6(w_{HL}+1)(1-\tilde\Omega_m)>0\label{eq:stab1}\end{equation}

To these stability conditions, arising from Eqs. (\ref{eq:qs}) and (\ref{eq:qt}), one should add the requirement that the friction term in the perturbation equations for $\delta\phi$, or equivalently, for the gravitational potentials $\Phi,\Psi$, is positive. This condition is quite milder than those from Eqs. (\ref{eq:qs}) and (\ref{eq:qt}). While a negative $c_s^2$, for instance, even for a short period, induces a unbounded growth for $k\to\infty$, a negative friction term typically leads to a power-law growth $a^p$, which might be a problem only if it lasts for too long. 
However, in order to obtain the friction instability condition one should carefully investigate the existence of growing modes also when the various coefficient are time-dependent and no simple criteria have been identified so far. Therefore we just quote the condition for negative friction (i.e. stability) for the gravitational waves, best obtained by writing down the equation in conformal time, since in this case the $k^2$ term is time-independent (provided $\alpha_T =const$). The condition is simply $\alpha_M>-2$.


From the $h_i-\alpha_i$ relations \eqref{eq:h-alpha}  we can derive  the Yukawa strengths
\begin{eqnarray}
\alpha_s=\frac{h_{4}-h_{3}}{h_3} & = & \frac{\alpha_{1}\alpha_{B}}{(2-\alpha_B)\alpha_1+ 2\alpha_2}\nonumber\\
\alpha_t=\frac{h_{5}-h_{3} }{h_3}& = & \frac{\alpha_{1}^{2}}{((2-\alpha_B)\alpha_1+ 2\alpha_2)\left(\alpha_{T}+1\right)}\label{streng}
\end{eqnarray}
The Yukawa strength $\alpha_t$ is always positive, and therefore  the fifth force is attractive, if $Q_T,Q_S,c_S^2>0$. We also notice that if $\alpha_M=\alpha_T=0$, then $\alpha_1=\alpha_B$ and the two strengths become equal, and $h_2=1$. Therefore  $\Psi=-\Phi$ and, finally, $\eta=1$, even if both potentials do actually have a non-vanishing Yukawa correction, so that $Y\not=1$. In order for the parameters $\alpha_M,\alpha_T$ to vanish, the gravity sector of the HL must be standard,  $G_4=const,G_5=0$, barring the case of accidental dynamical cancellation for some particular background evolution. Therefore, we conclude that  $\eta\not=1$ implies, and is implied by, modified gravity, at least when matter is represented by a perfect fluid \cite{Saltas:2014dha}. One cannot make a similar statement for $Y$. This is a crucial statement for what follows. Notice however that, as we show below, although modified gravity implies $\eta\not=1$, a value $\eta=1$ does not necessarily implies standard gravity, but only scale-free gravity, at least at the quasi-static level. In ref. \cite{Sawicki:2016klv} it has been shown that $\eta=1$ at all scales implies indeed standard gravity.

We can draw more conclusions from eqs. (\ref{streng}).
\begin{itemize}
\item
The two strengths $\alpha_t,\alpha_s$ are equal also if $\alpha_M=\alpha_T$. In this case, $\eta=(1+\alpha_T)^{-1}$ and has no scale dependence.

\item
The $k\to\infty$ limit
of the modified gravity parameters (provided we are still in the linear regime) is
\begin{eqnarray}
 Y_{\infty}&=&
 \frac{h_1 h_5}{h_3}=\frac{2}{M_{\star}^2}\frac
 {\alpha_1(1+\alpha_M)+\alpha_2(1+\alpha_T)}{2\alpha_2+\alpha_1(2-\alpha_B)}\\
\eta_{\infty} &=&  \frac{h_2 h_4}{h_5}
=\frac{\alpha_1+\alpha_2}{\alpha_1(1+\alpha_M)+\alpha_2(1+\alpha_T)}
\end{eqnarray}
This coincides with eqs. (4.9) of \cite{Bellini:2014fua}. If $\alpha_T=0$ then 
\begin{eqnarray}
 Y_{\infty}&=&
\frac{1}{M_{\star}^2}\left[1+\frac
 {(2\alpha_M+\alpha_B)^2}{2\alpha_2+(2\alpha_M+\alpha_B)(2-\alpha_B)}\right]\\
\eta_{\infty} &=& 
1-\frac{(2\alpha_M+\alpha_B)\alpha_M}{(2\alpha_M+\alpha_B)(1+\alpha_M)+\alpha_2}
\end{eqnarray}
It turns out that if one imposes stability, $c_s^2>0$, then $Y$ is always larger than, or equal to,  $1/M_{\star}^2$, so that matter perturbations in Horndeski with $\alpha_T=0$ always grow faster, in the quasi-static regime, than any standard gravity model with the same $M_{\star}$ and the same background. It also follows that the lensing combination that appears in Eq. (\ref{eq:lensing-psi}) amounts to
\begin{equation}
\Sigma\equiv Y(1+\eta)=2\left[1+\frac{(2\alpha_M+\alpha_B)(\alpha_B+\alpha_M)}{2\alpha_2+(2-\alpha_B)(2\alpha_M+\alpha_B)}\right]
\end{equation}
Since the denominator has to be positive for stability, the sign of the effect on the gravitational lensing depends only on $\alpha_M,\alpha_B$.

\item
The  Yukawa corrections disappear completely if $\alpha_1=0$, i.e. for
\begin{equation}
\alpha_B=2\frac{\alpha_T-\alpha_M}{1+\alpha_T} 
\end{equation} 
This is therefore the general condition to have a scale-free gravity, corresponding to $h_3 =h_4=h_5$\footnote{We recently noticed that this relation was first provided in an unpublished draft by Mariele Motta in early 2016}. If we also assume $\alpha_T=0$ and consequently $G_{4X}=G_5=0$ ({\it conformal coupling}) in the HL, as required by the GW speed constraints we discuss in Sec. \ref{sec:GW}, it follows $\alpha_B=-2\alpha_M$ \cite{2017PhLB..765..382L,2018JCAP...03..005L}\footnote{Note that in Ref. \cite{2017PhLB..765..382L} $\alpha_B$ is defined as our $-\alpha_B/2$} and
\begin{equation}
G_{4\phi} =-XG_{3X}
\end{equation}
which gives an algebraic cancellation for $G_3=-f'(\phi)\log X$ and $G_4=f(\phi)$. In this particular model, the local gravity experiments would not detect a Yukawa correction even if gravity  actually couples to the scalar field. Gravity becomes then {\it scale free}. The Planck mass would still vary with time, though. So in this model $\eta\to 1$ even if gravity is actually modified. Assuming a $\Lambda$CDM background, for this model to be stable, $c_s^2>0$ implies the condition $(\alpha_B H)'>0$. For $\alpha_B$ constant or slowly-varying, the stability condition amounts to $\alpha_B<0$, so $\alpha_M>0$ and therefore $Y$, or the effective Newton's constant, will decrease with time. A larger $Y$ in the past means faster perturbation growth for the same $\Omega_m$. Once again, however, since $\Omega_m$ is not a MI observable quantity, whether this means that perturbations grow faster than in $\Lambda$CDM or not is a model-dependent statement.
\item
From eq. (\ref{eq:lensing-psi})  we find also that the lensing potential lacks a Yukawa term whenever $A_L=0$, defined in \eqref{eq:AL}, i.e. $h_2\alpha_s+\alpha_t=0$, which amounts to
\begin{equation}
\alpha_1\alpha_B+\alpha_1^2=0
\end{equation}
Then we see that $A_L=0$ not only when  $\alpha_1=0$, but also for $\alpha_1=-\alpha_B$. Again imposing the GW speed constraint, this becomes  $\alpha_B=-\alpha_M$. On the HL functions, this implies
\begin{equation}
G_{3X}=0
\end{equation}
which actually means that the $G_3$ sector, after an integration by parts, can be absorbed in $K(\phi,X)$. So for the conformal coupling and when the $G_3$ term is absent or does not depend on $X$, the lensing potential becomes simply twice the standard Newtonian potential
\begin{equation}\label{gwpot}
\Psi(r)-\Phi(r)=2h_{1}\frac{G_{N}M}{r}\,.
\end{equation}
This means radiation, being conformally-invariant,\footnote{The electromagnetic Lagrangian $\sqrt{-g}F_{\alpha\beta}g^{\beta\mu}g^{\alpha\nu}F_{\mu\nu}$ does not change for $g_{\mu\nu}\to f(\phi)g_{\mu\nu}$.} does not feel the modification of gravity, except for the overall factor $h_1$ which, if time dependent, induces a time-dependent mass or Newton's constant.
\item
In the same case as above, $\alpha_B=-\alpha_M$ and $\alpha_T=0$, one has
\begin{equation}\label{gwpot2}
\alpha_t=\frac{\alpha_M^2}{c_s^2 (2\alpha_K+3\alpha_M^2)}\,
\end{equation}
which becomes
\begin{equation}\label{gwpot3}
\alpha_t=\frac{1}{3 c_s^2 }\,
\end{equation}
when the kinetic component $\alpha_K$ is small. Similarly, $\alpha_s=-1/(3c_s^2)$. For $c_s=1$ one obtains a Yukawa strength of $1/3\,(-1/3)$ for the $\Psi \,(\Phi)$ potential. This case is exactly realized for the $f(R)$ models.
\item
Finally, in the uncoupled case $\alpha_M=\alpha_T=0$, in which only the kinetic sector of the scalar field is modified, one has that $\alpha_s=\alpha_t>0$, so that there is a Yukawa correction, but $\eta=1$ at all quasi-static scales.
\end{itemize}

\section{Local tests of gravity}

Gravity has been tested since a long time in the laboratory and within
the solar system (see e.g. \cite{Will:1993ns}). The generic outcome of these
experiments is that Einsteinian gravity works well at all the scales
that have been probed so far. In many experiments one assumes the existence
of the same type of "fifth-force" Yukawa correction to the static Newtonian potential predicted by the HL model,
\begin{equation}
\Psi(r)=-\frac{G_{N}M}{r}(1+\alpha e^{-mr})\label{eq:yuk}
\end{equation}
(here we drop the subscript from $\alpha_t$ since we need consider only $\Psi$; moreover, any overall parameter can be absorbed in $G_NM$). Current limits on $\alpha$ and $\lambda=1/m$ have been obtained in a range of scales from micrometers to astronomical units. The constraints on the strength $\alpha$  obviously weakens for very small $\lambda$.
To give an idea, at the smallest scales probed in laboratory, one has \cite{2003ARNPS..53...77A} $|\alpha| \le 10^{6}$ at $\lambda \sim 10^{-5}$m
and $|\alpha|\le 10^{-2}$ at $\lambda \sim 10^{-3}$m (Casimir-force experiments probe even shorter scales, but the constraints on $|\alpha|$ get correspondingly weaker). At planetary scales, one has $|\alpha|\le 10^{-6}$ for $\lambda\sim 10^6$m (Earth-Moon distance), and $|\alpha|\le 10^{-8}$  at $\lambda\sim 10^{11}$m (planetary orbits). Beyond this distance, the constraints from direct tests of the Newtonian $1/r$ potential weaken again. 

However, the scalar field  responsible for the Yukawa term induces also two post-Newtonian corrections to the Minkowski metric. For a mass distribution with velocity field  $v_i(\mathbf{x},t)$ and density distribution $\rho(\mathbf{x},t)$, we define  $U$  as the potential that solves the standard Poisson equation for non-relativistic particles, i.e. \cite{Will:1993ns}
\begin{equation}
U=\int \frac{\rho(\mathbf{x},t)}{|\mathbf{x}-\mathbf{x}'|}d^3x'
\end{equation}
and $V_i$ as  a velocity-weighted potential 
\begin{equation}
V_i=\int \frac{\rho(\mathbf{x},t)v_i}{|\mathbf{x}-\mathbf{x}'|}d^3x'
\end{equation}
Then we can write down the parametrized post-Newtonian metric as follows
\begin{eqnarray}
g_{00}=& -1+2U-2(1+\beta) U^2\\
g_{0i}=& -\frac{1}{2}(3+4(1+\gamma))V_i\\
g_{ij}=& (1+2(1+\gamma) U)\delta_{ij}
\end{eqnarray}
(the full post-Newtonian metric includes several other terms which however are not excited by a conformally coupled scalar field, see e.g. \cite{2012PhR...513....1C}). Clearly, $\gamma=\beta=0$ produces the standard weak-field metric.
Taking the extreme case of $\lambda\to\infty$, one has 
\begin{eqnarray}
\beta=& \frac{1}{2}\frac{\beta_0\alpha }{(1+\alpha)^2}\\
\gamma=&-2\frac{\alpha}{(1+\alpha)}
\end{eqnarray}
where  $\beta_0=d\sqrt{\alpha(\phi)}/d\phi$. The parameter $1+\gamma$ can be seen as the local-gravity analogue of the anisotropic stress $\eta$, both being the ratio of $(g_{ii}-1)/(g_{00}+1)$ at linear level.

Local tests of gravity can therefore measure the Yukawa correction for both $\Phi,\Psi$, i.e.  $\alpha_t,\alpha_s$ and $\lambda$, and the ratio $\Phi/\Psi$, in a model-independent way. 
The parameter $|\gamma|$, for instance, is constrained to be less than $10^{-5}$ \cite{Patrignani:2016xqp}, inducing a similar constraint on $\alpha$ at large scales. A similar constraint applies also to $\beta$. With such a small strength, there would hardly be any interesting effect in cosmology.

However, all these tests are performed within a limited range of scales,
both spatial and temporal. Moreover, the tests are performed with (some of the) standard
matter particles and not with, say, dark matter. Therefore, they are
completely escaped if standard model particles do not feel modified
gravity, for instance because the scalar field that carries the modification
of gravity does not couple to them or because of screening effects, as we discuss next.

So far we considered only linear scales. At strongly non-linear scales,
e.g. in the galaxy or in the solar system, the effects of modified
gravity depend on the actual configuration of the scalar field. If
such a configuration is static and homogeneous within a scale $r_{s}$,
then the effects of modified gravity can be screened within $r_{s}$,
since they are proportional to the variation of $\phi$. This is the
so-called chameleon effect \cite{Khoury:2003rn,Khoury:2003aq}. On the other hand, screening
can occur also because of non-linearities in the kinetic part of the
Klein-Gordon equation: this is the Vainshtein effect \cite{Vainshtein:1972sx,Babichev:2013usa}.
Finally, a third mechanism appears if the coupling $\alpha$ sets on a vanishing value in structures (high density regions), via a symmetry restoration, while being different from zero at the background (low density) \cite{Pietroni:2005pv,Olive:2007aj,Hinterbichler:2010es,Hinterbichler:2011ca}.
In all cases, the strong deviation from standard gravity that we
might see in cosmology are no longer visible by local experiments.
In this sense, one can always build models that escape the local gravity
constraints. This can be achieved also by assuming the baryons are 
completely decoupled from the scalar field.

In the light of these arguments, let us consider for instance in more detail the constraint on $G_N$ associated with the big bang nucleosynthesis (BBN), sometimes quoted as one of the most stringent cosmological bound. The yields of light elements during the primordial expansion depends on the baryon-to-photon constant ratio $\eta_b$ and on the cosmic expansion rate during nucleosynthesis, which in turn depend on $G_N$ at that time and on various standard model parameters. Fixing the standard model parameters and estimating $\eta_b$  by CMB measurements, one can find constraints on $G_N(t_{\textrm{BBN}})/G_N(t_0)\approx 1\pm 0.2$ \cite{2004PhRvL..92q1301C} by comparing the predicted abundances with the observed ones, for instance deuterium in quasar absorption systems.  This  means that $G_N$ at nucleosynthesis was close to $G_N$ on Earth today. The easiest explanation, that $G_N$ did not vary at all or anyway less than 0.2 throughout the expansion,  implies $|\alpha_M(t_0)|<0.2(H_0 T_0)^{-1}$ (equal to $\approx 0.2$ in $\Lambda$CDM), where $T_0$ is the cosmic age.  However, $G_N$ in the solar system might be screened, as we have mentioned, and therefore equal to the "bare" $G_N$ of standard gravity. Therefore,
any model which is standard general relativity  in the early Universe, like essentially all models built to explain present day's acceleration, will automatically pass the BBN constraint.
Moreover, one should  notice that this constraint depends on a estimate of $\Omega_b h^2$ from CMB that assumed $\Lambda$CDM. Also, $G_N$ is in fact degenerate with the number $g_*$ of relativistic degrees of freedom at nucleosynthesis, so that the bound applies to $G_N g_*$ rather than to $G_N$ alone.  Finally, a simultaneous change of the other standard-model parameters might weaken considerably the constraint, see \cite{2007PhRvD..76f3513D,2009PhR...472....1I}. 

\section{The impact of gravitational waves}

\label{sec:GW}
The Horndeski model predict an anomalous propagation speed $c_{T}$
for gravitational waves (or rather, $c_{T}/c$), since the scalar
field is coupled in a non-conformal way. As already mentioned, one has \cite{Amendola:2012ky,Bettoni:2016mij}, 
\begin{equation}
c_{T}^{2}=1+\alpha_T\label{eq:ct}
\end{equation}
The almost simultaneous detection of GWs and the electromagnetic counterparts
tells us that within $40$ Mpc (at $z\sim0.008$) from us, GWs propagate essentially
at the speed of light \cite{LIGO:2017qsa}. Since the signals arrived
within $1$s difference and light took $10^{15}$ s to reach us, we
have that $|c_{T}^{2}/c^{2}-1|<10^{-15}$. Such tight constraint immediately
ruled out most of the scalar-tensor theories containing derivative
couplings to gravity or at least those models which show this effect
in the nearby universe (in cosmological scales) \cite{Lombriser:2015sxa,Lombriser:2016yzn,Ezquiaga:2017ekz,Creminelli:2017sry,Baker:2017hug,Sakstein:2017xjx}.
That is, we need to have $G_{5}=0$ and $G_{4,X}=0$. In other words, the surviving
Lagrangian has arbitrary $K,G_{3}$ but vanishing $G_{5}$ and $X$-independent
$G_{4}$. This kind of Lagrangian is just a form of Brans-Dicke gravity (plus a scalar field potential and a non-canonical kinetic term). It is also equivalent to standard gravity with matter conformally coupled to a scalar field, i.e. coupled to a metric $\hat g_{\mu\nu}=f(\phi)g_{\mu\nu}$. 
A dynamical cancellation among the terms depending on $G_{5}$ and
$G_{4,X}$ appears extremely fine-tuned. A possible way out is to design a model with
an attractor on which the conformal coupling holds, as proposed in \cite{Amendola:2018ltt}. In this case after the attractor is reached we measure $c_T=1$, but this does not have to be true in the past. Deviations from the speed of light in the past could be detected in B-mode CMB polarization \cite{Amendola:2014wma}.

The constraints on $c_{T}$ also affect directly $\eta$. From eq. (\ref{eq:h2}) one has
in fact 
\begin{equation}
h_{2}=\frac{1}{c_{T}^{2}}
\end{equation}
Hence, the GW constraint $h_{2}=1$ implies that $\eta$ should also
be equal to unity for sufficiently large scales (small $k$)~\cite{Amendola:2017orw},
i.e. it should recover its General Relativity value. The obvious
exception are theories \emph{without} a mass scale beside the Planck
mass \cite{Nersisyan:2018auj}, in which case $\eta=h_{4}/h_{5}$ at all
scales. On the other hand, no obvious GW constraint affects $Y$.

Gravitational waves might in principle measure another HL parameter, the running of the Planck mass, $\alpha_M$. In fact, as it has been shown for instance in \cite{Saltas:2014dha}, the GW amplitude $h$  obeys the equation 
\begin{equation}
\ddot{h}+(3+\alpha_{M})H\dot{h}+c_T^2 \frac{k^{2}}{a^2}h=0.\label{gweq}
\end{equation}
Assuming $c_T=1$, this equation in the sub-horizon limit is solved by \cite{Amendola:2017ovw}
\begin{equation}
h_a = \left(\frac{M_{*\text{,em}}}{M_{*\text{,obs}}}\right) \times h_s \,,
\end{equation}
where the prefactor is the ratio of the Planck mass values at emission and at observation, and  $h_s$  is the standard amplitude expression that, 
for merging binaries, can be approximated as
(see e.g. \cite{maggiore2008gravitational}, eq. 4.189)
\begin{equation}
h_s=\frac{4}{d_L}\left(\frac{G\mathcal{M}_c}{c^2}\right)^{5/3}\left(\frac{\pi f_{\rm GW}}{c}\right)^{2/3}. \label{hgw2}
\end{equation}
Here,  $d_L$ is the luminosity distance, $\mathcal{M}_c$ the so-called chirp mass and $f_{\rm GW}$ the GW frequency measured by the observer.

GWs in standard gravity can measure the luminosity distance $d_L$ because the chirp mass and the frequency can be independently measured by the interferometric signal. In modified gravity, what is really measured is therefore a GW distance \cite{Nishizawa:2017nef,Amendola:2017ovw,Belgacem:2017ihm}
\begin{equation}
d_{GW}=\left(\frac{M_{*\text{,obs}}}{M_{*\text{,em}}}\right)d_L
\end{equation}
Comparing this with an optical determination of $d_L$ leads to a direct measurement of $M_{\star}$ at various epochs, and therefore of $\alpha_M$.

It is however likely that both the emission and the observation occur in heavily screened environments. In this case, $M_{\star}$ is the same at both ends, and no deviation from $d_L$ would be observed. If emission  occurs in a partially unscreened environment, then one should see instead some deviation, although not necessarily connected to the cosmological, unscreened, value of $\alpha_M$.

\section{Model dependence}

\global\long\def\amt{}
 \global\long\def\lcdm{\Lambda CDM}
 \global\long\def\sct{}
 \label{sec:mod-dep}
 The standard model of cosmology, $\Lambda$CDM, is amazingly simple. It consists
of a flat, homogeneous and isotropic background space with perturbations
that, at scales above some Megaparsec, have been evolving linearly
until recently. The initial conditions for perturbations are set by the inflationary mechanism, and provide an initially linear and scale-invariant  spectrum of scalar, vector, and tensor perturbations, that is,  power-law spectra $k^{n_x}$, where $x$ stands for the three types of perturbations that can be excited in General Relativity. These are encoded in a spin-2 massless field, that mediates gravitational interactions via Einstein's
equations. The energy content is shared
among relativistic particles, photons, and quasi-relativistic ones,
neutrinos, pressureless ``cold'' dark matter particles, standard
model particles (``baryons''), and a cosmological constant.
The density of photons can be directly measured via the CMB temperature:
it amounts to $0.005\%$; the density of neutrinos depends on their mass
 and is known to be less than 1\% of the total content today.
Therefore, today, only the last three components are important. The
density of baryons can be fixed by the primordial Big Bang Nucleosynthesis \cite{Cyburt:2015mya}. Since the space curvature has been measured (although so far only in a model-dependent way) to be negligible, only a single parameter is left free,
 the present fraction of the total energy density in pressureless  matter, $\Omega_{m0}$. The fraction in the cosmological constant is then $\Omega_{\Lambda 0}=1-\Omega_{m0}$.

With this one free parameter, the fraction of energy in the cosmological
constant, $\Omega_{\Lambda}\approx0.7$, $\Lambda$CDM fits all the
current cosmological data: the cosmic microwave background (CMB),
the weak lensing data, the redshift distortion data (RSD),
the distance indicators (supernovae Ia, SNIa; baryon acoustic oscillations,
BAO; cosmic chronometers, CCH; gravitational waves, GW).

There are actually a few discrepancies. Two in particular seem to
be more robust. The first is the value of $H_{0}$ obtained through
local measurements, in particular through Cepheids, $H_{0}=(73.45\pm1.66)$
km/s/Mpc \cite{2018ApJ...855..136R}, independent of cosmology, that
deviates from the Planck \cite{2016A&A...594A..13P} value obtained
through an extrapolation from the last scattering epoch performed
assuming $\Lambda$CDM, $H_{0}=(67.51\pm0.64)$ km/s/Mpc \cite{2016A&A...594A..13P}.
The second one is the level of linear matter clustering embodied in
the normalization parameter $\sigma_{8}$: here again, the value from
CMB ($\sigma_{8}=0.82\pm0.014)$ \cite{2016A&A...594A..13P} differs
from the late-universe value delivered by weak lensing, $\sigma_{8}=0.745\pm0.039$
\cite{2017MNRAS.465.1454H}, and by RSD data \cite{Barros:2018efl}
$\sigma_{8}=0.75\pm0.024$.

Another source of discrepancies is related to the dark matter clustering \cite{Weinberg:2013aya}. Dark matter-only simulations fail at reproducing some of the observed properties of the DM distribution. Although the inclusion of baryon physics may solve this, so far there is no conclusive statement and some of these issues may in fact be due to modification of gravity.

These conflicting results already display a basic problem of cosmological
parameter estimation, namely the fact that it is very often model-dependent.
The Planck satellite estimates of the cosmological parameters, from $\Omega_{m}$
to $h$, from the equation of state of dark energy $w_{0}$ to the
clustering amplitude $\sigma_{8}$, can be obtained only by assuming,
among others, a particular model of initial conditions (inflation)
and of later evolution ($\Lambda$CDM). For instance, if we assume
$w_{0}=-0.9$ instead of the cosmological constant value $w_{0}=-1$,
one obtains $H_{0}\approx65.5$ km/sec/Mpc(\cite{2016A&A...594A..13P},
fig. 27), outside the error range given above.

Another example of model-dependency comes from distance indicators
and the dark energy equation of state. Cosmological distance indicators,
whether based on SNIa, BAO or other, measure basically the comoving
distance
\begin{equation}
r(z)=\frac{1}{H_{0}\sqrt{\Omega_{k0}}}\sinh\left(\sqrt{\Omega_{k0}}\int_{0}^{z}\frac{d\hat{z}}{E(\hat{z})}\right)\label{eq:rz}
\end{equation}
where $\Omega_{k0}=-k/H_{0}^{2}$ is the present amount of spatial
curvature $k$ expressed as a fraction of total energy density. We
see that $r(z)$ depends only on $H_{0},\Omega_{k0}$ and $E(z)=H(z)/H_{0}$.
However, since distance indicators depend on the assumption of a standard
candle or ruler or clock, whose absolute value we do not know, the
absolute scale of $r(z)$, that is $H_{0}$, cannot be measured (except for "standard sirens", i.e. gravitational waves, \cite{Abbott:2017xzu}). Assuming
for simplicity that $\Omega_{k0}=0$, the only direct observable is
$E(z)=H(z)/H_{0}$. If we neglect also radiation (a very good approximation
for observations at redshift less than a few) and assume that beside
pressureless matter we have a dark energy component with EOS $w(z)$,
we have 
\begin{equation}
E^{2}(z)=\Omega_{m0}(1+z)^{3}+(1-\Omega_{m0})(1+z)^{3(1+\bar{w})}\label{eq:ez}
\end{equation}
where 
\begin{equation}
\bar{w}(z)=\frac{1}{\ln(1+z)}\int_{0}^{z}\frac{w(\hat{z})}{1+\hat{z}}d\hat{z}
\end{equation}
We can then invert the relation (\ref{eq:ez}) and obtain 
\begin{equation}
w(z)=\frac{2(1+z)EE_{,z}-3E^{2}}{3E^{2}-3\Omega_{m0}(1+z)^{3}}
\end{equation}
(here $E_{,z}$ means differentiation with respect to redshift). It
appears then that in order to reconstruct $w(z)$ one needs to know
$\Omega_{m0}$, beside $E(z)$. For instance, if the true cosmology
is $\Lambda$CDM with $\Omega_{m0}=0.3,$ and we assume erroneously
that $\Omega_{m0}=0.31$, we would infer $w(z=0)=-0.986$ and $w(z=1)=-0.897$,
way different from the true value $-1$.

The problem is that $\Omega_{m0}$ is not a model-independent observable.
Whenever an estimate of $\Omega_{m0}$ is given, e.g. from CMB or
lensing or SNIa, it always depends on assuming a model. The reason
is that there is no way, with phenomena based on gravity alone (clustering
and velocity of galaxies, lensing, integrated Sachs-Wolfe, etc) to distinguish between
various components of matter, since matter responds to gravity in a
universal manner, unless one breaks the equivalence principle (see
the ``dark degeneracy'' of Ref. \cite{Kunz2009}). So in order to
measure $w(z)$ one has to assume a model, that is, a parametrization,
even with extremely precise measurements. For instance, if $w(z)=w_{0}+w_{a}(1-a)$,
then we reduce the complexity to just two parameters, and a measurement
of $E(z)$ at at least three different redshifts can fix simultaneously
$w_{0},w_{a},\Omega_{m0}$. Without a parametrization, $w(z)$ cannot
be reconstructed. With a parametrization, the result depends on the
parametrization itself.

On the other hand, it is clear that we can always perform {\it null tests} on $w(z)$, as for most other cosmological parameters. That is, we can assume a specific $w(z)$, e.g. $w=-1$, and test whether it is consistent with the data. In this case, in flat space, one needs just three distance measurements at three different redshifts, since there are only two parameters, $H_0$ and $\Omega_{m0}$. If the system of three equations in two parameters has no solution, the $\Lambda$CDM model is falsified.  While it is relatively easy to test, i.e. falsify, a model of gravity, it is much more complicate to measure the properties of gravity in a way that does not demand too many assumptions. This explains why the title of this paper mentions "measuring", and not "testing,"  gravity.

The rest of the paper will discuss what kind of model-independent
measurements we can perform in cosmology, with emphasis on 
parameters of modified gravity. As it is obvious, one cannot claim absolute model independence.
The point is rather to isolate clearly which are the assumptions,
and see how far can one reach with a minimum amount of them. In the
following, we will assume this set of assumptions:

\emph{a}) the Universe is well-described by a homogeneous and isotropic
geometry with small (linear) perturbations;

\emph{b}) gravity is universal;

\emph{c}) standard model particles behave from inflation onwards in
the same way as we test them in our laboratories;

\emph{d}) dark matter is ``cold''.

One can replace the last statement with the assumption that we know
the equation of state and sound speed of dark matter, provided it
is not relativistic, and that the fluid remains barotropic, i.e. $p=p(\rho)$, as we will show later on. Unless otherwise specified, however, for the rest of this work we assume pressureless,
cold dark matter.

Notice that we are not assuming any particular form of gravity, standard or otherwise: in fact, we refer
to ``gravity'' as to one or more forces that act universally and
without screening, at least beyond a certain scale. So we include
in our treatment also gravity plus one or more scalar, vector and
tensor fields. Later on we will use the Horndeski generalized scalar-tensor
model for a specific example but the methods discussed here are not restricted to this case.

\section{Model-independent determination of the homogeneous and isotropic
geometry}

What we observe in cosmology are redshifts and angular positions of sources.  What we need to build and test models, however, are distances.
Can we convert redshifts and angles into distances in a model-independent
(MI) way? If this turns out not to be possible then there is no reason
to continue our investigation to the perturbation level. Fortunately
it appears we can.

The FLRW metric of a homogeneous and isotropic Universe in spherical
coordinates is 
\begin{equation}
ds^{2}=-dt^{2}+a^{2}(t)\left[\frac{dr^{2}}{1-kr^{2}}+r^{2}d\theta^{2}+r^{2}\sin^{2}\theta d\phi^{2}\right]
\end{equation}
where $a(t)$ is the scale factor normalized at present time to $a(t_{0})=1$.
If we measure $s,t,r$ in units of the natural scale length $H_{0}^{-1}$,
the metric can be rewritten as 
\begin{equation}
ds^{2}=-dt^{2}+a^{2}(t)\left[\frac{dr^{2}}{1+\Omega_{k0}r^{2}}+r^{2}d\theta^{2}+r^{2}\sin^{2}\theta d\phi^{2}\right]
\end{equation}
The value of $\Omega_{k0}$ has been estimated by Planck to be extremely
small, $|\Omega_{k0}|<0.004$ (\cite{2016A&A...594A..13P}, tab. 5,
last column) but, again, this is a model-dependent estimate, and for
now we consider it as a free parameter. We see than up to an overall
scale, the FLRW metric depends only on $\Omega_{k0}$ and on $E(z)=\dot{a}/a$,
from which $a(t)$ is obtained by inverting 
\begin{equation}
t-t_{0}=\int_{1}^{a}\frac{d\bar{a}}{\bar{a}E(\bar{a})}
\end{equation}
(where again $t$ is in units of $H_{0}^{-1}$).

Baryon-acoustic oscillations are the remnant of the primordial pressure
waves propagating through the plasma of baryons and photons before
their decoupling. By assumption $c)$, we assume their interaction
at all times is the same as in our laboratories. Therefore, we can
predict that the comoving scale of the BAO today is a constant $R$
independent of the redshift at which it is observed. For instance,
in $\Lambda$CDM, $R$ (in units of $H_{0}^{-1}$) is equal to
\begin{equation}
R=\frac{4}{3}\sqrt{\frac{\Omega_{\gamma0}}{\Omega_{m0}\Omega_{b0}}}\ln\left(\frac{\sqrt{R_{s}(z_{drag})+R_{s}(z_{eq})}+\sqrt{1+R_{s}(z_{drag})}}{1+\sqrt{R_{s}(z_{eq})}}\right)
\end{equation}
where the indexes $\gamma,b,m$ refer to radiation, baryons, and dark matter,
respectively; moreover, $R_{s}(z)=(3\Omega_{b0}/4\Omega_{\gamma0})/(1+z)$,
$z_{drag}\approx1000$ is the redshifts of the \emph{drag epoch} (see
the numerical formula given in \cite{1998ApJ...496..605E}), and $z_{eq}=2.396\times10^{4}\Omega_{m0}h^{2}\approx3200$
is the redshift at equivalence. The value $R$ can be used as a standard
ruler: as for SNIa, we do not need to know $R$, but just to assume
that it is constant. Therefore, we can search in the clustering of
galaxies for such a scale, in particular by identifying a peak in
the correlation function. The angle under which we observe $R$ gives
us the 'transverse BAO'. In turn, this angle gives us the dimensionless
angular diameter distance 
\begin{equation}
H_{0}d_{A}(z)\equiv\frac{R}{\theta}=\frac{1}{(1+z)\sqrt{\Omega_{k0}}}\sinh(\sqrt{\Omega_{k0}}\int_{0}^{z}\frac{d\hat{z}}{E(\hat{z})})
\end{equation}
The correlation function however depends both on the angle between
sources and on their redshift difference. That is, one can observe
also a 'longitudinal BAO' scale which, for a small redshift separation
$dz$, amounts to 
\begin{equation}
E(z)=\frac{dz}{R}
\end{equation}
This means that BAO can estimate at every redshift two combinations involving $E(z)$
and $\Omega_{k0}$, and therefore determine both in a MI way. Therefore
the FLRW metric can in principle be reconstructed, within the range
covered by BAO observations, without assumptions beside \emph{a}).
Clearly, SNIa and other distance indicators can contribute to the
statistics, but do not offer information on alternative combinations
of cosmological parameters. Once we have the FLRW metric, the redshifts
and angles can be converted to distance by solving $ds=0$ . Given
two sources at redshifts $z_{1},z_{2}$ separated by an angle $\theta$,
their relative distance $r_{12}$
is \cite{2000MNRAS.319..557L}
\begin{equation}
r_{12}^{2}(z_{1},z_{2},\theta)=r^{2}\left(\frac{z_{1}+z_{2}}{2}\right)\sin^{2}\frac{\theta}{2}+r^{2}\left(\frac{z_{2}-z_{1}}{2}\right)\cos^{2}\frac{\theta}{2}
\end{equation}
where the comoving distance $r(z)$ is defined in (\ref{eq:rz}). The background geometry
is then recoverable in a MI way. But this is not a test of gravity.

We move then to the next layer, perturbations.

\section{Measuring gravity: the anisotropic stress}

We have seen that the gravitational slip $\eta$ is defined as the
ratio between the two gravitational potentials 
\begin{equation}
\eta=-\Phi/\Psi\,\,.
\end{equation}
The lensing potential $\Psi-\Phi$ is the combination that exerts
a force on the relativistic particles (i.e., for our purposes, light),
while $\Psi$ exerts a force on non-relativistic particles (i.e.,
for our purposes, galaxies). The explicit form of the equation of
motion for a generic particle moving with velocity $\boldsymbol{v}$
and relativistic factor $\gamma^{2}=(1-v^{2})^{-1}$ in a weak-field
Minkowski metric is in fact \cite{Baldi:2011es} 
\begin{equation}
\gamma^{2}\dot{\boldsymbol{v}}=\gamma^{2}[2\boldsymbol{v}(\boldsymbol{v}\cdot\boldsymbol{\nabla}(\Psi-\Phi)-v^{2}\boldsymbol{\nabla}(\Psi-\Phi)]-\boldsymbol{\nabla}\Psi
\end{equation}
For small velocities, only the last term on the rhs survives; for
relativistic velocities, only the square bracket term. This means
that in order to test gravity at cosmological scales we need to combine
observations of lensing and of clustering and velocity of galaxies. 

The linear gravitational perturbation theory gives the growth of the
matter density contrast $\delta_{m}(k,z)$ at any redshift $z$ and
any wavenumber $k$, given a background cosmology and a gravity model.
It is convenient to define also the growth function 
\begin{equation}
G(k,z)=\frac{\delta_{m}(k,z)}{\delta_{m}(k,0)}
\end{equation}
and the growth rate 
\begin{equation}
f(k,z)=\frac{\delta_{m}'(k,z)}{\delta_{m}(k,z)}
\end{equation}
where, as usual, the prime stands for derivative with respect to $N=\log_{e}a$.

However, what we observe is the galaxy number density contrast in
redshift space, usually expressed in terms of the galaxy power spectrum
as a function of wavenumber $k$ and redshift, $P_{gal}(k,z)$. Since
galaxies are expected to be a biased tracer of mass, we need to introduce
a bias function 
\begin{equation}
b(k,z)=\frac{\delta_{gal}(k,z)}{\delta_{m}(k,z)}
\end{equation}
that in general depends on time and space (that is, on $k,z$). If
$b=1$ then the number density of galaxies in a given place is proportional
to the amount of underlying total matter, $\rho_{gal}=const\times\rho_{m}$.
If $b>1$ ($<1$) then galaxies are more (less) clustered than matter.
Moreover, since we observe in redshift space, which means we observe
sum of cosmic expansion and the radial component of the local peculiar
velocity, to convert to real space we need the Kaiser transformation
\cite{kaiser_clustering_1987}, which induces a correction factor $(1+f\mu^{2}/b)^{2}$
that depends on the cosine $\mu$ of the angle between the line of
sight and the wavevector $\vec{k}$.

This means that the relation between what we observe, namely the galaxy
power spectrum in redshift space, and what we need to test gravity,
namely $\delta_{m}$, can be written as \cite{2003ApJ...598..720S} 
\begin{equation}
P_{gal}(k,z,\mu)=(A+R\mu^{2})^{2}\label{gcpower}
\end{equation}
where 
\begin{align}
A(k,z) & =Gb\delta_{\text{m0}}\,,\qquad R(k,z)=Gf\delta_{\text{m0}}\,,\label{eq:DirectObs}
\end{align}
($A,R$ are mnemonics for amplitude and redshift, respectively)  where $\delta_{\text{m0}}(k)=\delta_{m}(k,0)$ is the root-mean-square
matter density contrast today. With this definition, $\delta_{m0}$ is normalized as
\begin{equation}
\frac{1}{2\pi^2}\int \delta_{\text{m0}}^2W^2_8(k R_8)d^3 k=\sigma_8^2 \label{eq:sigma8-of-pk}
\end{equation}
where $W_8(k R_8)$ is the window function for a 8 $h^{-1}$Mpc sphere, $W(x) = 3(\sin{x} - x \cos{x})/x^3$. Sometimes one defines $\hat\delta _{m0}=\delta_{m0}/\sigma_8$, which is then normalized to unity. $\hat\delta _{m0}$ can be referred to as the shape of the present power spectrum.
Eq.  (\ref{gcpower}) shows that $A,R$ are the only
two observables one can derive from linear galaxy clustering. This dataset is often collectively called redshift distortion, RSD.

There is then a third observable that one can obtain from weak lensing. From eq. (\ref{eq:lensing-psi}) we see that by estimating the shear distortion one can measure the quantity
\begin{equation}\label{eq:lensing}
Y(1+\eta)\rho_{m}\delta_{m}=Y(1+\eta)\frac{\Omega_{m0}(1+z)^{3}}{E^2}G\delta_{m0}
\end{equation}
Since $E(z)$ can be estimated independently, we define another observable, 
to be denoted $L$ \cite{Amendola:2012ky}, as follows
\begin{equation} 
L(k,z)=\Omega_{m0}Y(1+\eta)G\delta_{m0}\,\,
\end{equation}

Together with $E=H(z)/H_{0}$, the quantities $A,R,L$ are the only
cosmological information one can directly gather  at the linear level\footnote{As we have seen, also $\Omega_{k0}$ is a direct observable, but for simplicity we have assumed that is negligible at all relevant epochs.}. Other
observations, like the integrated Sachs-Wolfe or velocity fields, only give combinations
of $A,R,L,E$, rather than new information. A direct measurement of the peculiar velocity field and its time derivative, for instance, would produce through the Euler equation (\ref{eq:euler}) an estimation of the combination $V=\Omega_{m0}YG\delta_{m0}$, which however is equivalent to $2RE^2(2+ (\log ER)')/3(1+z)^3$.
That is, at least
at the linear level, one can add more statistics, but will always
end up with these four quantities rather than, say, a direct estimate
of $\Omega_{m0}$ or $Y$. A preliminary non-linear analysis \cite{Rampf:2017srp}
shows that employing higher-order statistics we can obtain more MI
information, but we will not consider this here.

We can now write down the lensing equation in Fourier space in the
following way (see \cite{Motta2013})
\begin{align}
-\hat{k}^{2}(\Psi-\Phi) & =\frac{3(1+z)^{3}L}{2E^{2}}\,\label{eq:lensing-arle}
\end{align}
where $\hat{k}=k/aH$. The linearized matter conservation equations,
i.e. the continuity equation and the Euler equation, can be combined
in a single second-order equation
\begin{equation}
\delta_{m}''+\delta_{m}'(2+\frac{E'}{E})=-\hat{k}^{2}\Psi\label{eq:growth}
\end{equation}
that depends only on the pressureless assumption $d$) and not on
the gravitational model. In terms of our observational variables and for slowly varying potentials, this
becomes
\begin{align}
-\hat{k}^{2}\Psi & =R'+R(2+\frac{E'}{E})\, \quad .\label{eq:fgrowth-arle} 
\end{align}
These equations show clearly that lensing and matter growth can measure
some combination of $R,L,E$ and their derivatives, as will be seen explicitly below.
For now, let us just rewrite eq. (\ref{eq:fgrowth-arle}), employing also eq. (\ref{eq:poisson-2-1}) as
\begin{align}
\frac{3}{2}\frac{\Omega_{m0}(1+z)^3 Y}{fE^2} & =\frac{R'}{R}+(2+\frac{E'}{E})\,\label{eq:fgrowth-arle2}
\end{align}
We see then that $Y$ is not, unfortunately, a MI quantity. Even if we have precise information on $R,E$, we would still need at any $k,z$ the combination $\Omega_{m0}/f$, which is not an observable. 
Only a null test of standard gravity {\it plus} a specific cosmological model, say $\Lambda$CDM, is possible: in this case in fact $Y=1$, and $f\approx \Omega_m^{0.55}$ are known, and we have that $\Omega_{m0}$ is uniquely measured by a combination of $R,R',E,E'$. Any two measures at different $k$ or $z$ must then give the same $\Omega_{m0}$.
We show below that $\eta$, in contrast to $Y$, is a MI quantity.

Although $A,R,L$ might be interesting statistics on their own, our
goal here is to test gravity. Now, the bias function depends on complicate,
possibly non-linear and hydrodynamical processes, so that, even if
$b$ depends on gravity, we do not know how. Also the shape $\delta_{m0}$
of the power spectrum depends on initial conditions (inflation) and,
possibly, on processes that distorted the initial spectrum during
the cosmic evolution. In fact, even if we could exactly measure the
power spectrum shape from CMB without a parametrization like $n_{s}$
or its ``running'', nothing prevents that an unknown process, for
instance the presence of early dark energy or early modified gravity,
distorted the spectrum at some intermediate redshift between last
scattering and today. Therefore, in order to obtain model-independent
measures, we should get rid of both $b$ and $\delta_{m0}$. It was
shown in \cite{Amendola2012,Motta2013} that one can obtain only three
 statistics where the effects of the shape of the primordial
power spectrum is canceled out, namely 
\begin{align}
P_{1} & \equiv\frac{R}{A}=\frac{f}{b},\\
P_{2} & \equiv\frac{L}{R}=\frac{\Omega_{\text{m}0}Y(1+\eta)}{f},\label{p2def}\\
P_{3} & \equiv\frac{R'}{R}=f+\frac{f'}{f}=\frac{(f\sigma_{8}(k,z))'}{f\sigma_{8}(k,z)}\,\,\,.
\end{align}
In the last equation we introduced the often-employed quantity
\begin{equation}
f\sigma_{8}(k,z)=\sigma_{8}G(k,z)f(k,z)=R\frac{\sigma_8}{\delta_{m0}}\,,
\end{equation}
Notice that we are not defining $\sigma_8(k,z)$ as an integral over the power spectrum at $z$, 
as in eq. (\ref{eq:sigma8-of-pk}),
because we are interested in the $k$-dependence.  These quantities depend in general on $k,z$ in an arbitrary way. Every other ratio of $A,R,L$
or their derivatives can be obtained through $P_{1-3}$ or their derivatives.

Let us discuss the three statistics $P_{1-3}$ in turn.
The first quantity, $P_{1}$, often called $\beta$ in the literature,
contains the bias function. Since we do not know how to extract gravitational information, if any, from the bias,  we do not consider it any longer.

Concerning $P_3$, we notice that, although related,  what is observed is $R$ and not $f\sigma_8(k,z)$. In order to determine the latter from the observable $R$, one has to
assume a value of $\delta_{m0}/\sigma_8=\hat\delta_{m0}$ (typically chosen to be given by $\Lambda$CDM), so that it is
 not a model-independent observable. 
 One could imagine that $P_{3}$ alone is instead a direct
test of gravity, since it depends only on $f$. However, in order
to predict the theoretical value of $R$ (or $f$) as a function of the gravity parameter $Y$ from eq. (\ref{eq:fgrowth-arle2})
one needs to choose a value of  $\Omega_{m0}$ and the initial condition $f(k,z_{i})$ at some epoch
$z_{i}$ for every $k$. In almost all the papers on this topic since \cite{lahav_1991},
this initial condition is assumed to be given by a purely matter-dominated
universe at some high redshift (this is, for instance, how the well-known
approximated formula $f\approx\Omega_{m}^{\gamma}(z)$ is obtained). 
However, in models of early dark energy or early modified gravity,
this assumption is broken. Therefore, once again, $P_{3}$ alone cannot
provide a MI measurement of gravity. Clearly, exactly as we have seen for the dark energy EOS $w(z)$, if one parametrizes $Y(k,z)$ with a sufficiently small number of free parameters, then the RSD data alone, which provide $R(k,z)$, can fix both $\Omega_{m0}$ and $Y(k,z)$.

We can also see that $P_{2}$ is trivially related to the $E_{G}$
statistics, whose expected value at a scale $k$ is (see \cite{Leonard2015}
and references therein) as 
\begin{equation}
E_{g}=\left\langle \frac{a\nabla^{2}(\Psi-\Phi)}{3H_{0}^{2}f\delta_m}\right\rangle_{k} \,.\label{eg-ideal}
\end{equation}
In $\Lambda$CDM and with Planck 2015 parameters, its present value is $E_{g0}\approx \Omega_{m0}/f_0\approx 0.58$.
With  our definitions,
the relation with $P_{2}$ is given by 
\begin{equation}
P_{2}=2E_{g}\,.\label{eq.P2Eg}
\end{equation}

The $E_{g}$ statistics has been used several times as a test of modified
gravity \cite{Zhang2007,2010Natur.464..256R,delaTorre:2016rxm,Leonard2015}.
However, it is not \emph{per se} a model-independent test. In fact,
the theoretical value of $E_{g}$ depends on $\Omega_{m0}$ and on
$f$. As already stressed, $\Omega_{m0}$ is not an observable quantity.
Moreover,
the growth rate $f$ is estimated by solving the differential equation
of the perturbation growth and this requires initial conditions and $Y$. As
a consequence of this, when we compare $E_{g}$ to the predicted value (\ref{p2def}),
we can never know whether any discrepancy 
is due to a different value of $\Omega_{m0}$ or different initial
conditions, or  non-standard modified gravity parameters $Y,\eta$. As previously, one can employ $E_g$ only to perform a null test of standard gravity plus $\Lambda$CDM, or other specific models. This is of course a task of primary importance, but is different  from measuring the properties of gravity in a model-independent way.

In contrast, we can define a MI statistics to measure gravity, in particular
the parameter $\eta,$ by combining the equation for the growth of
structure formation eq. (\ref{eq:fgrowth-arle}) with the lensing
equation (\ref{eq:lensing-arle}), and with eqs. (\ref{eq:poisson-2-1},\ref{eq:3ho-1-1}).
We see then that the gravitational slip as a function of model-independent
observables is given by
\begin{equation}
\eta_{\textrm{obs}}\equiv\frac{3P_{2}(1+z)^{3}}{2E^{2}\left(P_{3}+2+\frac{E'}{E}\right)}-1=\eta\,.\label{eq:etaobsp2p3}
\end{equation}
In order to distinguish the observables from the theoretical expectations,
we denoted the combination on the left-hand-side of this equation
as $\eta_{\textrm{obs}}$. The statistics $\eta_{\textrm{obs}}$ is model-independent
because it estimates directly $\eta$ without any need to assume a
model for the bias, nor to guess $\sigma_{8}$ or $\Omega_{m0}$,
nor to assume initial conditions for $f$. So if observationally one
finds $\eta_{\textrm{obs}}\not=$1, then $\Lambda$CDM and all the models in
standard gravity and in which dark energy is a  perfect
fluid are ruled out. As a consequence,  cautionary remarks like those in \cite{Amon2017},
namely that their results about $E_{g}$ cannot be employed until
the tension between $\Omega_{m0}$ in different observational dataset
is resolved, do not apply to $\eta_{\textrm{obs}}$. The price to pay is
that eq. (\ref{eq:etaobsp2p3}) depends on derivatives of $E$ and,
through $P_{3}$, of $f\sigma_{8}(z)$. Derivatives of random variables
are notoriously very noisy. In the next sections we will compare several
methods to extract the signal.

If we abandon the linear regime then of course new observables can be devised, see e.g. \cite{Rampf:2017srp}. One interesting case is provided by relaxed galaxy clusters, for which we can reasonable expect that the virial theorem is at least approximately respected. In this case, we can directly measure the potential $\Psi$ by the Jeans equation, i.e. the equilibrium equation between the motion of the member galaxies and the gravitational force (note that the potential remains linear for galaxies and clusters, even for a non-linear distribution of matter). The lensing potential can instead be mapped through weak and strong lensing of background galaxies. In this case, one can gather much more information on the modified gravity parameters than in the linear regime \cite{2017JCAP...07..023P}. However, the validity of this approach relies entirely on two important assumptions. First, we must assume the validity of the virial theorem, which can be more or less reasonable, but cannot be proved independently. Second, since we have access only to the radial component of the member galaxy velocities, we must assume a model for the velocity anisotropy, i.e. how the other components are distributed within the cluster.

Concluding this section, we recap and emphasize the main points. $A,R,L,E$
are the only independent linear observables in cosmology. The ratios
$P_{1-3}$ are independent of the initial conditions (i.e., of the
power spectrum shape). $P_{2},P_{3}$ are also independent of the
galaxy bias. The combination $\eta_{\textrm{obs}}(P_{2,}P_{3,}E)$ is therefore
a model-independent test of gravity: it does not depend on bias, on
initial conditions, nor on other unobservable quantities like $\Omega_{m0}$
or $\sigma_{8}$. If $\eta_{\textrm{obs}}\not=1$, gravity is not Einsteinian;
if $\eta_{\textrm{obs}}$ does not have the same $k^{2}$ dependence as the
Horndeski theory, the entire Horndeski model is rejected. All this, of course, provided our conditions $a)-d)$ are verified.

\section{General perfect fluid}

What happens if we remove condition $d$), namely, that matter is pressureless? If  matter is a perfect
fluid and we know or hypothesize a different equation of state and
sound speed, then eq. (\ref{eq:growth}) is modified since the continuity and Euler equations, which come directly from the conservation of the energy-momentum tensor, now read
\begin{align}
 \delta' &= -\frac{1+w}{a H}(\theta - 3 a H\Phi' ) - 3 aH (c^2_s - w) \delta \label{eq:deltadot-general} \\
\theta' &= - (1-3w)\theta - \frac{ w'}{1+w} \theta +\frac{c^2_s}{1+w} aH \hat{k}^2 \delta -  aH \hat{k}^2 \sigma + aH \hat{k}^2 \Psi \label{eq:thetadot-general} \quad, 
\end{align}
where the sound speed is $c^2_s \equiv \delta p / \delta \rho $ and $\sigma$ is the matter anisotropic stress. Here we are assuming that $\delta$ represents the density contrast of  matter, both baryons and dark matter, whose microphysical properties are described by with some effective parameters $\sigma,c_s,w$.
Assuming a zero anisotropic stress, since we are dealing with non-relativistic matter, and for small and constant $w$ and $c_{s}^2$, we obtain the following second order differential equation 
\begin{equation}
\frac{\delta''}{1+w} + \left(2+\frac{H'}{H} + 3(c_s^2 - 2w) \right)\frac{\delta'}{1+w} + 6(c_s^2 - w)\left(1+\frac{H'}{H}\right)\frac{\delta}{1+w} = -\hat{k}^2\Psi
\end{equation}
which reduces to eq. (\ref{eq:deltapp-qsa}) for cold dark matter, 
where $\sigma = w = c_s^2 = 0$.
In the case of a constant $w$, the matter would not follow  an $a^{-3}$ behavior as a function of time, but it would scale with $(1+z)^{3(1+w)}$, so that the lensing equation (\ref{eq:lensing-arle}) would now read
\begin{align}
-\hat{k}^{2}(\Psi-\Phi) & =\frac{3(1+z)^{3(1+w)}L}{2E^{2}} \quad.\label{eq:lensing-withw}
\end{align}
Taking the appropriate ratios of the two  equations above, we can obtain $\eta$ as we did for eq. \ref{eq:etaobsp2p3}, but this time some extra term appears
\begin{equation}
\frac{3(1+w)P_{2}(1+z)^{3(1+w)}}{2E^{2}\left(P_{3}+2+\frac{E'}{E} + \mathcal{W}_1 + \frac{\mathcal{W}_2}{f}(1+\frac{E'}{E})\right)}-1=\eta_{\textrm{obs}} \quad,\label{eq:etaobswithw}
\end{equation}
where $\mathcal{W}_1 = 3(c_s^2 - 2w)$ and $\mathcal{W}_2 = 6(c_s^2 - w)$. Both  $\mathcal{W}_1$ and  $\mathcal{W}_2$ reduce to zero for standard cold dark matter, such that we recover eq. (\ref{eq:etaobsp2p3}) exactly in that case. 
For a barotropic fluid such that $p=p(\rho)$, $c_s^2=w$ and $\mathcal{W}_2=0$. In this case, we have again a MI estimator for $\eta$, provided we know $c_s,w$.
On the other hand, if $\mathcal{W}_2\not=0$, we see that 
this estimation of $\eta$ contains the growth rate $f$, which we argued not to be  a model-independent observable in the linear regime. However, an extension of this formalism to the quasilinear scales \cite{Rampf:2017srp} has shown that $f$ can indeed be recovered in a model-independent way, using observations of the bispectrum.

\section{The linear, scalar, quasi-static, model-independent Horndeski observables}
\label{sec:obs}
For the previous sections we can draw a remarkable conclusion. Since $\eta$ is the only linear, quasi-static, MI cosmological observable, we see that, among the HL parameters, only the time-dependent functions  $h_2,h_4,h_5$ (see eq. \ref{eq:etay-1})  share the same property.
The GW speed constraint has already measured $h_2(t_0)=1$. Assuming this can be extended at all times, so that $\alpha_T=0$, and assuming $H$ is also measured in a MI way, we see that what can still be measured at the linear perturbation level are the combinations
\begin{eqnarray}
O_1 &=& \frac{\alpha_1+\alpha_2}{\mu^2}\\
O_2 &=& \frac{\alpha_1\alpha_M}{\mu^2}
\end{eqnarray}
that corresponds to the two scales one can measure in $\eta$. If  $O_{2}$ vanishes, $h_4=h_5$ and  $\eta=h_2=1$ as in the standard case. As we have already seen, this happens only in two cases, for $\alpha_M=0$ and for $\alpha_M=-\alpha_B/2$.

\section{Data}

\label{datapart}

\begin{figure}[h]
\includegraphics[width=0.32\textwidth]{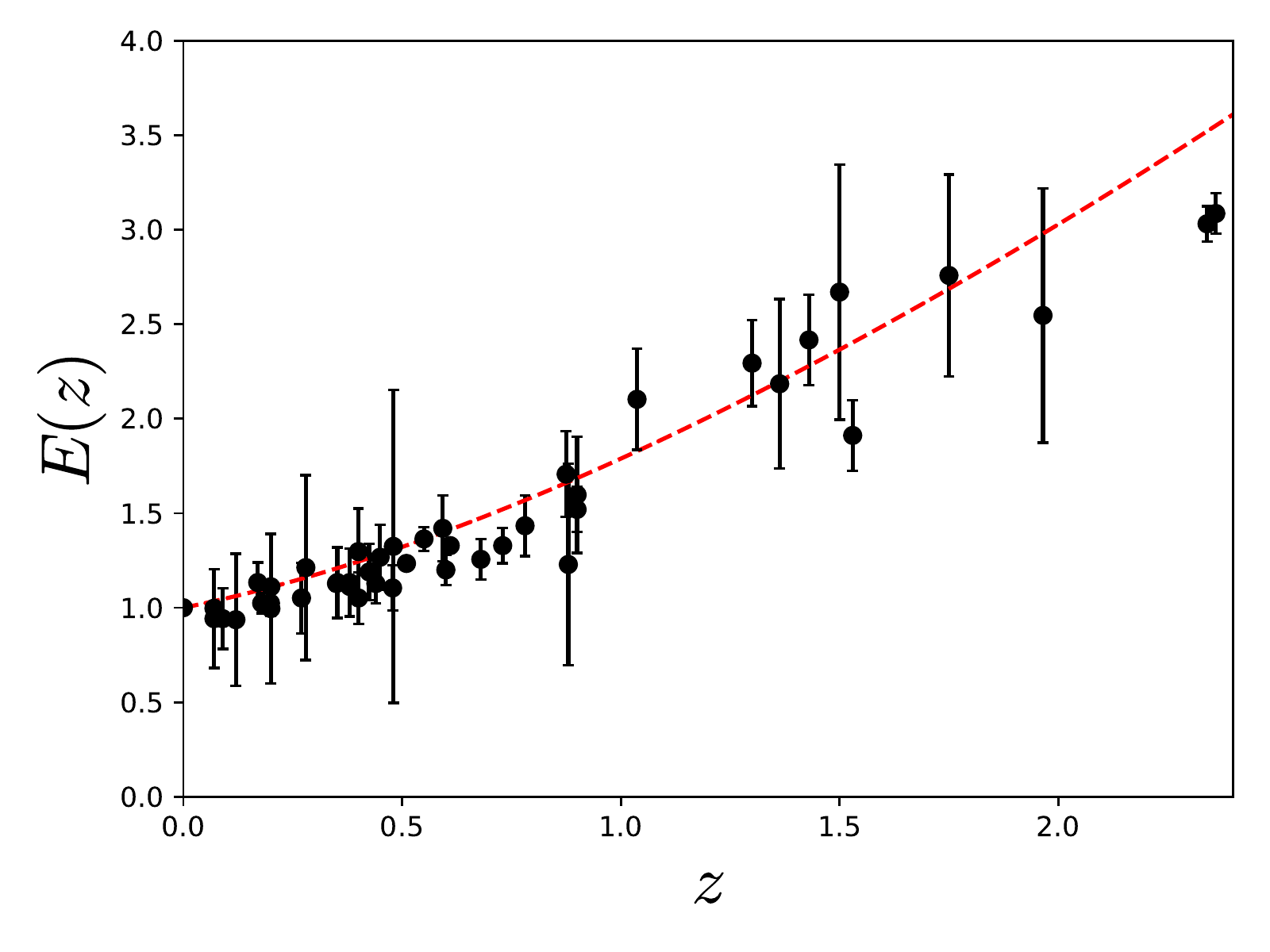}
\includegraphics[width=0.32\textwidth]{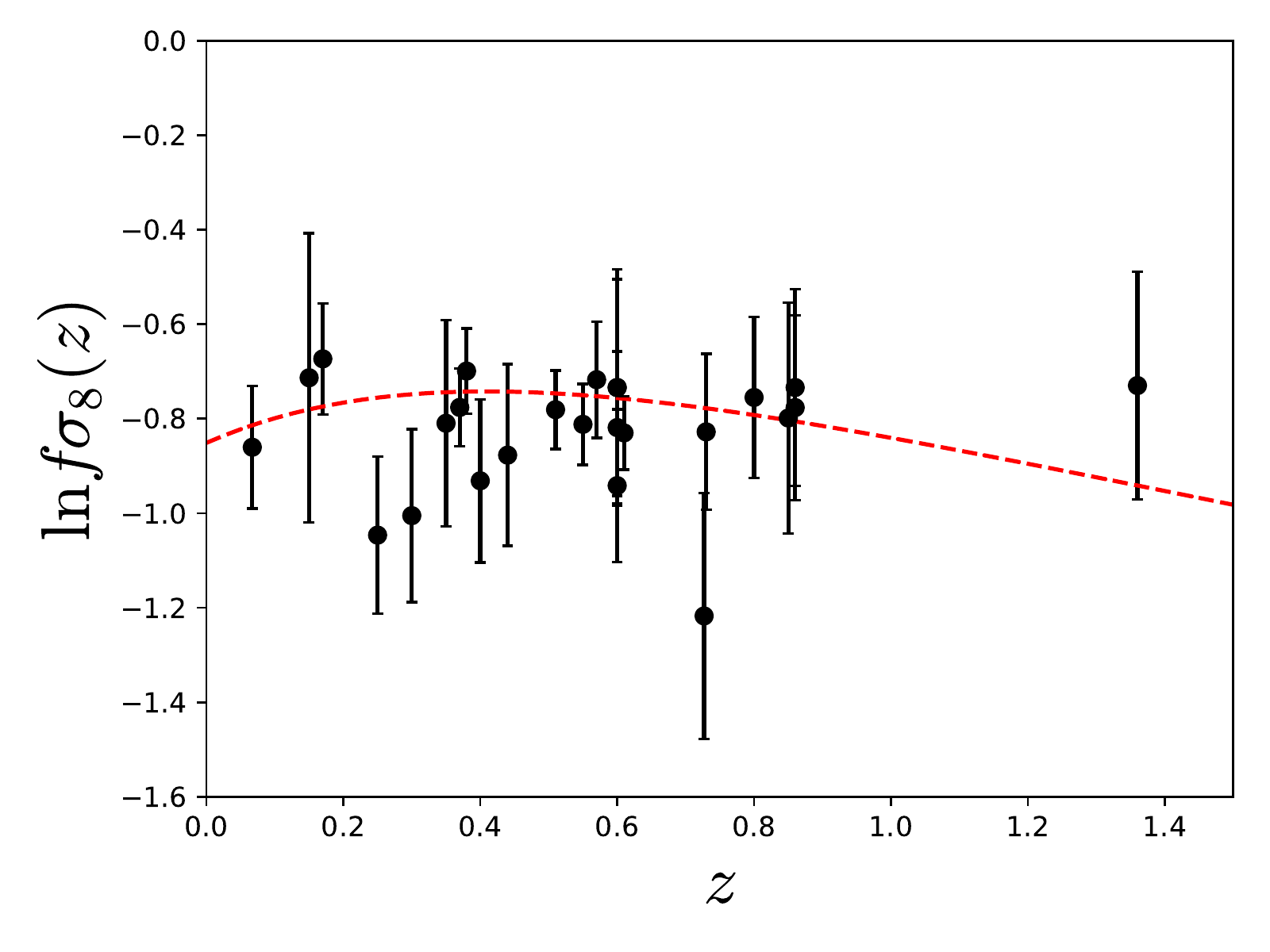}
\includegraphics[width=0.32\textwidth]{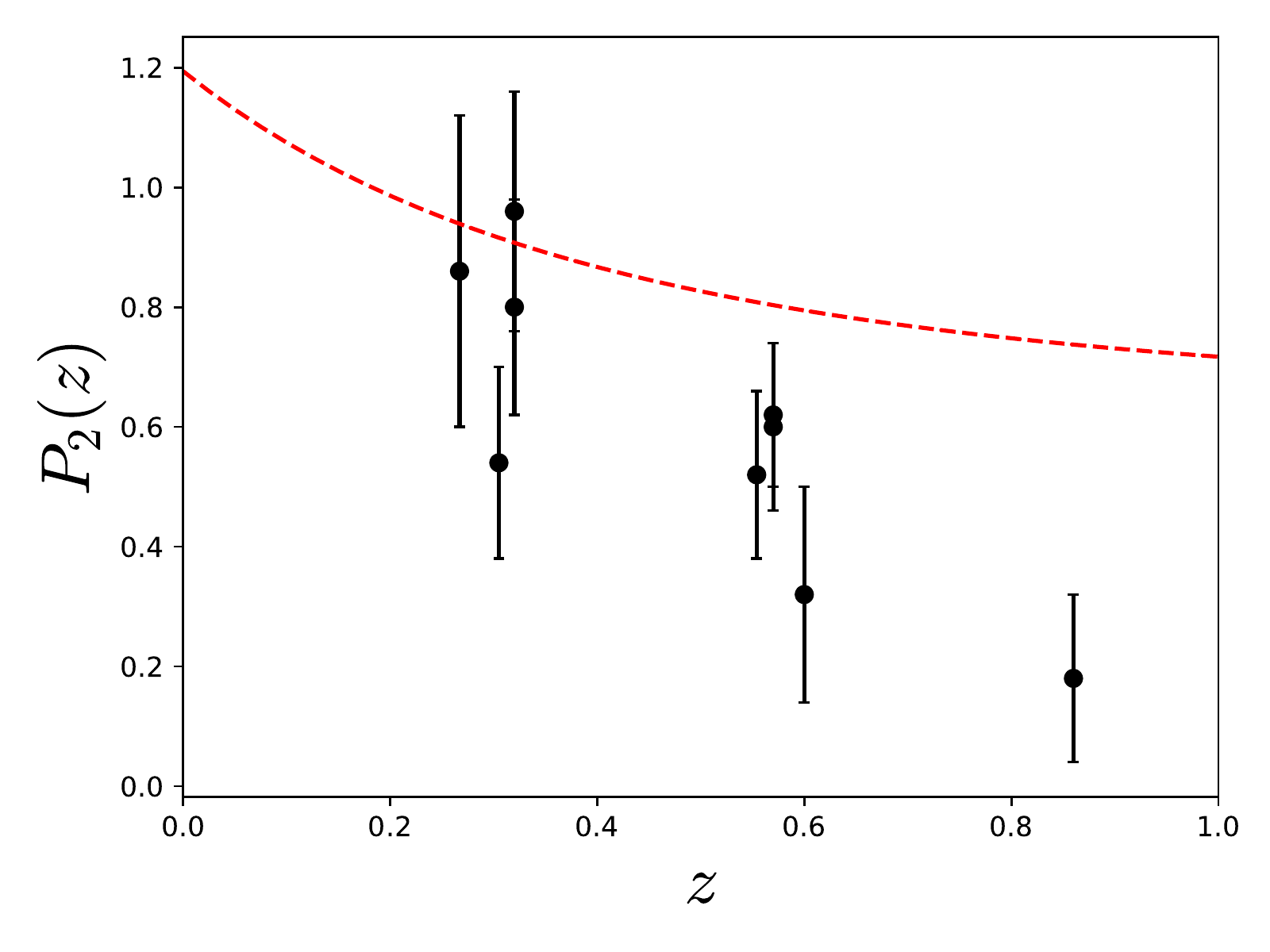}
\caption{\label{fig:Data-sets-used}Data sets used in this work (black dots
with error bars), plotted with the corresponding theoretical $\Lambda$CDM
prediction as a function of redshift (solid red line), using a Planck
2015 cosmology. \textbf{Left panel:} $E(z)$ data. We used the Planck
2015 value of $H_{0}$ to convert some of the data points from $H(z)$
to $E(z)$ (see main text). \textbf{Central panel:} Plot of the logarithm
of the $f\sigma_{8}$ data points. \textbf{Right panel:} Data set
for $P_{2}$, obtained using $E_{g}$ data and the relation \ref{eq.P2Eg}
that converts between different notations. For $z>0.5$ we see a larger
discrepancy between $\Lambda$CDM and the data points, which was also
noted in \cite{Amon2017} and references therein.}
\end{figure}

In the next sections we obtain an estimate of $\eta_{\textrm{obs}}$ using all the currently available data\footnote{This section and the next two are a summary of the following published paper, A. M. Pinho, S. Casas, and L. Amendola, Model-independent reconstruction of the linear anisotropic stress $\eta$, arXiv:1805.00025, JCAP11(2018)027}.
The first step is to
reconstruct $E(z)$ (and therefore $E'(z)$), $P_{2}(z)$ and $P_{3}(z)$ using the data all the currently relevant available data, shown in Fig. \ref{fig:Data-sets-used}, where we also plot the $\Lambda$CDM
curves of the different functions using the cosmological parameters 
from the TT+lowP+lensing Planck 2015 best-fits \cite{PlanckCollaboration2015}. 
A similar analysis, with the much smaller dataset then available, was carried out also in Ref. \cite{AlejandroThesis}.

For the Hubble parameter measurements, we have used the most
recent compilation of $H(z)$ data from \cite{Yu2017}, including
the measurements from \cite{Simon2004,Stern2009,Moresco:2012jh,Moresco:2015cya}, Baryon Oscillation Spectroscopic Survey (BOSS)
\cite{Delubac:2014aqe,Font-Ribera:2013wce,Moresco2016}
and the Sloan Digital Sky Survey (SDSS) \cite{Zhang2012,Alam2016}.
In this compilation, the majority of the measurements was obtained
using the cosmic chronometric technique. This method infers the expansion
rate $dz/dt$ by taking the difference in redshift of a pair passively-evolving
galaxies. The remaining measurements were obtained  through the
position of the BAO peaks in the power
spectrum of a galaxy distribution for a given redshift. For this case, the measurements from \cite{Delubac:2014aqe}
and \cite{Font-Ribera:2013wce} are obtained using the BAO signal in the
Lyman-$\alpha$ forest distribution alone or cross correlated with
Quasi-Stellar Objects (QSO) (for the details of the method, we refer
the reader to the original papers). Ref. \citep{Alam2016} provides the
covariance matrix of  three $H(z)$ measurements from the radial BAO
galaxy distribution. To this compilation we add the results from WiggleZ
\cite{Blake2012}.
In addition to these, we use the recent results from \cite{Riess2017}
where a compilation of Type Ia Supernovae from CANDELS and CLASH Multi-cycle
Treasury programs were analyzed providing a few tight measurements
of the expansion rate $E(z)$.

The $E_{g}$ data include the results from KiDS+2dFLenS+GAMA \cite{Amon2017},
i.e, a joint analysis of weak gravitational lensing, galaxy clustering
and redshift space distortions. We also include image and spectroscopic
measurements of the Red Cluster Sequence Lensing Survey (RCSLenS)
\cite{Blake2016} where the analysis combines the the Canada-France-Hawaii
Telescope Lensing Survey (CFHTLenS), the WiggleZ Dark Energy Survey
and the Baryon Oscillation Spectroscopic Survey (BOSS). Finally the
work of VIMOS Public Extragalactic Redshift Survey (VIPERS) \cite{DelaTorre2016}
is also accounted for in our data. The latter reference uses redshift-space
distortions and galaxy-galaxy lensing.

These sources provide measurements in real space within the scales $ 3 < R_p < 60 h^{-1} $Mpc 
and in the linear
regime, which is the one we are interested in. They have been obtained
over a relatively narrow range of scales $\lambda$ meaning that we
can consider them relative to the $k=2\pi/\lambda$-th Fourier component,
as a first approximation. In any case, the discussion about the $k$-dependence
of $\eta$ is beyond the scope of this work, so the final result can
be seen as an average over the range of scales effectively employed
in the observations.
Moreover, in the estimation of $E_g$, based on \cite{Leonard2015}, one assumes that  the redshift of the lens galaxies can be approximated by a single value. With these approximations, indeed $E_g$ is equivalent to $P_2/2$, otherwise $E_g$ represents some sort of average value along the line of sight.
We caution that these approximations can have a systematic effect both on the measurement of $E_g$ and on our derivation of $\eta$. In a future work we will quantify the level of bias possibly introduced by these approximations in our estimate.

Finally, the quantity $f\sigma_{8}(z)$ is connected to the $P_{3}$ parameter.
Our data include measurements from the 6dF Galaxy Survey \cite{Beutler2012},
the Subaru FMOS galaxy redshift survey (FastSound) \cite{Okumura2015},
WiggleZ \cite{Blake2012}, VIMOS-VLT Deep Survey (VVDS) \cite{Song2008},
VIMOS Public Extragalactic Redshift Survey (VIPERS) \cite{DelaTorre2016,Hawken2016,DelaTorre2013,Mohammad2017} and
the Sloan Digital Sky Survey (SDSS) \cite{Howlett2015,Samushia2012,Tojeiro:2012rp,Chuang2012,Alam2016,Gil-Marin:2015sqa,Gil-Marin2016,Chuang2016}. The values from \cite{Cabre2009}
and \cite{Guzzo2008} will not be considered since the $f\sigma_{8}(z)$
value is not directly reported.

\section{Reconstruction of functions from data}

\label{recopart}

The only difficulty in obtaining $\eta_{\textrm{obs}}$ is that we need to take the ratios $P_2,P_3$ at the same redshift, while we have datapoints at different redshifts, and that we need to take derivatives of $E(z)$ and $f\sigma_8(z)$. This essentially means
we need to have a reliable way to interpolate the data to reconstruct the underlying behavior. 

There is no universally accepted method to interpolate data. Depending on how many assumptions one makes regarding the theoretical model, e.g. whether the reconstructed functions need just to be continuous, or smooth, depending on few or many parameters, etc., one gets unavoidably different results, especially in the final errors.
Here, we consider and compare three methods to obtain the value of $\eta_{\textrm{obs}}$: binning, Gaussian Process (GP), and generalized linear
regression. 

The first, and simplest, method assembles the data into bins. This consists in dividing
the data into particular redshift interval (bin) and for each of these
intervals one calculates the average value of the subset of
the data contained in that bin. The corresponding redshift and error of each bin are computed as weighted averages.  

Another way to reconstruct a continuous function from a dataset is
using a Gaussian Process algorithm as explained in \cite{GPbook}. This
process can be regarded as the generalization of Gaussian distributions
to function space since it provides a distribution over functions
instead of a distribution of a random variable. Considering a dataset
$\mathcal{D}=\{(x_{i},y_{i})\vert i=1,...n\}$, where $x_{i}$ are
deterministic variables and $y_{i}$ random variables, the goal is
to obtain a continuous function $f(x)$ that best describes the dataset.
A function $f$ evaluated at a point $x$ is a Gaussian random variable
with mean $\mu(x)$ and variance $\mathrm{Var}(x)$. The $f(x)$ values
depend on the function value evaluated at other $\bar{x}$ point 
(particularly if they are close points). The relation between these can be 
given by a covariance function $\mathrm{cov}(f(x),f(\bar{x}))=k(x,\bar{x})$.
The covariance function $k(x,\bar{x})$ is in principle arbitrary. Since
we are interested in reconstruct the derivative of data, a Gaussian covariance function as 
\begin{equation}
k(x,\bar{x})=\sigma_{f}^{2}\exp\bigg[-\frac{(x-\bar{x})^{2}}{2\ell^{2}}\bigg].
\end{equation}
is the chosen function since it is the most common having the least number 
of additional parameters. This function depends on the hyperparameters 
$\sigma_{f}$ and $\ell$ that allow to set the strength of the covariance 
function. These hyperparameters can be regarded as the typical scale and 
change in the $x$ and $y$ direction. The full covariance function takes
the data covariance matrix $C$ into account by $M(x,\bar{x})=k(x,\bar{x})+C$. The log  likelihood is then
\begin{equation}
\ln\mathcal{L}=-\frac{1}{2}\sum_{i,j=1}^{N}\Bigg\{\big[y_{i}-\mu(x_{i})\big](M^{-1})_{ij}\big[y_{j}-\mu(x_{j})\big]\Bigg\}+\ln\vert M\vert+N\ln2\pi\label{eq.logL}
\end{equation}
where $\vert M\vert$ is the determinant of $M(x_{i},x_{j})$.  The
distribution eq. (\ref{eq.logL}) is usually sharply peaked and so we 
maximize the distribution to optimize the hyperparameters, although this is  an approximation to the marginalization process and it may not be the best approach for all datasets.
We employ the Python publicly available 
GaPP code from Seikel et al. (2012) \cite{Seikel2012}.

As a third method, we use a generalized linear regression. Let us
assume we have $N$ data $y_{i}$, one for each value of the \emph{independent}
variable $x_{i}$ and that
\begin{equation}
y_{i}=f_{i}+e_{i}
\end{equation}
where $e_{i}$ are errors (random variables) which are assumed to
be distributed as Gaussian variables. Here $f_{i}$ are theoretical
functions that depend linearly on a number of parameters $A_{\alpha}$
\begin{equation}
f_{i}=\sum_{\alpha=0}^n A_{\alpha}g_{i\alpha}
\end{equation}
where $g_{i\alpha}(x_{i})$ are functions of the variable $x_{i}$,
chosen to be simple powers, $g_{i\alpha}=x_i^\alpha$, so that $f_i$ are polynomials of order $n$. 

The order of the polynomial is in principle arbitrary, up to the number $N$
of datapoints. However, it is clear that with too many free parameters
the resulting $\chi^{2}$ will be very close to zero, that is, statistically
unlikely. At the same time, too many parameters also render the numerical
Fisher matrix computationally unstable (producing, e.g., a non-positive
definite matrix) and the polynomial wildly oscillating. On the other
hand, too few parameters restrict the allowed family of functions.
Therefore, we select the order of the polynomial function by choosing
the degree  for which the reduced chi-squared $\chi_{red}^{2}=\frac{\chi_{min}^{2}}{N-n-1}$,
is closest to unity and such that the Fisher matrix is positive definite. 

\begin{figure}[htp]
\includegraphics[width=0.44\textwidth]{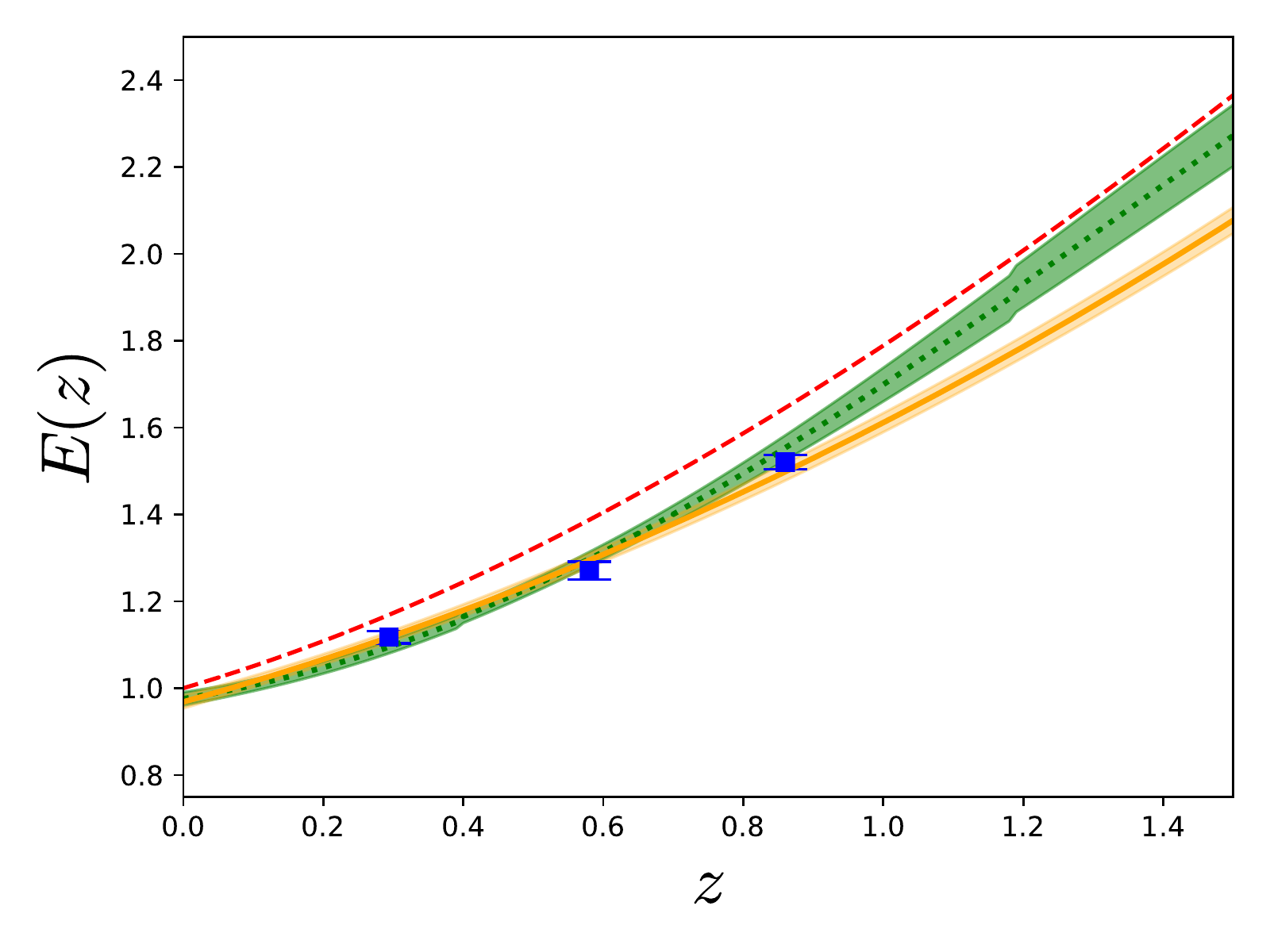}
\includegraphics[width=0.44\textwidth]{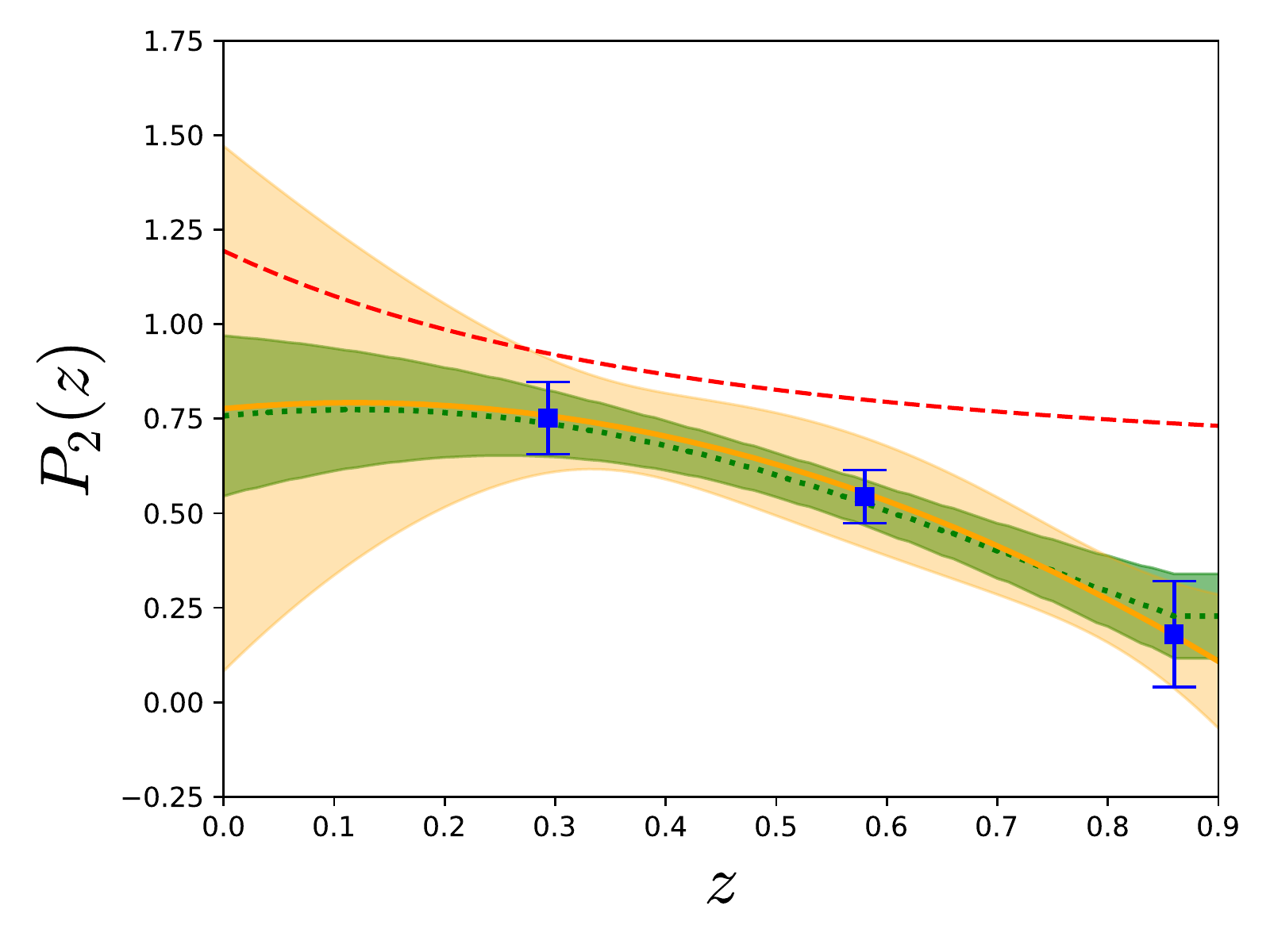}\\
\includegraphics[width=0.44\textwidth]{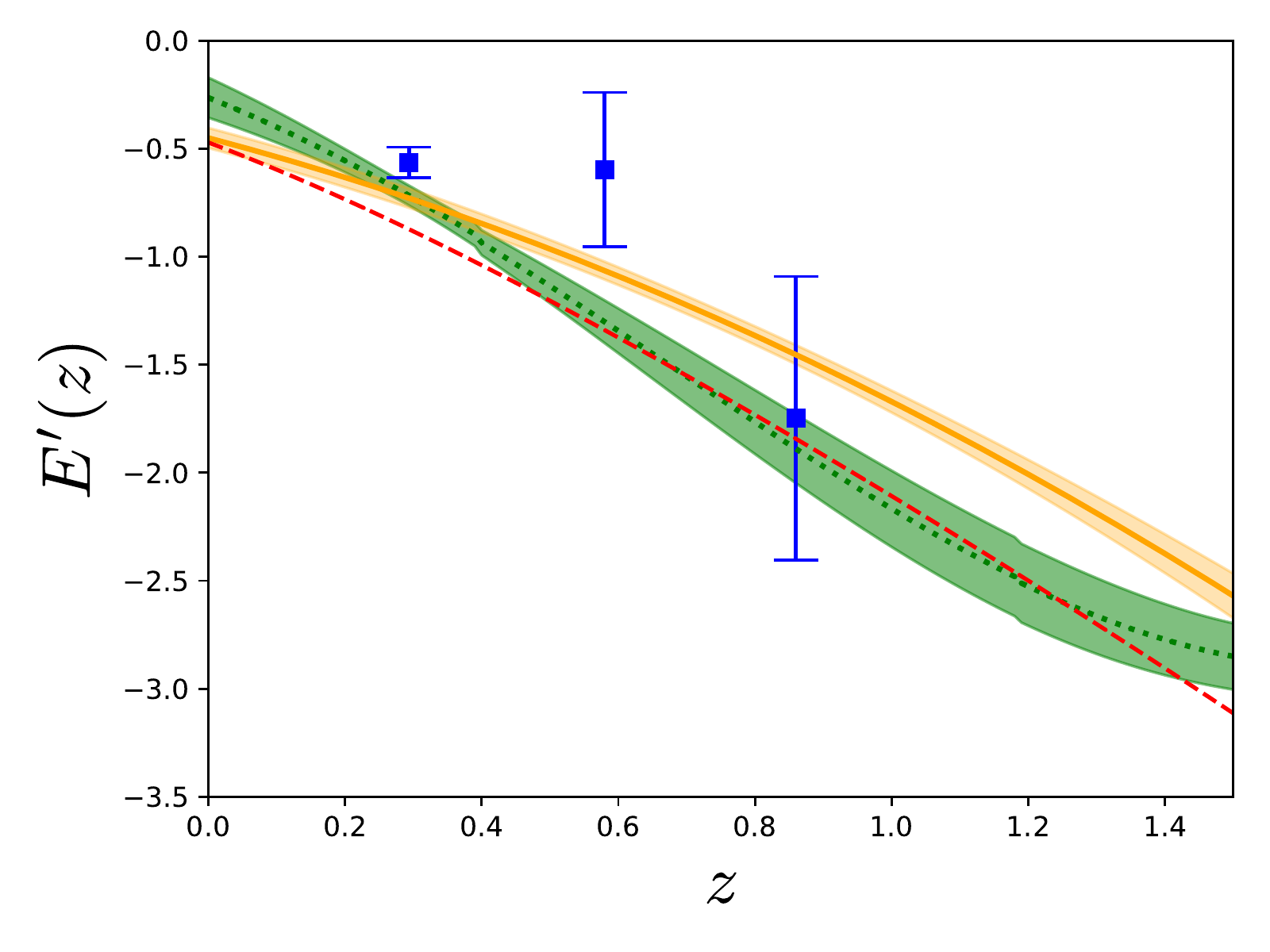}
\includegraphics[width=0.44\textwidth]{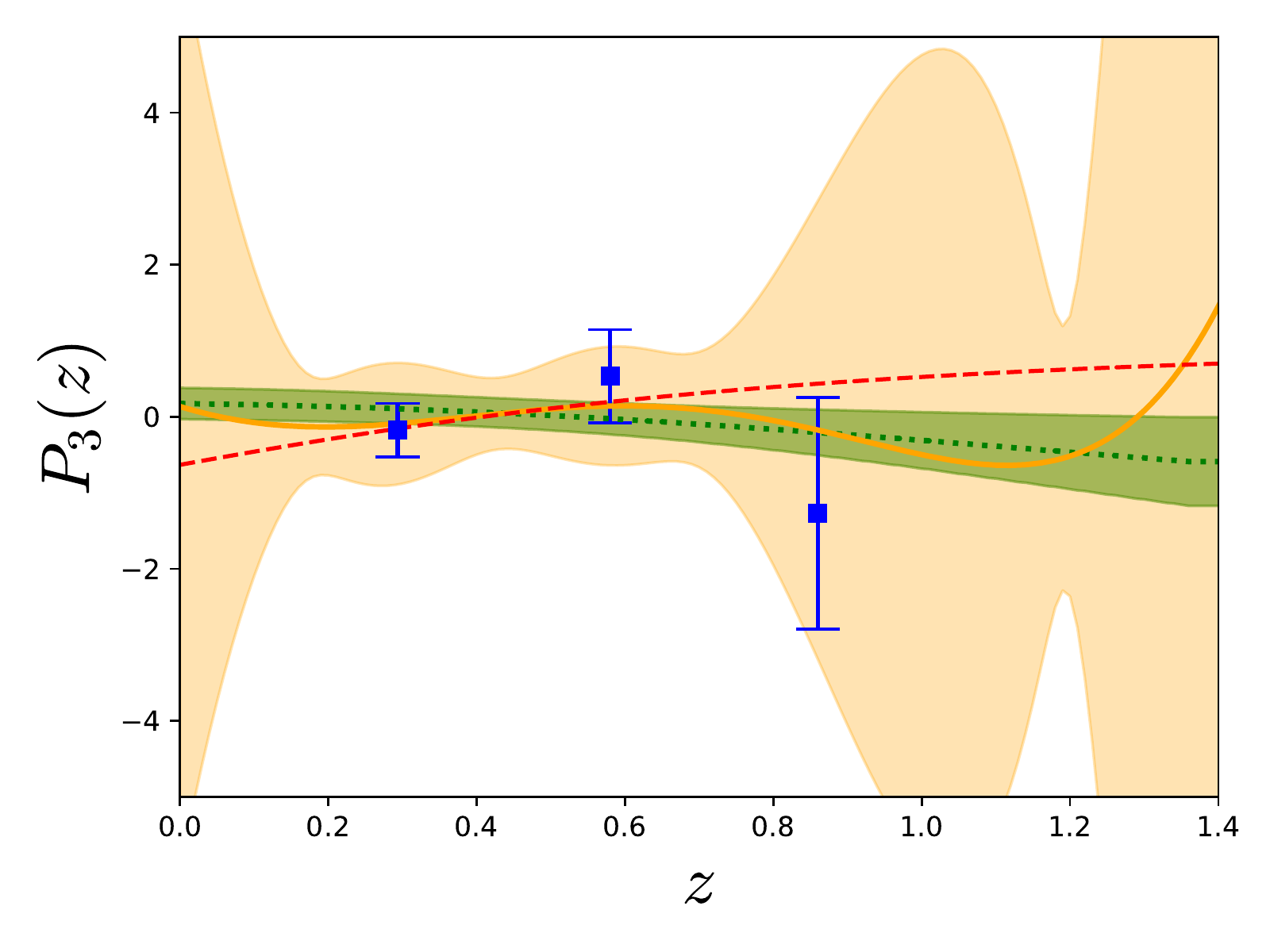}
\caption{Comparison of the three reconstruction methods for each of the model-independent variables. The binning method in blue squares
with error bars, Gaussian Process as a green dotted line with green bands, polynomial regression as a solid yellow line with yellow bands. All of them depicting the $1\sigma$ uncertainty.
\textbf{Left panel:} Plot of the reconstructed $E(z)$ function on
the top and its derivative $E'(z)$ on the bottom.
\textbf{Right panel:} Plot of the reconstructed $P_{2}(z)$ function
on the top and the reconstructed $P_3(z)$ function on
the bottom. For each case, we show the theoretical prediction of our reference $\Lambda$CDM model
as a red dashed curve.}
\label{plt.rec} 
\end{figure}

\section{Results}

\label{resupart}

Let us now discuss the results of the final observable $\eta_{{\rm obs}}$
for each of these methods.
The binning method contains the least number of assumptions compared to the polynomial regression
or the Gaussian Process method. 
It is essentially a weighted average over the data points and its
error bars at each redshift bin. Since we need to take
derivatives in order to calculate $P_{3}$ and $E'$, and we
have few data points, we opt to compute finite difference derivatives.
This has the caveat that it introduces correlations among the errors
of the function and its derivatives, that we cannot take into account
with this simple method. Moreover, for the binning method, we do not take into account possible
non-diagonal covariance matrices for the data, which we do for polynomial regression and the Gaussian Process reconstruction.

Figure \ref{plt.rec} shows the
reconstructed functions obtained by the binning method, the Gaussian Process  and with polynomial regression,
alongside with the theoretical prediction of the standard $\Lambda$CDM
model. In all cases the error bars or the bands
represent the $1\sigma$ uncertainty.

With the binning method, the number of bins is limited by the maximum
number of existing data redshifts from the smallest data set corresponding
to one of our model-independent observables. In this case, this is
the quantity $E_{g}$, for which we have effectively only three redshift
bins. There are
nine $E_g$ data points, but most of them are very close to each other in redshift, due to being
measured by different collaborations or at different scales in real
space for the same $z$. As explained in the data section above, we just regard this
data as an average over different scales, assuming that non-linear
corrections have been correctly taken into account by the respective
experimental collaboration. Since we do not have to take derivatives
of $E_{g}$, or equivalently $P_{2}$, this leaves us with three possible
redshift bins, centered at $z_{1}=0.294$, $z_{2}=0.580$ and $z_{3}=0.860$, all of them with
an approximate bin width of $\Delta z \approx 0.29$.
At these redshifts we obtain $\eta_{{\rm obs}}(z_{1})=0.48 \pm 0.45$,
$\eta_{{\rm obs}}(z_{2})=-0.03 \pm 0.34$ and $\eta_{{\rm obs}}(z_{3})=-2.78 \pm 6.84$.
These values and the estimation of the intermediate model-independent
quantities can be seen in Table \ref{tab.results}.

Regarding the Gaussian Process method, we have computed the normalized Hubble function and its derivative,
$E(z)$ and $E'(z)$ with the \textit{dgp} module of the GaPP code.
We reconstructed the $E(z)$ and $E'(z)$ for the redshift interval
of the data using the Gaussian function as the covariance function and initial
values of the hyperparameters $\theta=[\sigma_{f}=0.5,\ell_{f}=0.5]$ that later are estimated
by the code. The same procedure was done for the $P_{2}(z)$ data.
We obtain for $E(z)$ and $E'(z)$ functions the hyperparameters $\sigma_{f}=2.12$ and $\ell_{f}=2.06$ and
for the $P_{2}$ function, $\sigma_{f}=0.58$ and $\ell_{f}=0.67$.

For the $P_3(z)$ observable, the hyperparameters obtained by the GaPP code led to 
a very flat and unrealistic reconstruction, that suggested us to take another approach for obtaining the optimal hyperparameters. 
We sampled the logarithm of the marginal likelihood on a grid of hyperparameters $\sigma_{f}$, $\ell_{f}$ 
from 0.01 to 2, setting this way a prior with the redshift range of the dataset,
and  300 points equally separated in log-space for each dimension. 
Remember that the hyperparameter $\ell_{f}$ constrains the typical scale on the independent variable $z$. 
Thus, as an additional prior, we impose that $\ell_{f}$ needs to be smaller than the redshift range of the data, which was not guaranteed by the default GaPP code.
Then we chose the pair of hyperparameters corresponding to the maximum
of the log-marginal likehood.
Therefore, for the $\ln (f\sigma_{8}(z))$ data, we obtain $\sigma_{f}=0.549$ and $\ell_{f}=1.361$.
Its reconstructed derivative $P_3$ can be seen in the lower right panel of Figure \ref{plt.rec}.
The function remains relatively flat, compared to the one given by other methods, but this approach has improved the 
determination of this observable.

Regarding the choice of the kernel function, several functions
were compared, each of them with a different number of parameters to see the impact
on the output. We tested the Gaussian kernel with two parameters, ($\sigma_{f},\ell_f$);
the rational quadratic kernel with three parameters and the double Gaussian
kernel with four parameters (see the original reference for the explicit implemented
formula \cite{Seikel2012}). We performed tests using the $H(z)$
data obtained with the cosmic chronometer technique and the $f\sigma_{8}(z)$
data. Our tests show that the different choices shift the reconstructed function
up to $6\%$ on its central value compared
to the Gaussian kernel function. This happens for $H(z)$ while the effect is negligible
for $f\sigma_{8}(z)$. 
Taking into account the above choices and procedure, we report that with the Gaussian Process method 
we obtain $\eta_{{\rm obs}}(z_{1})=  0.38 \pm 0.23 $,
$\eta_{{\rm obs}}(z_{2})=0.91 \pm 0.36 $ and $\eta_{{\rm obs}}(z_{3})=0.58 \pm 0.93$.

For the polynomial regression method, we find $\eta_{{\rm obs}}(z_{1})=0.57\pm1.05$,
$\eta_{{\rm obs}}(z_{2})=0.48\pm0.96$ and $\eta_{{\rm obs}}(z_{3})=-0.11\pm3.21$.
Note that we applied the criteria of a $\chi_{red}^{2}$ closest to
one and a positive definite Fisher matrix to chose the order of the
polynomial for each of the datasets. These criteria led to a choice
of a polynomial of order 3 for the $E(z)$ and $E_{g}(z)$ data and
order 6 for the $\ln (f\sigma_{8}(z))$ data. These polynomials can be seen in 
Figure \ref{plt.rec} as solid yellow lines, together with their $1\sigma$ uncertainty bands.
The higher order of the polynomial of  $\ln (f\sigma_{8}(z))$ explains the "bumpiness" 
of the reconstruction of $P_3$, leading to larger errors on this observable in comparison to the GP method.

\begin{figure}[htp]
\includegraphics[width=0.7\textwidth]{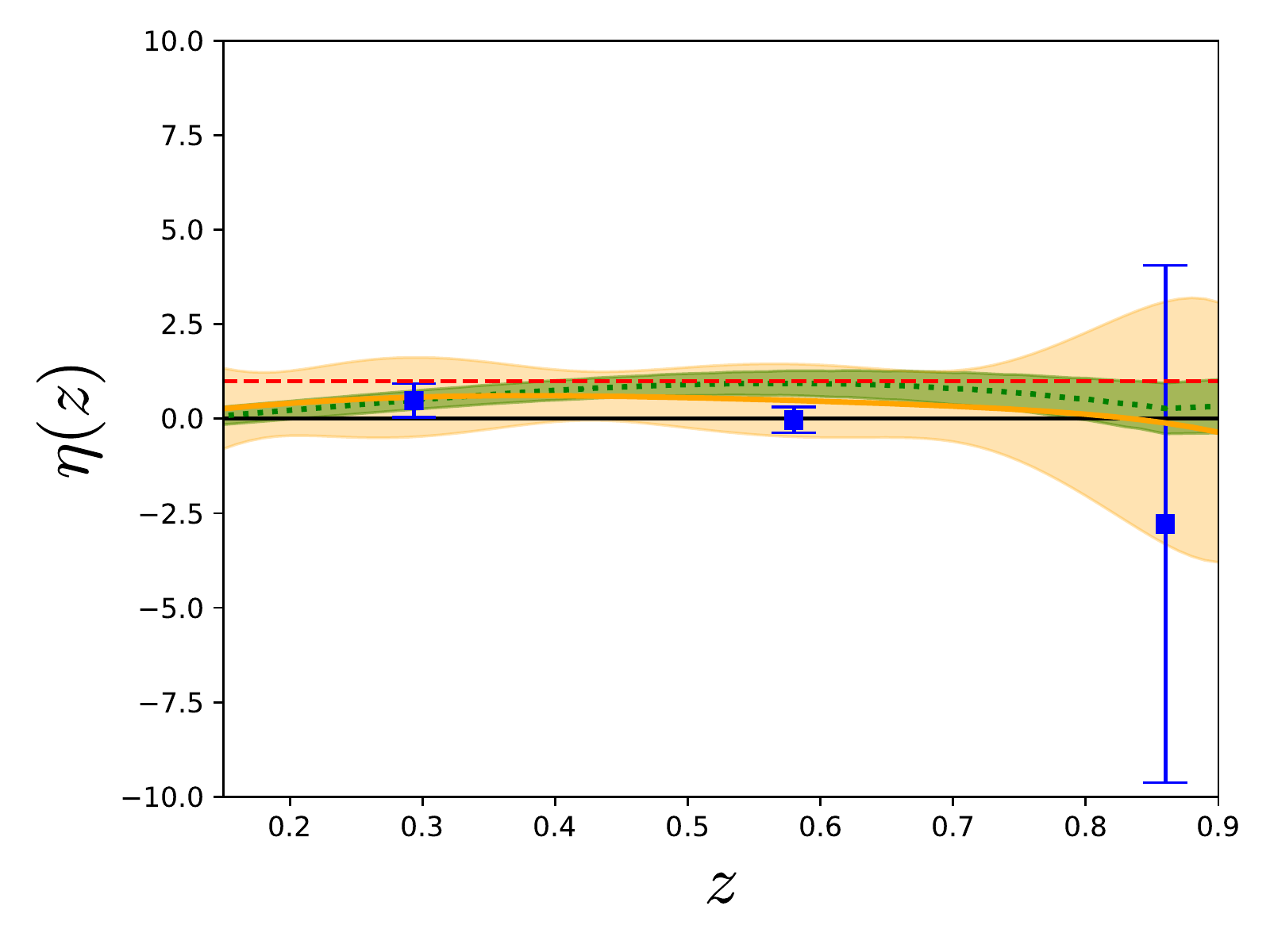}
\caption{\label{fig:Plot-of-the} Plot of the reconstructed $\eta_{\textrm{obs}}$ as a function of redshift, using the binning method (blue squares), Gaussian Process
(green dotted line) and linear regression (yellow solid line). The corresponding error bands (error bars for the binning method),
represent the 1 estimated error on the reconstruction. As a reference, we show in a dashed red line, the value in a standard
gravity scenario. }
\end{figure}

In Fig. \ref{fig:Plot-of-the} we show the reconstructed $\eta_{\rm obs}$
as a function of redshift with the three different methods, again with GP in a green dashed line,
polynomial regression in a yellow solid line and the binning method in blue squares with error bars. 
It is possible to
conclude that the methods are consistent with each other, within their  $1\sigma$ uncertainties and that
in most bins the results are consistent with the standard gravity scenario.
We find that the error bars of the Gaussian Process reconstruction are generally
smaller than the other methods, such that at the lowest redshift, 
GP is not compatible with $\eta_{\rm obs}=1$ at nearly $2\sigma$, 
while in the case of the binning method at the intermediate redshift, $z=0.58$, the tension is nearly $3\sigma$.

Finally, we can combine the estimates at three redshifts
of Table \ref{tab.results} into a single value. Assuming a constant
$\eta_{\rm obs}$ in this entire observed range and performing a simple
weighted average, we find finally $\eta_{\rm obs}=0.15 \pm 0.27$ (binning),
$\eta_{\rm obs}=0.53 \pm 0.19$ (Gaussian Process) and $\eta_{\rm obs}=0.49 \pm 0.69$
(polynomial regression).
The Gaussian Process method yields the smallest error and would exclude standard gravity. 
However, despite being sometimes advertised as ``model-independent'', we believe that 
this method actually makes a strong assumption, since it compresses the ignorance
about the reconstruction into a kernel function that depends on two or a small number of parameters, 
which are often not even fully marginalized over, as we did in our case. 
Also the binning method taken at face value would rule out standard gravity. However, as already mentioned, 
we did not take into account the correlation induced by the finite differences, and this might have decreased
the overall error.
Overall, we think the polynomial regression method is the most satisfactory one, 
providing the best compromise between the least number of assumptions and the best 
estimation of the data derivative. Therefore, we consider it as our ``fiducial'' result.

\section{Conclusion}

Measuring the properties of gravity at large scales is one of the main tasks of cosmology for the next years. Several large observational campaigns that are underway, or will soon be \cite{Laureijs:2011gra,Amendola:2012ys,SKAmission,2009arXiv0912.0201L,2013arXiv1308.0847L}, will collect enough data on galaxy clustering and lensing to render this task possible to a high level of accuracy.

In order to test gravity one has to provide an alternative, either a full model or at least some parametrization that goes beyond Einstein's gravity. Here we chose to consider the Horndeski Lagrangian because, although based on a single scalar field, displays most of the properties that make the field of modified gravity models such a rich area of research. We connected a more theoretical-oriented parameterization, the $\alpha_i$ parameters of Ref. \cite{Bellini:2014fua} with more phenomenologically-oriented ones, the $h_i$ parameters. To gain a deeper physical understanding, we discuss also some interesting limiting cases,  how the Newtonian potentials look in real space, and the impact of the constraints from gravitational wave speed.

The first practical goal in cosmology is to test and possibly rule out specific models of gravity and background expansion. For instance, one can rule out $\Lambda$CDM in a number of way, the simplest of which being measuring a deviation from the predicted $H(z)$ behavior (which is not equivalent, as we have seen, to simply finding deviations from a $w=-1$ equation of state). Models in which gravity is modified can often be designed to have a perfect $\Lambda$CDM background, so it is necessary to test them at perturbation level. 

Here, however, a problem arises, namely that many more assumptions need generally to be made. Some of them, listed in sec. \ref{sec:mod-dep}, are in some sense of fundamental character, and we follow them in this work. However, most cosmological analyses that test gravity assume in addition one or more of the following assumptions: 1) that the initial conditions are given by a simple inflationary spectrum described by one or two parameters; 2) that the cosmological evolution at $z$ larger than a few is given by a pure CDM dominated Universe living in standard gravity; 3) that the linear bias depends only on time and not on scale; and 4) that the value of some parameters, like $\Omega_{m0}$ obtained from CMB analyses assuming $\Lambda$CDM, can be applied also to different models.

We have shown that a statistics called $\eta_{\textrm{obs}}$ can be measured without any of the assumptions 1-4. This statistics is an estimator of the anisotropic stress parameter $\eta$, one of the two phenomenological functions of linearized, scalar, sub-horizon modified gravity. In this sense, we say that $\eta_{\textrm{obs}}$ is (relatively) model independent. If $\eta_{\textrm{obs}}$ differs from unity, either gravity is modified, or at least one of the four "fundamental" assumptions of sec. \ref{sec:mod-dep} are false.

We provided a preliminary estimate of $\eta_{\textrm{obs}}$ based on currently available data, $\eta_{\textrm{obs}}=0.49\pm0.69$ in the redshift range $z=(0.2-0.8)$. The full $k$- and $z$-dependence is still unaccessible with current data.
According to \cite{Amendola:2012ky}, the Euclid mission will be able to measure $\eta$ to a few percent, so almost two orders of magnitude better than current values, and begin to put interesting limits on the $k$- and $z$-dependence. As the philosopher of science Alexandre Koyré said concerning the emergence of modern science\footnote{Koyré, A. (1948). {\it Du monde de l’à peu près à l’univers de la précision}. In A. Koyré (Ed.), Etudes d’histoire de la pensée philosophique (pp. 341-362). Paris: Gallimard. }, also in measuring gravity at cosmological scales, we will finally move "from the world of approximation to the Universe of precision".

\section*{Acknowledgments}
We acknowledge DFG for support through the TR33 project "The Dark Universe". We thank Alejandro Guarnizo Trilleras, Jiaming Zhao, Adri{\`a} G{\'o}mez, Guillem Dom{\`e}nech, Martin Kunz, Ippocratis Saltas, Mariele Motta, and Ignacy Sawicki for discussions and collaboration on these topic. We also wish to thank Jiaming Zhao for his valuable comments. DB acknowledges support from the Attracc\'ion  del Talento  Cient\'ifico  en  Salamanca programme 
 during the last stages of this work. A.M.P. gratefully acknowledges the support by the Landesgraduiertenförderung
(LGF) grant of the Graduiertenakademie Universität Heidelberg. S.C. acknowledges support from CNES and CNRS grants.

\appendix \section{Background equations of motion of the Horndeski Lagrangian}\label{sec:appendix}

The equations of motion for a flat Friedmann-Lemaître-Robertson-Walker (FLRW) metric are  \cite{DeFelice:2011hq,Bellini:2014fua}
\begin{align}
 & 3M_{\star}^2H^{2}=\rho_m+\rho_{HL}\label{eq:Friedmann}\\
 & M_{\star}^2(2\dot{H}+3H^{2})=-p_m-p_{HL}\nonumber 
\end{align}
Here $\rho_m,p_m$ are the conserved background energy density and the
pressure of matter, respectively, and analogously ${\rho}_{HL}$
and ${p}_{HL}$ are the background energy density and pressure
of the Horndeski field, defined as 
\begin{eqnarray}
{\rho}_{HL} & \equiv & -K+2X\left(K_{X}-G_{3\phi}\right)+6\dot{\phi}H\left(XG_{3X}-G_{4\phi}-2XG_{4\phi X}\right)\\
 &  & +12H^{2}X\left(G_{4X}+2XG_{4XX}-G_{5\phi}-XG_{5\phi X}\right)+4\dot{\phi}H^{3}X\left(G_{5X}+XG_{5XX}\right)\,,\nonumber \\
{p}_{HL} & = & K-2X\left(G_{3\phi}-2G_{4\phi\phi}\right)+4\dot{\phi}H\left(G_{4\phi}-2XG_{4\phi X}+XG_{5\phi\phi}\right)-M_{*}^{2}\alpha_{\text{B}}H\frac{\ddot{\phi}}{\dot{\phi}}\\
 &  & +2\dot{\phi}H^{3}XG_{5X}-4H^{2}X^{2}G_{5\phi X}\,,\nonumber 
\end{eqnarray}
where $M_{\star}$ is defined in eq. (\ref{eq:planckmass}) and $\alpha_B$  in eq. (\ref{eq:ab}). Since in the literature there appear various definitions of the energy density associated to the scalar field, we report in table \ref{tab:notation} the relation between the one adopted in this work and others.\footnote{We thank Jiaming Zhao for pointing out potential issues related to this.} Following \cite{Bellini:2014fua}, we write the equation of motion of the scalar field $\phi$ as a (non-)conservation equation of the
 "shift-charge density" $n$,
\[
\dot{n}+3H n={p}_{HL,\phi}
\]
with 
\begin{eqnarray}
n & \equiv & \dot{\phi}\left(K_{X}-2G_{3\phi}\right)+6HX\left(G_{3X}-2G_{4\phi X}\right)+\label{consshift}\\
 &  & +6H^{2}\dot{\phi}\left(G_{4X}+2XG_{4XX}-G_{5\phi}-XG_{5\phi X}\right)+\nonumber \\
 &  & +2H^{3}X\left(3G_{5X}+2XG_{5XX}\right)\,,\nonumber 
\end{eqnarray}
and the non-conservation term, driven by a violation of the shift symmetry $\phi\to\phi+\mathrm{const}$, given
by the $\phi-$derivative of  $p_{HL}$
\begin{align}
p_{HL,\phi}\equiv & K_{\phi}-2XG_{3\phi\phi}+2\ddot{\phi}\left(XG_{3\phi X}+3H\dot{\phi}G_{4\phi X}\right)+6\dot{H}G_{4\phi}+\label{eq:non-shift}\\
 & +6H^{2}\left(2G_{4\phi}+2XG_{4\phi X}-XG_{5\phi\phi}\right)+2H^{3}\dot{\phi}XG_{5\phi X\,.}\nonumber 
\end{align}

\begin{table}[]
    \centering
    \begin{tabular}{c|c|c|c}
    \hline
         & This work & Ref. \cite{Zumalacarregui:2016pph}  & Ref.  \cite{Bellini:2014fua}\\
         \hline \hline
       Energy density  & $\rho_{HL}$ & $\rho_{DE}=\frac{\rho_{HL}}{3}-H^2(M^2_*-1)$   & $\tilde{\mathcal{E}}=\frac{\rho_{HL}}{M^2_*}\equiv \tilde \rho_{HL}$\\
       \hline
       Pressure & $p_{HL}$ & $p_{DE} = \frac{p_{HL}}{3}+(3H^2+2 \dot H)(M^2_*-1)$ & $\tilde{\mathcal{P}}= \frac{p_{HL}}{M^2_*}\equiv \tilde p_{HL}$\\
      \hline
    \end{tabular}
    \caption{Comparison table for the different definition of energy density and pressure of the scalar field}
    \label{tab:notation}
\end{table}

\begin{table}[h]
	\centering
	\begin{tabular}{c|c|ccc|c}
		\hline
		Method  	&	Parameter		& 							& Redshift bins					& 					& Weighted mean \\
		\qquad		&					& $z_1 = 0.294$ 			& $z_2 = 0.58$ 				& $z_3 = 0.86$ 			& \\ 
		\hline \hline		
					& $E(z)$ 			& $1.12 \pm 0.01$ 			& $1.27 \pm 0.02$ 			& $1.51 \pm 0.02$ 		& \\
					& $E'(z)$ 			& $-0.56 \pm 0.07$ 			& $-0.60 \pm 0.36$ 			& $-1.75 \pm 0.66$ 		& \\
		Binning		& $P_2(z)$ 			& $0.75 \pm 0.10$ 			& $0.54 \pm 0.07$ 			& $0.18 \pm 0.14$ 		& \\
					& $P_3(z)$	 		& $-0.17 \pm 0.35$ 			& $0.53 \pm 0.61$ 			& $-1.27 \pm 1.52$ 		& \\
					& $\eta_{obs}(z)$ 	& $0.48 \pm 0.45$ 			& $-0.03 \pm 0.34$			& $-2.78 \pm 6.84$ 		& $0.15 \pm 0.27$ \\
		\hline
					& $E(z)$ 			& $1.10 \pm 0.01 $ 			& $1.30 \pm 0.02 $ 			& $1.55 \pm 0.03 $ 		& \\
					& $E'(z)$ 			& $-0.73 \pm 0.05 $ 		& $-1.30 \pm 0.10 $ 		& $-1.89 \pm 0.16 $ 	& \\
		Gaussian Process& $P_2(z)$ 		& $ 0.74 \pm 0.09 $ 		& $ 0.53 \pm 0.06 $ 		& $ 0.23 \pm 0.11 $ 	& \\
					& $P_3(z)$	 		& $ -0.10 \pm 0.20 $ 		& $ -0.03 \pm 0.21 $ 		& $ -0.21 \pm 0.30 $ 	& \\
					& $\eta_{obs}(z)$ 	& $ 0.38 \pm 0.23 $ 		& $ 0.91 \pm 0.36 $			& $ 0.58 \pm 0.93 $ 	& $0.53 \pm 0.19$ \\
		\hline
							& $E(z)$ 			& $1.12 \pm 0.01$ 		& $1.29 \pm 0.02 $ 		& $1.50 \pm 0.02 $ 		& \\
							& $E'(z)$ 			& $-0.73 \pm 0.04$ 		& $-1.06 \pm 0.04 $ 	& $-1.45 \pm 0.04 $ 	& \\
		Polynomial Regression	& $P_2(z)$ 		& $0.76 \pm 0.15$ 		& $ 0.55 \pm 0.15 $ 	& $ 0.18 \pm 0.14 $ 	& \\
							& $P_3(z)$			& $-0.09 \pm 0.80$ 		& $ 0.14 \pm 0.78 $		& $ -0.17 \pm 3.02 $ 	& \\
							& $\eta_{obs}(z)$ 	& $0.57 \pm 1.05$ 		& $ 0.48 \pm 0.96 $ 	& $-0.11 \pm 3.21 $ 	& $0.49 \pm 0.69$\\
		\hline
	\end{tabular}
	\caption{The reconstructed or measured model-independent variables $E, E', P_2, P_3, \eta(z)$ at three different redshifts $z=(0.294, 0.58, 0.86)$, together with their $1\sigma$ errors, for each of the reconstruction methods. The polynomial regression method is compabitle with the $\Lambda$CDM scenario while the other two methods show some tension at lower redshift.}
	\label{tab.results} 
\end{table}

\section*{}

\bibliography{scaling_bib,refs,AnisoRefs,AnisoRefs2}

\begin{thebibliography}{129}%
\makeatletter
\providecommand \@ifxundefined [1]{%
 \@ifx{#1\undefined}
}%
\providecommand \@ifnum [1]{%
 \ifnum #1\expandafter \@firstoftwo
 \else \expandafter \@secondoftwo
 \fi
}%
\providecommand \@ifx [1]{%
 \ifx #1\expandafter \@firstoftwo
 \else \expandafter \@secondoftwo
 \fi
}%
\providecommand \natexlab [1]{#1}%
\providecommand \enquote  [1]{``#1''}%
\providecommand \bibnamefont  [1]{#1}%
\providecommand \bibfnamefont [1]{#1}%
\providecommand \citenamefont [1]{#1}%
\providecommand \href@noop [0]{\@secondoftwo}%
\providecommand \href [0]{\begingroup \@sanitize@url \@href}%
\providecommand \@href[1]{\@@startlink{#1}\@@href}%
\providecommand \@@href[1]{\endgroup#1\@@endlink}%
\providecommand \@sanitize@url [0]{\catcode `\\12\catcode `\$12\catcode
  `\&12\catcode `\#12\catcode `\^12\catcode `\_12\catcode `\%12\relax}%
\providecommand \@@startlink[1]{}%
\providecommand \@@endlink[0]{}%
\providecommand \url  [0]{\begingroup\@sanitize@url \@url }%
\providecommand \@url [1]{\endgroup\@href {#1}{\urlprefix }}%
\providecommand \urlprefix  [0]{URL }%
\providecommand \Eprint [0]{\href }%
\providecommand \doibase [0]{http://dx.doi.org/}%
\providecommand \selectlanguage [0]{\@gobble}%
\providecommand \bibinfo  [0]{\@secondoftwo}%
\providecommand \bibfield  [0]{\@secondoftwo}%
\providecommand \translation [1]{[#1]}%
\providecommand \BibitemOpen [0]{}%
\providecommand \bibitemStop [0]{}%
\providecommand \bibitemNoStop [0]{.\EOS\space}%
\providecommand \EOS [0]{\spacefactor3000\relax}%
\providecommand \BibitemShut  [1]{\csname bibitem#1\endcsname}%
\let\auto@bib@innerbib\@empty
\bibitem [{\citenamefont {{Lovelock}}(1972)}]{1972JMP....13..874L}%
  \BibitemOpen
  \bibfield  {author} {\bibinfo {author} {\bibfnamefont {D.}~\bibnamefont
  {{Lovelock}}},\ }\bibfield  {title} {\enquote {\bibinfo {title} {{The
  Four-Dimensionality of Space and the Einstein Tensor}},}\ }\href {\doibase
  10.1063/1.1666069} {\bibfield  {journal} {\bibinfo  {journal} {Journal of
  Mathematical Physics}\ }\textbf {\bibinfo {volume} {13}},\ \bibinfo {pages}
  {874--876} (\bibinfo {year} {1972})}\BibitemShut {NoStop}%
\bibitem [{\citenamefont {Clifton}\ \emph {et~al.}(2012)\citenamefont
  {Clifton}, \citenamefont {Ferreira}, \citenamefont {Padilla},\ and\
  \citenamefont {Skordis}}]{Clifton:2011jh}%
  \BibitemOpen
  \bibfield  {author} {\bibinfo {author} {\bibfnamefont {Timothy}\ \bibnamefont
  {Clifton}}, \bibinfo {author} {\bibfnamefont {Pedro~G.}\ \bibnamefont
  {Ferreira}}, \bibinfo {author} {\bibfnamefont {Antonio}\ \bibnamefont
  {Padilla}}, \ and\ \bibinfo {author} {\bibfnamefont {Constantinos}\
  \bibnamefont {Skordis}},\ }\bibfield  {title} {\enquote {\bibinfo {title}
  {{Modified Gravity and Cosmology}},}\ }\href {\doibase
  10.1016/j.physrep.2012.01.001} {\bibfield  {journal} {\bibinfo  {journal}
  {Phys. Rept.}\ }\textbf {\bibinfo {volume} {513}},\ \bibinfo {pages} {1--189}
  (\bibinfo {year} {2012})},\ \Eprint {http://arxiv.org/abs/1106.2476}
  {arXiv:1106.2476 [astro-ph.CO]} \BibitemShut {NoStop}%
\bibitem [{\citenamefont {Abbott}\ \emph
  {et~al.}(2017{\natexlab{a}})\citenamefont {Abbott} \emph
  {et~al.}}]{LIGO:2017qsa}%
  \BibitemOpen
  \bibfield  {author} {\bibinfo {author} {\bibfnamefont {B.~P.}\ \bibnamefont
  {Abbott}} \emph {et~al.} (\bibinfo {collaboration} {Virgo, LIGO
  Scientific}),\ }\bibfield  {title} {\enquote {\bibinfo {title} {{GW170817:
  Observation of Gravitational Waves from a Binary Neutron Star Inspiral}},}\
  }\href {\doibase 10.1103/PhysRevLett.119.161101} {\bibfield  {journal}
  {\bibinfo  {journal} {Phys. Rev. Lett.}\ }\textbf {\bibinfo {volume} {119}},\
  \bibinfo {pages} {161101} (\bibinfo {year} {2017}{\natexlab{a}})},\ \Eprint
  {http://arxiv.org/abs/1710.05832} {arXiv:1710.05832 [gr-qc]} \BibitemShut
  {NoStop}%
\bibitem [{\citenamefont {{Papantonopoulos}}(2015)}]{2015LNP...892.....P}%
  \BibitemOpen
  \bibinfo {editor} {\bibfnamefont {E.}~\bibnamefont {{Papantonopoulos}}},\
  ed.,\ \href {\doibase 10.1007/978-3-319-10070-8} {\emph {\bibinfo {title}
  {Lecture Notes in Physics, Berlin Springer Verlag}}},\ \bibinfo {series}
  {Lecture Notes in Physics, Berlin Springer Verlag}, Vol.\ \bibinfo {volume}
  {892}\ (\bibinfo {year} {2015})\BibitemShut {NoStop}%
\bibitem [{\citenamefont {Horndeski}(1974)}]{Horndeski:1974wa}%
  \BibitemOpen
  \bibfield  {author} {\bibinfo {author} {\bibfnamefont {Gregory~Walter}\
  \bibnamefont {Horndeski}},\ }\bibfield  {title} {\enquote {\bibinfo {title}
  {{Second-order scalar-tensor field equations in a four-dimensional space}},}\
  }\href {\doibase 10.1007/BF01807638} {\bibfield  {journal} {\bibinfo
  {journal} {Int. J. Theor. Phys.}\ }\textbf {\bibinfo {volume} {10}},\
  \bibinfo {pages} {363--384} (\bibinfo {year} {1974})}\BibitemShut {NoStop}%
\bibitem [{\citenamefont {Deffayet}\ \emph {et~al.}(2009)\citenamefont
  {Deffayet}, \citenamefont {Esposito-Farese},\ and\ \citenamefont
  {Vikman}}]{Deffayet:2009wt}%
  \BibitemOpen
  \bibfield  {author} {\bibinfo {author} {\bibfnamefont {C.}~\bibnamefont
  {Deffayet}}, \bibinfo {author} {\bibfnamefont {Gilles}\ \bibnamefont
  {Esposito-Farese}}, \ and\ \bibinfo {author} {\bibfnamefont {A.}~\bibnamefont
  {Vikman}},\ }\bibfield  {title} {\enquote {\bibinfo {title} {{Covariant
  Galileon}},}\ }\href {\doibase 10.1103/PhysRevD.79.084003} {\bibfield
  {journal} {\bibinfo  {journal} {Phys. Rev.}\ }\textbf {\bibinfo {volume}
  {D79}},\ \bibinfo {pages} {084003} (\bibinfo {year} {2009})},\ \Eprint
  {http://arxiv.org/abs/0901.1314} {arXiv:0901.1314 [hep-th]} \BibitemShut
  {NoStop}%
\bibitem [{\citenamefont {Zumalacarregui}\ and\ \citenamefont
  {García-Bellido}(2014)}]{Zumalacarregui:2013pma}%
  \BibitemOpen
  \bibfield  {author} {\bibinfo {author} {\bibfnamefont {Miguel}\ \bibnamefont
  {Zumalacarregui}}\ and\ \bibinfo {author} {\bibfnamefont {Juan}\ \bibnamefont
  {García-Bellido}},\ }\bibfield  {title} {\enquote {\bibinfo {title}
  {{Transforming gravity: from derivative couplings to matter to second-order
  scalar-tensor theories beyond the Horndeski Lagrangian}},}\ }\href {\doibase
  10.1103/PhysRevD.89.064046} {\bibfield  {journal} {\bibinfo  {journal} {Phys.
  Rev.}\ }\textbf {\bibinfo {volume} {D89}},\ \bibinfo {pages} {064046}
  (\bibinfo {year} {2014})},\ \Eprint {http://arxiv.org/abs/1308.4685}
  {arXiv:1308.4685 [gr-qc]} \BibitemShut {NoStop}%
\bibitem [{\citenamefont {Gleyzes}\ \emph
  {et~al.}(2015{\natexlab{a}})\citenamefont {Gleyzes}, \citenamefont
  {Langlois}, \citenamefont {Piazza},\ and\ \citenamefont
  {Vernizzi}}]{Gleyzes:2014qga}%
  \BibitemOpen
  \bibfield  {author} {\bibinfo {author} {\bibfnamefont {Jérôme}\
  \bibnamefont {Gleyzes}}, \bibinfo {author} {\bibfnamefont {David}\
  \bibnamefont {Langlois}}, \bibinfo {author} {\bibfnamefont {Federico}\
  \bibnamefont {Piazza}}, \ and\ \bibinfo {author} {\bibfnamefont {Filippo}\
  \bibnamefont {Vernizzi}},\ }\bibfield  {title} {\enquote {\bibinfo {title}
  {{Exploring gravitational theories beyond Horndeski}},}\ }\href {\doibase
  10.1088/1475-7516/2015/02/018} {\bibfield  {journal} {\bibinfo  {journal}
  {JCAP}\ }\textbf {\bibinfo {volume} {1502}},\ \bibinfo {pages} {018}
  (\bibinfo {year} {2015}{\natexlab{a}})},\ \Eprint
  {http://arxiv.org/abs/1408.1952} {arXiv:1408.1952 [astro-ph.CO]} \BibitemShut
  {NoStop}%
\bibitem [{\citenamefont {Gleyzes}\ \emph
  {et~al.}(2015{\natexlab{b}})\citenamefont {Gleyzes}, \citenamefont
  {Langlois}, \citenamefont {Piazza},\ and\ \citenamefont
  {Vernizzi}}]{Gleyzes:2014dya}%
  \BibitemOpen
  \bibfield  {author} {\bibinfo {author} {\bibfnamefont {Jérôme}\
  \bibnamefont {Gleyzes}}, \bibinfo {author} {\bibfnamefont {David}\
  \bibnamefont {Langlois}}, \bibinfo {author} {\bibfnamefont {Federico}\
  \bibnamefont {Piazza}}, \ and\ \bibinfo {author} {\bibfnamefont {Filippo}\
  \bibnamefont {Vernizzi}},\ }\bibfield  {title} {\enquote {\bibinfo {title}
  {{Healthy theories beyond Horndeski}},}\ }\href {\doibase
  10.1103/PhysRevLett.114.211101} {\bibfield  {journal} {\bibinfo  {journal}
  {Phys. Rev. Lett.}\ }\textbf {\bibinfo {volume} {114}},\ \bibinfo {pages}
  {211101} (\bibinfo {year} {2015}{\natexlab{b}})},\ \Eprint
  {http://arxiv.org/abs/1404.6495} {arXiv:1404.6495 [hep-th]} \BibitemShut
  {NoStop}%
\bibitem [{\citenamefont {Crisostomi}\ \emph
  {et~al.}(2016{\natexlab{a}})\citenamefont {Crisostomi}, \citenamefont
  {Koyama},\ and\ \citenamefont {Tasinato}}]{Crisostomi:2016czh}%
  \BibitemOpen
  \bibfield  {author} {\bibinfo {author} {\bibfnamefont {Marco}\ \bibnamefont
  {Crisostomi}}, \bibinfo {author} {\bibfnamefont {Kazuya}\ \bibnamefont
  {Koyama}}, \ and\ \bibinfo {author} {\bibfnamefont {Gianmassimo}\
  \bibnamefont {Tasinato}},\ }\bibfield  {title} {\enquote {\bibinfo {title}
  {{Extended Scalar-Tensor Theories of Gravity}},}\ }\href {\doibase
  10.1088/1475-7516/2016/04/044} {\bibfield  {journal} {\bibinfo  {journal}
  {JCAP}\ }\textbf {\bibinfo {volume} {1604}},\ \bibinfo {pages} {044}
  (\bibinfo {year} {2016}{\natexlab{a}})},\ \Eprint
  {http://arxiv.org/abs/1602.03119} {arXiv:1602.03119 [hep-th]} \BibitemShut
  {NoStop}%
\bibitem [{\citenamefont {Crisostomi}\ \emph
  {et~al.}(2016{\natexlab{b}})\citenamefont {Crisostomi}, \citenamefont {Hull},
  \citenamefont {Koyama},\ and\ \citenamefont {Tasinato}}]{Crisostomi:2016tcp}%
  \BibitemOpen
  \bibfield  {author} {\bibinfo {author} {\bibfnamefont {Marco}\ \bibnamefont
  {Crisostomi}}, \bibinfo {author} {\bibfnamefont {Matthew}\ \bibnamefont
  {Hull}}, \bibinfo {author} {\bibfnamefont {Kazuya}\ \bibnamefont {Koyama}}, \
  and\ \bibinfo {author} {\bibfnamefont {Gianmassimo}\ \bibnamefont
  {Tasinato}},\ }\bibfield  {title} {\enquote {\bibinfo {title} {{Horndeski:
  beyond, or not beyond?}}}\ }\href {\doibase 10.1088/1475-7516/2016/03/038}
  {\bibfield  {journal} {\bibinfo  {journal} {JCAP}\ }\textbf {\bibinfo
  {volume} {1603}},\ \bibinfo {pages} {038} (\bibinfo {year}
  {2016}{\natexlab{b}})},\ \Eprint {http://arxiv.org/abs/1601.04658}
  {arXiv:1601.04658 [hep-th]} \BibitemShut {NoStop}%
\bibitem [{\citenamefont {Ben~Achour}\ \emph {et~al.}(2016)\citenamefont
  {Ben~Achour}, \citenamefont {Langlois},\ and\ \citenamefont
  {Noui}}]{Achour:2016rkg}%
  \BibitemOpen
  \bibfield  {author} {\bibinfo {author} {\bibfnamefont {Jibril}\ \bibnamefont
  {Ben~Achour}}, \bibinfo {author} {\bibfnamefont {David}\ \bibnamefont
  {Langlois}}, \ and\ \bibinfo {author} {\bibfnamefont {Karim}\ \bibnamefont
  {Noui}},\ }\bibfield  {title} {\enquote {\bibinfo {title} {{Degenerate higher
  order scalar-tensor theories beyond Horndeski and disformal
  transformations}},}\ }\href {\doibase 10.1103/PhysRevD.93.124005} {\bibfield
  {journal} {\bibinfo  {journal} {Phys. Rev.}\ }\textbf {\bibinfo {volume}
  {D93}},\ \bibinfo {pages} {124005} (\bibinfo {year} {2016})},\ \Eprint
  {http://arxiv.org/abs/1602.08398} {arXiv:1602.08398 [gr-qc]} \BibitemShut
  {NoStop}%
\bibitem [{\citenamefont {Langlois}\ and\ \citenamefont
  {Noui}(2016)}]{Langlois:2015cwa}%
  \BibitemOpen
  \bibfield  {author} {\bibinfo {author} {\bibfnamefont {David}\ \bibnamefont
  {Langlois}}\ and\ \bibinfo {author} {\bibfnamefont {Karim}\ \bibnamefont
  {Noui}},\ }\bibfield  {title} {\enquote {\bibinfo {title} {{Degenerate higher
  derivative theories beyond Horndeski: evading the Ostrogradski
  instability}},}\ }\href {\doibase 10.1088/1475-7516/2016/02/034} {\bibfield
  {journal} {\bibinfo  {journal} {JCAP}\ }\textbf {\bibinfo {volume} {1602}},\
  \bibinfo {pages} {034} (\bibinfo {year} {2016})},\ \Eprint
  {http://arxiv.org/abs/1510.06930} {arXiv:1510.06930 [gr-qc]} \BibitemShut
  {NoStop}%
\bibitem [{\citenamefont {Kobayashi}(2019)}]{Kobayashi:2019hrl}%
  \BibitemOpen
  \bibfield  {author} {\bibinfo {author} {\bibfnamefont {Tsutomu}\ \bibnamefont
  {Kobayashi}},\ }\bibfield  {title} {\enquote {\bibinfo {title} {{Horndeski
  theory and beyond: a review}},}\ }\href@noop {} {\  (\bibinfo {year}
  {2019})},\ \Eprint {http://arxiv.org/abs/1901.07183} {arXiv:1901.07183
  [gr-qc]} \BibitemShut {NoStop}%
\bibitem [{\citenamefont {Bettoni}\ and\ \citenamefont
  {Liberati}(2013)}]{Bettoni:2013diz}%
  \BibitemOpen
  \bibfield  {author} {\bibinfo {author} {\bibfnamefont {Dario}\ \bibnamefont
  {Bettoni}}\ and\ \bibinfo {author} {\bibfnamefont {Stefano}\ \bibnamefont
  {Liberati}},\ }\bibfield  {title} {\enquote {\bibinfo {title} {{Disformal
  invariance of second order scalar-tensor theories: Framing the Horndeski
  action}},}\ }\href {\doibase 10.1103/PhysRevD.88.084020} {\bibfield
  {journal} {\bibinfo  {journal} {Phys. Rev.}\ }\textbf {\bibinfo {volume}
  {D88}},\ \bibinfo {pages} {084020} (\bibinfo {year} {2013})},\ \Eprint
  {http://arxiv.org/abs/1306.6724} {arXiv:1306.6724 [gr-qc]} \BibitemShut
  {NoStop}%
\bibitem [{\citenamefont {Sotiriou}\ and\ \citenamefont
  {Faraoni}(2010)}]{Sotiriou:2008rp}%
  \BibitemOpen
  \bibfield  {author} {\bibinfo {author} {\bibfnamefont {Thomas~P.}\
  \bibnamefont {Sotiriou}}\ and\ \bibinfo {author} {\bibfnamefont {Valerio}\
  \bibnamefont {Faraoni}},\ }\bibfield  {title} {\enquote {\bibinfo {title}
  {{f(R) Theories Of Gravity}},}\ }\href {\doibase 10.1103/RevModPhys.82.451}
  {\bibfield  {journal} {\bibinfo  {journal} {Rev. Mod. Phys.}\ }\textbf
  {\bibinfo {volume} {82}},\ \bibinfo {pages} {451--497} (\bibinfo {year}
  {2010})},\ \Eprint {http://arxiv.org/abs/0805.1726} {arXiv:0805.1726 [gr-qc]}
  \BibitemShut {NoStop}%
\bibitem [{\citenamefont {{De Felice}}\ \emph {et~al.}(2011)\citenamefont {{De
  Felice}}, \citenamefont {Kobayashi},\ and\ \citenamefont
  {Tsujikawa}}]{DeFelice2011}%
  \BibitemOpen
  \bibfield  {author} {\bibinfo {author} {\bibfnamefont {Antonio}\ \bibnamefont
  {{De Felice}}}, \bibinfo {author} {\bibfnamefont {Tsutomu}\ \bibnamefont
  {Kobayashi}}, \ and\ \bibinfo {author} {\bibfnamefont {Shinji}\ \bibnamefont
  {Tsujikawa}},\ }\bibfield  {title} {\enquote {\bibinfo {title} {{Effective
  gravitational couplings for cosmological perturbations in the most general
  scalar–tensor theories with second-order field equations}},}\ }\href
  {\doibase 10.1016/J.PHYSLETB.2011.11.028} {\bibfield  {journal} {\bibinfo
  {journal} {Physics Letters B}\ }\textbf {\bibinfo {volume} {706}},\ \bibinfo
  {pages} {123--133} (\bibinfo {year} {2011})}\BibitemShut {NoStop}%
\bibitem [{\citenamefont {De~Felice}\ and\ \citenamefont
  {Tsujikawa}(2012)}]{DeFelice:2011bh}%
  \BibitemOpen
  \bibfield  {author} {\bibinfo {author} {\bibfnamefont {Antonio}\ \bibnamefont
  {De~Felice}}\ and\ \bibinfo {author} {\bibfnamefont {Shinji}\ \bibnamefont
  {Tsujikawa}},\ }\bibfield  {title} {\enquote {\bibinfo {title} {{Conditions
  for the cosmological viability of the most general scalar-tensor theories and
  their applications to extended Galileon dark energy models}},}\ }\href
  {\doibase 10.1088/1475-7516/2012/02/007} {\bibfield  {journal} {\bibinfo
  {journal} {JCAP}\ }\textbf {\bibinfo {volume} {1202}},\ \bibinfo {pages}
  {007} (\bibinfo {year} {2012})},\ \Eprint {http://arxiv.org/abs/1110.3878}
  {arXiv:1110.3878 [gr-qc]} \BibitemShut {NoStop}%
\bibitem [{\citenamefont {Ostrogradski}(1850)}]{Ostrogradski:1850}%
  \BibitemOpen
  \bibfield  {author} {\bibinfo {author} {\bibfnamefont {M.}~\bibnamefont
  {Ostrogradski}},\ }\bibfield  {title} {\enquote {\bibinfo {title} {{}},}\
  }\href@noop {} {\bibfield  {journal} {\bibinfo  {journal} {Mem. Ac. St.
  Petersbourg}\ }\textbf {\bibinfo {volume} {VI}},\ \bibinfo {pages} {385}
  (\bibinfo {year} {1850})}\BibitemShut {NoStop}%
\bibitem [{\citenamefont {Woodard}(2015)}]{Woodard:2015zca}%
  \BibitemOpen
  \bibfield  {author} {\bibinfo {author} {\bibfnamefont {Richard~P.}\
  \bibnamefont {Woodard}},\ }\bibfield  {title} {\enquote {\bibinfo {title}
  {{Ostrogradsky's theorem on Hamiltonian instability}},}\ }\href {\doibase
  10.4249/scholarpedia.32243} {\bibfield  {journal} {\bibinfo  {journal}
  {Scholarpedia}\ }\textbf {\bibinfo {volume} {10}},\ \bibinfo {pages} {32243}
  (\bibinfo {year} {2015})},\ \Eprint {http://arxiv.org/abs/1506.02210}
  {arXiv:1506.02210 [hep-th]} \BibitemShut {NoStop}%
\bibitem [{\citenamefont {Chen}\ \emph {et~al.}(2013)\citenamefont {Chen},
  \citenamefont {Fasiello}, \citenamefont {Lim},\ and\ \citenamefont
  {Tolley}}]{Chen:2012au}%
  \BibitemOpen
  \bibfield  {author} {\bibinfo {author} {\bibfnamefont {Tai-jun}\ \bibnamefont
  {Chen}}, \bibinfo {author} {\bibfnamefont {Matteo}\ \bibnamefont {Fasiello}},
  \bibinfo {author} {\bibfnamefont {Eugene~A.}\ \bibnamefont {Lim}}, \ and\
  \bibinfo {author} {\bibfnamefont {Andrew~J.}\ \bibnamefont {Tolley}},\
  }\bibfield  {title} {\enquote {\bibinfo {title} {{Higher derivative theories
  with constraints: Exorcising Ostrogradski's Ghost}},}\ }\href {\doibase
  10.1088/1475-7516/2013/02/042} {\bibfield  {journal} {\bibinfo  {journal}
  {JCAP}\ }\textbf {\bibinfo {volume} {1302}},\ \bibinfo {pages} {042}
  (\bibinfo {year} {2013})},\ \Eprint {http://arxiv.org/abs/1209.0583}
  {arXiv:1209.0583 [hep-th]} \BibitemShut {NoStop}%
\bibitem [{\citenamefont {{Bellini}}\ and\ \citenamefont
  {{Sawicki}}(2014)}]{Bellini2014}%
  \BibitemOpen
  \bibfield  {author} {\bibinfo {author} {\bibfnamefont {E.}~\bibnamefont
  {{Bellini}}}\ and\ \bibinfo {author} {\bibfnamefont {I.}~\bibnamefont
  {{Sawicki}}},\ }\bibfield  {title} {\enquote {\bibinfo {title} {{Maximal
  freedom at minimum cost: linear large-scale structure in general
  modifications of gravity}},}\ }\href {\doibase 10.1088/1475-7516/2014/07/050}
  {\bibfield  {journal} {\bibinfo  {journal} {Journal of Cosmology and
  Astroparticle Physics}\ }\textbf {\bibinfo {volume} {7}},\ \bibinfo {eid}
  {050} (\bibinfo {year} {2014})},\ \Eprint {http://arxiv.org/abs/1404.3713}
  {arXiv:1404.3713} \BibitemShut {NoStop}%
\bibitem [{\citenamefont {{Amendola}}\ and\ \citenamefont
  {{Tsujikawa}}(2010)}]{2010deto.book.....A}%
  \BibitemOpen
  \bibfield  {author} {\bibinfo {author} {\bibfnamefont {L.}~\bibnamefont
  {{Amendola}}}\ and\ \bibinfo {author} {\bibfnamefont {S.}~\bibnamefont
  {{Tsujikawa}}},\ }\href@noop {} {\emph {\bibinfo {title} {Dark Energy :
  Theory and Observations by Luca Amendola and Shinji Tsujikawa.~Cambridge
  University Press, 2010.~ISBN: 9780521516006}}}\ (\bibinfo {year}
  {2010})\BibitemShut {NoStop}%
\bibitem [{\citenamefont {Weinberg}(2004)}]{Weinberg:2003ur}%
  \BibitemOpen
  \bibfield  {author} {\bibinfo {author} {\bibfnamefont {Steven}\ \bibnamefont
  {Weinberg}},\ }\bibfield  {title} {\enquote {\bibinfo {title} {{Damping of
  tensor modes in cosmology}},}\ }\href {\doibase 10.1103/PhysRevD.69.023503}
  {\bibfield  {journal} {\bibinfo  {journal} {Phys. Rev.}\ }\textbf {\bibinfo
  {volume} {D69}},\ \bibinfo {pages} {023503} (\bibinfo {year} {2004})},\
  \Eprint {http://arxiv.org/abs/astro-ph/0306304} {arXiv:astro-ph/0306304
  [astro-ph]} \BibitemShut {NoStop}%
\bibitem [{\citenamefont {De~Felice}\ \emph {et~al.}(2011)\citenamefont
  {De~Felice}, \citenamefont {Kobayashi},\ and\ \citenamefont
  {Tsujikawa}}]{DeFelice:2011hq}%
  \BibitemOpen
  \bibfield  {author} {\bibinfo {author} {\bibfnamefont {Antonio}\ \bibnamefont
  {De~Felice}}, \bibinfo {author} {\bibfnamefont {Tsutomu}\ \bibnamefont
  {Kobayashi}}, \ and\ \bibinfo {author} {\bibfnamefont {Shinji}\ \bibnamefont
  {Tsujikawa}},\ }\bibfield  {title} {\enquote {\bibinfo {title} {{Effective
  gravitational couplings for cosmological perturbations in the most general
  scalar-tensor theories with second-order field equations}},}\ }\href
  {\doibase 10.1016/j.physletb.2011.11.028} {\bibfield  {journal} {\bibinfo
  {journal} {Phys.Lett.}\ }\textbf {\bibinfo {volume} {B706}},\ \bibinfo
  {pages} {123--133} (\bibinfo {year} {2011})},\ \Eprint
  {http://arxiv.org/abs/1108.4242} {arXiv:1108.4242 [gr-qc]} \BibitemShut
  {NoStop}%
\bibitem [{\citenamefont {Amendola}\ \emph
  {et~al.}(2013{\natexlab{a}})\citenamefont {Amendola}, \citenamefont {Kunz},
  \citenamefont {Motta}, \citenamefont {Saltas},\ and\ \citenamefont
  {Sawicki}}]{Amendola:2012ky}%
  \BibitemOpen
  \bibfield  {author} {\bibinfo {author} {\bibfnamefont {Luca}\ \bibnamefont
  {Amendola}}, \bibinfo {author} {\bibfnamefont {Martin}\ \bibnamefont {Kunz}},
  \bibinfo {author} {\bibfnamefont {Mariele}\ \bibnamefont {Motta}}, \bibinfo
  {author} {\bibfnamefont {Ippocratis~D.}\ \bibnamefont {Saltas}}, \ and\
  \bibinfo {author} {\bibfnamefont {Ignacy}\ \bibnamefont {Sawicki}},\
  }\bibfield  {title} {\enquote {\bibinfo {title} {{Observables and
  unobservables in dark energy cosmologies}},}\ }\href {\doibase
  10.1103/PhysRevD.87.023501} {\bibfield  {journal} {\bibinfo  {journal} {Phys.
  Rev.}\ }\textbf {\bibinfo {volume} {D87}},\ \bibinfo {pages} {023501}
  (\bibinfo {year} {2013}{\natexlab{a}})},\ \Eprint
  {http://arxiv.org/abs/1210.0439} {arXiv:1210.0439 [astro-ph.CO]} \BibitemShut
  {NoStop}%
\bibitem [{\citenamefont {Könnig}\ and\ \citenamefont
  {Amendola}(2014)}]{Konnig:2014dna}%
  \BibitemOpen
  \bibfield  {author} {\bibinfo {author} {\bibfnamefont {Frank}\ \bibnamefont
  {Könnig}}\ and\ \bibinfo {author} {\bibfnamefont {Luca}\ \bibnamefont
  {Amendola}},\ }\bibfield  {title} {\enquote {\bibinfo {title} {{Instability
  in a minimal bimetric gravity model}},}\ }\href {\doibase
  10.1103/PhysRevD.90.044030} {\bibfield  {journal} {\bibinfo  {journal} {Phys.
  Rev.}\ }\textbf {\bibinfo {volume} {D90}},\ \bibinfo {pages} {044030}
  (\bibinfo {year} {2014})},\ \Eprint {http://arxiv.org/abs/1402.1988}
  {arXiv:1402.1988 [astro-ph.CO]} \BibitemShut {NoStop}%
\bibitem [{\citenamefont {De~Felice}\ \emph {et~al.}(2016)\citenamefont
  {De~Felice}, \citenamefont {Heisenberg}, \citenamefont {Kase}, \citenamefont
  {Mukohyama}, \citenamefont {Tsujikawa},\ and\ \citenamefont
  {Zhang}}]{DeFelice:2016uil}%
  \BibitemOpen
  \bibfield  {author} {\bibinfo {author} {\bibfnamefont {Antonio}\ \bibnamefont
  {De~Felice}}, \bibinfo {author} {\bibfnamefont {Lavinia}\ \bibnamefont
  {Heisenberg}}, \bibinfo {author} {\bibfnamefont {Ryotaro}\ \bibnamefont
  {Kase}}, \bibinfo {author} {\bibfnamefont {Shinji}\ \bibnamefont
  {Mukohyama}}, \bibinfo {author} {\bibfnamefont {Shinji}\ \bibnamefont
  {Tsujikawa}}, \ and\ \bibinfo {author} {\bibfnamefont {Ying-li}\ \bibnamefont
  {Zhang}},\ }\bibfield  {title} {\enquote {\bibinfo {title} {{Effective
  gravitational couplings for cosmological perturbations in generalized Proca
  theories}},}\ }\href {\doibase 10.1103/PhysRevD.94.044024} {\bibfield
  {journal} {\bibinfo  {journal} {Phys. Rev.}\ }\textbf {\bibinfo {volume}
  {D94}},\ \bibinfo {pages} {044024} (\bibinfo {year} {2016})},\ \Eprint
  {http://arxiv.org/abs/1605.05066} {arXiv:1605.05066 [gr-qc]} \BibitemShut
  {NoStop}%
\bibitem [{\citenamefont {{Vardanyan}}\ and\ \citenamefont
  {{Amendola}}(2015)}]{2015PhRvD..92b4009V}%
  \BibitemOpen
  \bibfield  {author} {\bibinfo {author} {\bibfnamefont {V.}~\bibnamefont
  {{Vardanyan}}}\ and\ \bibinfo {author} {\bibfnamefont {L.}~\bibnamefont
  {{Amendola}}},\ }\bibfield  {title} {\enquote {\bibinfo {title} {{How can we
  tell whether dark energy is composed of multiple fields?}}}\ }\href {\doibase
  10.1103/PhysRevD.92.024009} {\bibfield  {journal} {\bibinfo  {journal}
  {\prd}\ }\textbf {\bibinfo {volume} {92}},\ \bibinfo {eid} {024009} (\bibinfo
  {year} {2015})}\BibitemShut {NoStop}%
\bibitem [{\citenamefont {Navarro}\ \emph {et~al.}(1996)\citenamefont
  {Navarro}, \citenamefont {Frenk},\ and\ \citenamefont
  {White}}]{Navarro:1995iw}%
  \BibitemOpen
  \bibfield  {author} {\bibinfo {author} {\bibfnamefont {Julio~F.}\
  \bibnamefont {Navarro}}, \bibinfo {author} {\bibfnamefont {Carlos~S.}\
  \bibnamefont {Frenk}}, \ and\ \bibinfo {author} {\bibfnamefont {Simon D.~M.}\
  \bibnamefont {White}},\ }\bibfield  {title} {\enquote {\bibinfo {title} {{The
  Structure of cold dark matter halos}},}\ }\href {\doibase 10.1086/177173}
  {\bibfield  {journal} {\bibinfo  {journal} {Astrophys. J.}\ }\textbf
  {\bibinfo {volume} {462}},\ \bibinfo {pages} {563--575} (\bibinfo {year}
  {1996})},\ \Eprint {http://arxiv.org/abs/astro-ph/9508025}
  {arXiv:astro-ph/9508025 [astro-ph]} \BibitemShut {NoStop}%
\bibitem [{\citenamefont {{Pizzuti}}\ \emph {et~al.}(2017)\citenamefont
  {{Pizzuti}}, \citenamefont {{Sartoris}}, \citenamefont {{Amendola}},
  \citenamefont {{Borgani}}, \citenamefont {{Biviano}}, \citenamefont
  {{Umetsu}}, \citenamefont {{Mercurio}}, \citenamefont {{Rosati}},
  \citenamefont {{Balestra}}, \citenamefont {{Caminha}}, \citenamefont
  {{Girardi}}, \citenamefont {{Grillo}},\ and\ \citenamefont
  {{Nonino}}}]{2017JCAP...07..023P}%
  \BibitemOpen
  \bibfield  {author} {\bibinfo {author} {\bibfnamefont {L.}~\bibnamefont
  {{Pizzuti}}}, \bibinfo {author} {\bibfnamefont {B.}~\bibnamefont
  {{Sartoris}}}, \bibinfo {author} {\bibfnamefont {L.}~\bibnamefont
  {{Amendola}}}, \bibinfo {author} {\bibfnamefont {S.}~\bibnamefont
  {{Borgani}}}, \bibinfo {author} {\bibfnamefont {A.}~\bibnamefont
  {{Biviano}}}, \bibinfo {author} {\bibfnamefont {K.}~\bibnamefont {{Umetsu}}},
  \bibinfo {author} {\bibfnamefont {A.}~\bibnamefont {{Mercurio}}}, \bibinfo
  {author} {\bibfnamefont {P.}~\bibnamefont {{Rosati}}}, \bibinfo {author}
  {\bibfnamefont {I.}~\bibnamefont {{Balestra}}}, \bibinfo {author}
  {\bibfnamefont {G.~B.}\ \bibnamefont {{Caminha}}}, \bibinfo {author}
  {\bibfnamefont {M.}~\bibnamefont {{Girardi}}}, \bibinfo {author}
  {\bibfnamefont {C.}~\bibnamefont {{Grillo}}}, \ and\ \bibinfo {author}
  {\bibfnamefont {M.}~\bibnamefont {{Nonino}}},\ }\bibfield  {title} {\enquote
  {\bibinfo {title} {{CLASH-VLT: constraints on f(R) gravity models with galaxy
  clusters using lensing and kinematic analyses}},}\ }\href {\doibase
  10.1088/1475-7516/2017/07/023} {\bibfield  {journal} {\bibinfo  {journal}
  {\jcap}\ }\textbf {\bibinfo {volume} {7}},\ \bibinfo {eid} {023} (\bibinfo
  {year} {2017})},\ \Eprint {http://arxiv.org/abs/1705.05179}
  {arXiv:1705.05179} \BibitemShut {NoStop}%
\bibitem [{\citenamefont {Dodelson}(2003)}]{Dodelson:2003ft}%
  \BibitemOpen
  \bibfield  {author} {\bibinfo {author} {\bibfnamefont {Scott}\ \bibnamefont
  {Dodelson}},\ }\href
  {http://www.slac.stanford.edu/spires/find/books/www?cl=QB981:D62:2003} {\emph
  {\bibinfo {title} {{Modern Cosmology}}}}\ (\bibinfo  {publisher} {Academic
  Press},\ \bibinfo {address} {Amsterdam},\ \bibinfo {year} {2003})\BibitemShut
  {NoStop}%
\bibitem [{\citenamefont {Kreisch}\ and\ \citenamefont
  {Komatsu}(2018)}]{Kreisch:2017uet}%
  \BibitemOpen
  \bibfield  {author} {\bibinfo {author} {\bibfnamefont {C.~D.}\ \bibnamefont
  {Kreisch}}\ and\ \bibinfo {author} {\bibfnamefont {E.}~\bibnamefont
  {Komatsu}},\ }\bibfield  {title} {\enquote {\bibinfo {title} {{Cosmological
  Constraints on Horndeski Gravity in Light of GW170817}},}\ }\href {\doibase
  10.1088/1475-7516/2018/12/030} {\bibfield  {journal} {\bibinfo  {journal}
  {JCAP}\ }\textbf {\bibinfo {volume} {1812}},\ \bibinfo {pages} {030}
  (\bibinfo {year} {2018})},\ \Eprint {http://arxiv.org/abs/1712.02710}
  {arXiv:1712.02710 [astro-ph.CO]} \BibitemShut {NoStop}%
\bibitem [{\citenamefont {Bellini}\ and\ \citenamefont
  {Sawicki}(2014)}]{Bellini:2014fua}%
  \BibitemOpen
  \bibfield  {author} {\bibinfo {author} {\bibfnamefont {Emilio}\ \bibnamefont
  {Bellini}}\ and\ \bibinfo {author} {\bibfnamefont {Ignacy}\ \bibnamefont
  {Sawicki}},\ }\bibfield  {title} {\enquote {\bibinfo {title} {{Maximal
  freedom at minimum cost: linear large-scale structure in general
  modifications of gravity}},}\ }\href {\doibase 10.1088/1475-7516/2014/07/050}
  {\bibfield  {journal} {\bibinfo  {journal} {JCAP}\ }\textbf {\bibinfo
  {volume} {1407}},\ \bibinfo {pages} {050} (\bibinfo {year} {2014})},\ \Eprint
  {http://arxiv.org/abs/1404.3713} {arXiv:1404.3713 [astro-ph.CO]} \BibitemShut
  {NoStop}%
\bibitem [{\citenamefont {Saltas}\ \emph {et~al.}(2014)\citenamefont {Saltas},
  \citenamefont {Sawicki}, \citenamefont {Amendola},\ and\ \citenamefont
  {Kunz}}]{Saltas:2014dha}%
  \BibitemOpen
  \bibfield  {author} {\bibinfo {author} {\bibfnamefont {Ippocratis~D.}\
  \bibnamefont {Saltas}}, \bibinfo {author} {\bibfnamefont {Ignacy}\
  \bibnamefont {Sawicki}}, \bibinfo {author} {\bibfnamefont {Luca}\
  \bibnamefont {Amendola}}, \ and\ \bibinfo {author} {\bibfnamefont {Martin}\
  \bibnamefont {Kunz}},\ }\bibfield  {title} {\enquote {\bibinfo {title}
  {{Anisotropic Stress as a Signature of Nonstandard Propagation of
  Gravitational Waves}},}\ }\href {\doibase 10.1103/PhysRevLett.113.191101}
  {\bibfield  {journal} {\bibinfo  {journal} {Phys. Rev. Lett.}\ }\textbf
  {\bibinfo {volume} {113}},\ \bibinfo {pages} {191101} (\bibinfo {year}
  {2014})},\ \Eprint {http://arxiv.org/abs/1406.7139} {arXiv:1406.7139
  [astro-ph.CO]} \BibitemShut {NoStop}%
\bibitem [{\citenamefont {Sawicki}\ \emph {et~al.}(2017)\citenamefont
  {Sawicki}, \citenamefont {Saltas}, \citenamefont {Motta}, \citenamefont
  {Amendola},\ and\ \citenamefont {Kunz}}]{Sawicki:2016klv}%
  \BibitemOpen
  \bibfield  {author} {\bibinfo {author} {\bibfnamefont {Ignacy}\ \bibnamefont
  {Sawicki}}, \bibinfo {author} {\bibfnamefont {Ippocratis~D.}\ \bibnamefont
  {Saltas}}, \bibinfo {author} {\bibfnamefont {Mariele}\ \bibnamefont {Motta}},
  \bibinfo {author} {\bibfnamefont {Luca}\ \bibnamefont {Amendola}}, \ and\
  \bibinfo {author} {\bibfnamefont {Martin}\ \bibnamefont {Kunz}},\ }\bibfield
  {title} {\enquote {\bibinfo {title} {{Nonstandard gravitational waves imply
  gravitational slip: On the difficulty of partially hiding new gravitational
  degrees of freedom}},}\ }\href {\doibase 10.1103/PhysRevD.95.083520}
  {\bibfield  {journal} {\bibinfo  {journal} {Phys. Rev.}\ }\textbf {\bibinfo
  {volume} {D95}},\ \bibinfo {pages} {083520} (\bibinfo {year} {2017})},\
  \Eprint {http://arxiv.org/abs/1612.02002} {arXiv:1612.02002 [astro-ph.CO]}
  \BibitemShut {NoStop}%
\bibitem [{\citenamefont {{Lombriser}}\ and\ \citenamefont
  {{Lima}}(2017)}]{2017PhLB..765..382L}%
  \BibitemOpen
  \bibfield  {author} {\bibinfo {author} {\bibfnamefont {L.}~\bibnamefont
  {{Lombriser}}}\ and\ \bibinfo {author} {\bibfnamefont {N.~A.}\ \bibnamefont
  {{Lima}}},\ }\bibfield  {title} {\enquote {\bibinfo {title} {{Challenges to
  self-acceleration in modified gravity from gravitational waves and
  large-scale structure}},}\ }\href {\doibase 10.1016/j.physletb.2016.12.048}
  {\bibfield  {journal} {\bibinfo  {journal} {Physics Letters B}\ }\textbf
  {\bibinfo {volume} {765}},\ \bibinfo {pages} {382--385} (\bibinfo {year}
  {2017})},\ \Eprint {http://arxiv.org/abs/1602.07670} {arXiv:1602.07670}
  \BibitemShut {NoStop}%
\bibitem [{\citenamefont {{Linder}}(2018)}]{2018JCAP...03..005L}%
  \BibitemOpen
  \bibfield  {author} {\bibinfo {author} {\bibfnamefont {E.~V.}\ \bibnamefont
  {{Linder}}},\ }\bibfield  {title} {\enquote {\bibinfo {title} {{No slip
  gravity}},}\ }\href {\doibase 10.1088/1475-7516/2018/03/005} {\bibfield
  {journal} {\bibinfo  {journal} {\jcap}\ }\textbf {\bibinfo {volume} {3}},\
  \bibinfo {eid} {005} (\bibinfo {year} {2018})},\ \Eprint
  {http://arxiv.org/abs/1801.01503} {arXiv:1801.01503} \BibitemShut {NoStop}%
\bibitem [{\citenamefont {Will}(1993)}]{Will:1993ns}%
  \BibitemOpen
  \bibfield  {author} {\bibinfo {author} {\bibfnamefont {C.~M.}\ \bibnamefont
  {Will}},\ }\href@noop {} {\emph {\bibinfo {title} {{Theory and experiment in
  gravitational physics}}}}\ (\bibinfo {year} {1993})\BibitemShut {NoStop}%
\bibitem [{\citenamefont {{Adelberger}}\ \emph {et~al.}(2003)\citenamefont
  {{Adelberger}}, \citenamefont {{Heckel}},\ and\ \citenamefont
  {{Nelson}}}]{2003ARNPS..53...77A}%
  \BibitemOpen
  \bibfield  {author} {\bibinfo {author} {\bibfnamefont {E.~G.}\ \bibnamefont
  {{Adelberger}}}, \bibinfo {author} {\bibfnamefont {B.~R.}\ \bibnamefont
  {{Heckel}}}, \ and\ \bibinfo {author} {\bibfnamefont {A.~E.}\ \bibnamefont
  {{Nelson}}},\ }\bibfield  {title} {\enquote {\bibinfo {title} {{Tests of the
  Gravitational Inverse-Square Law}},}\ }\href {\doibase
  10.1146/annurev.nucl.53.041002.110503} {\bibfield  {journal} {\bibinfo
  {journal} {Annual Review of Nuclear and Particle Science}\ }\textbf {\bibinfo
  {volume} {53}},\ \bibinfo {pages} {77--121} (\bibinfo {year} {2003})},\
  \Eprint {http://arxiv.org/abs/hep-ph/0307284} {hep-ph/0307284} \BibitemShut
  {NoStop}%
\bibitem [{\citenamefont {{Clifton}}\ \emph {et~al.}(2012)\citenamefont
  {{Clifton}}, \citenamefont {{Ferreira}}, \citenamefont {{Padilla}},\ and\
  \citenamefont {{Skordis}}}]{2012PhR...513....1C}%
  \BibitemOpen
  \bibfield  {author} {\bibinfo {author} {\bibfnamefont {T.}~\bibnamefont
  {{Clifton}}}, \bibinfo {author} {\bibfnamefont {P.~G.}\ \bibnamefont
  {{Ferreira}}}, \bibinfo {author} {\bibfnamefont {A.}~\bibnamefont
  {{Padilla}}}, \ and\ \bibinfo {author} {\bibfnamefont {C.}~\bibnamefont
  {{Skordis}}},\ }\bibfield  {title} {\enquote {\bibinfo {title} {{Modified
  gravity and cosmology}},}\ }\href {\doibase 10.1016/j.physrep.2012.01.001}
  {\bibfield  {journal} {\bibinfo  {journal} {\physrep}\ }\textbf {\bibinfo
  {volume} {513}},\ \bibinfo {pages} {1--189} (\bibinfo {year} {2012})},\
  \Eprint {http://arxiv.org/abs/1106.2476} {arXiv:1106.2476} \BibitemShut
  {NoStop}%
\bibitem [{\citenamefont {Patrignani}\ \emph {et~al.}(2016)\citenamefont
  {Patrignani} \emph {et~al.}}]{Patrignani:2016xqp}%
  \BibitemOpen
  \bibfield  {author} {\bibinfo {author} {\bibfnamefont {C.}~\bibnamefont
  {Patrignani}} \emph {et~al.} (\bibinfo {collaboration} {Particle Data
  Group}),\ }\bibfield  {title} {\enquote {\bibinfo {title} {{Review of
  Particle Physics}},}\ }\href {\doibase 10.1088/1674-1137/40/10/100001}
  {\bibfield  {journal} {\bibinfo  {journal} {Chin. Phys.}\ }\textbf {\bibinfo
  {volume} {C40}},\ \bibinfo {pages} {100001} (\bibinfo {year}
  {2016})}\BibitemShut {NoStop}%
\bibitem [{\citenamefont {Khoury}\ and\ \citenamefont
  {Weltman}(2004{\natexlab{a}})}]{Khoury:2003rn}%
  \BibitemOpen
  \bibfield  {author} {\bibinfo {author} {\bibfnamefont {Justin}\ \bibnamefont
  {Khoury}}\ and\ \bibinfo {author} {\bibfnamefont {Amanda}\ \bibnamefont
  {Weltman}},\ }\bibfield  {title} {\enquote {\bibinfo {title} {{Chameleon
  cosmology}},}\ }\href {\doibase 10.1103/PhysRevD.69.044026} {\bibfield
  {journal} {\bibinfo  {journal} {Phys. Rev.}\ }\textbf {\bibinfo {volume}
  {D69}},\ \bibinfo {pages} {044026} (\bibinfo {year} {2004}{\natexlab{a}})},\
  \Eprint {http://arxiv.org/abs/astro-ph/0309411} {arXiv:astro-ph/0309411
  [astro-ph]} \BibitemShut {NoStop}%
\bibitem [{\citenamefont {Khoury}\ and\ \citenamefont
  {Weltman}(2004{\natexlab{b}})}]{Khoury:2003aq}%
  \BibitemOpen
  \bibfield  {author} {\bibinfo {author} {\bibfnamefont {Justin}\ \bibnamefont
  {Khoury}}\ and\ \bibinfo {author} {\bibfnamefont {Amanda}\ \bibnamefont
  {Weltman}},\ }\bibfield  {title} {\enquote {\bibinfo {title} {{Chameleon
  fields: Awaiting surprises for tests of gravity in space}},}\ }\href
  {\doibase 10.1103/PhysRevLett.93.171104} {\bibfield  {journal} {\bibinfo
  {journal} {Phys. Rev. Lett.}\ }\textbf {\bibinfo {volume} {93}},\ \bibinfo
  {pages} {171104} (\bibinfo {year} {2004}{\natexlab{b}})},\ \Eprint
  {http://arxiv.org/abs/astro-ph/0309300} {arXiv:astro-ph/0309300 [astro-ph]}
  \BibitemShut {NoStop}%
\bibitem [{\citenamefont {Vainshtein}(1972)}]{Vainshtein:1972sx}%
  \BibitemOpen
  \bibfield  {author} {\bibinfo {author} {\bibfnamefont {A.~I.}\ \bibnamefont
  {Vainshtein}},\ }\bibfield  {title} {\enquote {\bibinfo {title} {{To the
  problem of nonvanishing gravitation mass}},}\ }\href {\doibase
  10.1016/0370-2693(72)90147-5} {\bibfield  {journal} {\bibinfo  {journal}
  {Phys. Lett.}\ }\textbf {\bibinfo {volume} {39B}},\ \bibinfo {pages}
  {393--394} (\bibinfo {year} {1972})}\BibitemShut {NoStop}%
\bibitem [{\citenamefont {Babichev}\ and\ \citenamefont
  {Deffayet}(2013)}]{Babichev:2013usa}%
  \BibitemOpen
  \bibfield  {author} {\bibinfo {author} {\bibfnamefont {Eugeny}\ \bibnamefont
  {Babichev}}\ and\ \bibinfo {author} {\bibfnamefont {Cédric}\ \bibnamefont
  {Deffayet}},\ }\bibfield  {title} {\enquote {\bibinfo {title} {{An
  introduction to the Vainshtein mechanism}},}\ }\href {\doibase
  10.1088/0264-9381/30/18/184001} {\bibfield  {journal} {\bibinfo  {journal}
  {Class. Quant. Grav.}\ }\textbf {\bibinfo {volume} {30}},\ \bibinfo {pages}
  {184001} (\bibinfo {year} {2013})},\ \Eprint {http://arxiv.org/abs/1304.7240}
  {arXiv:1304.7240 [gr-qc]} \BibitemShut {NoStop}%
\bibitem [{\citenamefont {Pietroni}(2005)}]{Pietroni:2005pv}%
  \BibitemOpen
  \bibfield  {author} {\bibinfo {author} {\bibfnamefont {Massimo}\ \bibnamefont
  {Pietroni}},\ }\bibfield  {title} {\enquote {\bibinfo {title} {{Dark energy
  condensation}},}\ }\href {\doibase 10.1103/PhysRevD.72.043535} {\bibfield
  {journal} {\bibinfo  {journal} {Phys. Rev.}\ }\textbf {\bibinfo {volume}
  {D72}},\ \bibinfo {pages} {043535} (\bibinfo {year} {2005})},\ \Eprint
  {http://arxiv.org/abs/astro-ph/0505615} {arXiv:astro-ph/0505615 [astro-ph]}
  \BibitemShut {NoStop}%
\bibitem [{\citenamefont {Olive}\ and\ \citenamefont
  {Pospelov}(2008)}]{Olive:2007aj}%
  \BibitemOpen
  \bibfield  {author} {\bibinfo {author} {\bibfnamefont {Keith~A.}\
  \bibnamefont {Olive}}\ and\ \bibinfo {author} {\bibfnamefont {Maxim}\
  \bibnamefont {Pospelov}},\ }\bibfield  {title} {\enquote {\bibinfo {title}
  {{Environmental dependence of masses and coupling constants}},}\ }\href
  {\doibase 10.1103/PhysRevD.77.043524} {\bibfield  {journal} {\bibinfo
  {journal} {Phys. Rev.}\ }\textbf {\bibinfo {volume} {D77}},\ \bibinfo {pages}
  {043524} (\bibinfo {year} {2008})},\ \Eprint {http://arxiv.org/abs/0709.3825}
  {arXiv:0709.3825 [hep-ph]} \BibitemShut {NoStop}%
\bibitem [{\citenamefont {Hinterbichler}\ and\ \citenamefont
  {Khoury}(2010)}]{Hinterbichler:2010es}%
  \BibitemOpen
  \bibfield  {author} {\bibinfo {author} {\bibfnamefont {Kurt}\ \bibnamefont
  {Hinterbichler}}\ and\ \bibinfo {author} {\bibfnamefont {Justin}\
  \bibnamefont {Khoury}},\ }\bibfield  {title} {\enquote {\bibinfo {title}
  {{Symmetron Fields: Screening Long-Range Forces Through Local Symmetry
  Restoration}},}\ }\href {\doibase 10.1103/PhysRevLett.104.231301} {\bibfield
  {journal} {\bibinfo  {journal} {Phys. Rev. Lett.}\ }\textbf {\bibinfo
  {volume} {104}},\ \bibinfo {pages} {231301} (\bibinfo {year} {2010})},\
  \Eprint {http://arxiv.org/abs/1001.4525} {arXiv:1001.4525 [hep-th]}
  \BibitemShut {NoStop}%
\bibitem [{\citenamefont {Hinterbichler}\ \emph {et~al.}(2011)\citenamefont
  {Hinterbichler}, \citenamefont {Khoury}, \citenamefont {Levy},\ and\
  \citenamefont {Matas}}]{Hinterbichler:2011ca}%
  \BibitemOpen
  \bibfield  {author} {\bibinfo {author} {\bibfnamefont {Kurt}\ \bibnamefont
  {Hinterbichler}}, \bibinfo {author} {\bibfnamefont {Justin}\ \bibnamefont
  {Khoury}}, \bibinfo {author} {\bibfnamefont {Aaron}\ \bibnamefont {Levy}}, \
  and\ \bibinfo {author} {\bibfnamefont {Andrew}\ \bibnamefont {Matas}},\
  }\bibfield  {title} {\enquote {\bibinfo {title} {{Symmetron Cosmology}},}\
  }\href {\doibase 10.1103/PhysRevD.84.103521} {\bibfield  {journal} {\bibinfo
  {journal} {Phys. Rev.}\ }\textbf {\bibinfo {volume} {D84}},\ \bibinfo {pages}
  {103521} (\bibinfo {year} {2011})},\ \Eprint {http://arxiv.org/abs/1107.2112}
  {arXiv:1107.2112 [astro-ph.CO]} \BibitemShut {NoStop}%
\bibitem [{\citenamefont {{Copi}}\ \emph {et~al.}(2004)\citenamefont {{Copi}},
  \citenamefont {{Davis}},\ and\ \citenamefont
  {{Krauss}}}]{2004PhRvL..92q1301C}%
  \BibitemOpen
  \bibfield  {author} {\bibinfo {author} {\bibfnamefont {C.~J.}\ \bibnamefont
  {{Copi}}}, \bibinfo {author} {\bibfnamefont {A.~N.}\ \bibnamefont {{Davis}}},
  \ and\ \bibinfo {author} {\bibfnamefont {L.~M.}\ \bibnamefont {{Krauss}}},\
  }\bibfield  {title} {\enquote {\bibinfo {title} {{New Nucleosynthesis
  Constraint on the Variation of G}},}\ }\href {\doibase
  10.1103/PhysRevLett.92.171301} {\bibfield  {journal} {\bibinfo  {journal}
  {Physical Review Letters}\ }\textbf {\bibinfo {volume} {92}},\ \bibinfo {eid}
  {171301} (\bibinfo {year} {2004})},\ \Eprint
  {http://arxiv.org/abs/astro-ph/0311334} {astro-ph/0311334} \BibitemShut
  {NoStop}%
\bibitem [{\citenamefont {{Dent}}\ \emph {et~al.}(2007)\citenamefont {{Dent}},
  \citenamefont {{Stern}},\ and\ \citenamefont
  {{Wetterich}}}]{2007PhRvD..76f3513D}%
  \BibitemOpen
  \bibfield  {author} {\bibinfo {author} {\bibfnamefont {T.}~\bibnamefont
  {{Dent}}}, \bibinfo {author} {\bibfnamefont {S.}~\bibnamefont {{Stern}}}, \
  and\ \bibinfo {author} {\bibfnamefont {C.}~\bibnamefont {{Wetterich}}},\
  }\bibfield  {title} {\enquote {\bibinfo {title} {{Primordial nucleosynthesis
  as a probe of fundamental physics parameters}},}\ }\href {\doibase
  10.1103/PhysRevD.76.063513} {\bibfield  {journal} {\bibinfo  {journal}
  {\prd}\ }\textbf {\bibinfo {volume} {76}},\ \bibinfo {eid} {063513} (\bibinfo
  {year} {2007})},\ \Eprint {http://arxiv.org/abs/0705.0696} {arXiv:0705.0696}
  \BibitemShut {NoStop}%
\bibitem [{\citenamefont {{Iocco}}\ \emph {et~al.}(2009)\citenamefont
  {{Iocco}}, \citenamefont {{Mangano}}, \citenamefont {{Miele}}, \citenamefont
  {{Pisanti}},\ and\ \citenamefont {{Serpico}}}]{2009PhR...472....1I}%
  \BibitemOpen
  \bibfield  {author} {\bibinfo {author} {\bibfnamefont {F.}~\bibnamefont
  {{Iocco}}}, \bibinfo {author} {\bibfnamefont {G.}~\bibnamefont {{Mangano}}},
  \bibinfo {author} {\bibfnamefont {G.}~\bibnamefont {{Miele}}}, \bibinfo
  {author} {\bibfnamefont {O.}~\bibnamefont {{Pisanti}}}, \ and\ \bibinfo
  {author} {\bibfnamefont {P.~D.}\ \bibnamefont {{Serpico}}},\ }\bibfield
  {title} {\enquote {\bibinfo {title} {{Primordial nucleosynthesis: From
  precision cosmology to fundamental physics}},}\ }\href {\doibase
  10.1016/j.physrep.2009.02.002} {\bibfield  {journal} {\bibinfo  {journal}
  {\physrep}\ }\textbf {\bibinfo {volume} {472}},\ \bibinfo {pages} {1--76}
  (\bibinfo {year} {2009})},\ \Eprint {http://arxiv.org/abs/0809.0631}
  {arXiv:0809.0631} \BibitemShut {NoStop}%
\bibitem [{\citenamefont {Bettoni}\ \emph {et~al.}(2017)\citenamefont
  {Bettoni}, \citenamefont {Ezquiaga}, \citenamefont {Hinterbichler},\ and\
  \citenamefont {Zumalacarregui}}]{Bettoni:2016mij}%
  \BibitemOpen
  \bibfield  {author} {\bibinfo {author} {\bibfnamefont {Dario}\ \bibnamefont
  {Bettoni}}, \bibinfo {author} {\bibfnamefont {Jose~María}\ \bibnamefont
  {Ezquiaga}}, \bibinfo {author} {\bibfnamefont {Kurt}\ \bibnamefont
  {Hinterbichler}}, \ and\ \bibinfo {author} {\bibfnamefont {Miguel}\
  \bibnamefont {Zumalacarregui}},\ }\bibfield  {title} {\enquote {\bibinfo
  {title} {{Speed of Gravitational Waves and the Fate of Scalar-Tensor
  Gravity}},}\ }\href {\doibase 10.1103/PhysRevD.95.084029} {\bibfield
  {journal} {\bibinfo  {journal} {Phys. Rev.}\ }\textbf {\bibinfo {volume}
  {D95}},\ \bibinfo {pages} {084029} (\bibinfo {year} {2017})},\ \Eprint
  {http://arxiv.org/abs/1608.01982} {arXiv:1608.01982 [gr-qc]} \BibitemShut
  {NoStop}%
\bibitem [{\citenamefont {Lombriser}\ and\ \citenamefont
  {Taylor}(2016)}]{Lombriser:2015sxa}%
  \BibitemOpen
  \bibfield  {author} {\bibinfo {author} {\bibfnamefont {Lucas}\ \bibnamefont
  {Lombriser}}\ and\ \bibinfo {author} {\bibfnamefont {Andy}\ \bibnamefont
  {Taylor}},\ }\bibfield  {title} {\enquote {\bibinfo {title} {{Breaking a Dark
  Degeneracy with Gravitational Waves}},}\ }\href {\doibase
  10.1088/1475-7516/2016/03/031} {\bibfield  {journal} {\bibinfo  {journal}
  {JCAP}\ }\textbf {\bibinfo {volume} {1603}},\ \bibinfo {pages} {031}
  (\bibinfo {year} {2016})},\ \Eprint {http://arxiv.org/abs/1509.08458}
  {arXiv:1509.08458 [astro-ph.CO]} \BibitemShut {NoStop}%
\bibitem [{\citenamefont {Lombriser}\ and\ \citenamefont
  {Lima}(2017)}]{Lombriser:2016yzn}%
  \BibitemOpen
  \bibfield  {author} {\bibinfo {author} {\bibfnamefont {Lucas}\ \bibnamefont
  {Lombriser}}\ and\ \bibinfo {author} {\bibfnamefont {Nelson~A.}\ \bibnamefont
  {Lima}},\ }\bibfield  {title} {\enquote {\bibinfo {title} {{Challenges to
  Self-Acceleration in Modified Gravity from Gravitational Waves and
  Large-Scale Structure}},}\ }\href {\doibase 10.1016/j.physletb.2016.12.048}
  {\bibfield  {journal} {\bibinfo  {journal} {Phys. Lett.}\ }\textbf {\bibinfo
  {volume} {B765}},\ \bibinfo {pages} {382--385} (\bibinfo {year} {2017})},\
  \Eprint {http://arxiv.org/abs/1602.07670} {arXiv:1602.07670 [astro-ph.CO]}
  \BibitemShut {NoStop}%
\bibitem [{\citenamefont {Ezquiaga}\ and\ \citenamefont
  {Zumalacarregui}(2017)}]{Ezquiaga:2017ekz}%
  \BibitemOpen
  \bibfield  {author} {\bibinfo {author} {\bibfnamefont {Jose~María}\
  \bibnamefont {Ezquiaga}}\ and\ \bibinfo {author} {\bibfnamefont {Miguel}\
  \bibnamefont {Zumalacarregui}},\ }\bibfield  {title} {\enquote {\bibinfo
  {title} {{Dark Energy After GW170817: Dead Ends and the Road Ahead}},}\
  }\href {\doibase 10.1103/PhysRevLett.119.251304} {\bibfield  {journal}
  {\bibinfo  {journal} {Phys. Rev. Lett.}\ }\textbf {\bibinfo {volume} {119}},\
  \bibinfo {pages} {251304} (\bibinfo {year} {2017})},\ \Eprint
  {http://arxiv.org/abs/1710.05901} {arXiv:1710.05901 [astro-ph.CO]}
  \BibitemShut {NoStop}%
\bibitem [{\citenamefont {Creminelli}\ and\ \citenamefont
  {Vernizzi}(2017)}]{Creminelli:2017sry}%
  \BibitemOpen
  \bibfield  {author} {\bibinfo {author} {\bibfnamefont {Paolo}\ \bibnamefont
  {Creminelli}}\ and\ \bibinfo {author} {\bibfnamefont {Filippo}\ \bibnamefont
  {Vernizzi}},\ }\bibfield  {title} {\enquote {\bibinfo {title} {{Dark Energy
  after GW170817 and GRB170817A}},}\ }\href {\doibase
  10.1103/PhysRevLett.119.251302} {\bibfield  {journal} {\bibinfo  {journal}
  {Phys. Rev. Lett.}\ }\textbf {\bibinfo {volume} {119}},\ \bibinfo {pages}
  {251302} (\bibinfo {year} {2017})},\ \Eprint
  {http://arxiv.org/abs/1710.05877} {arXiv:1710.05877 [astro-ph.CO]}
  \BibitemShut {NoStop}%
\bibitem [{\citenamefont {Baker}\ \emph {et~al.}(2017)\citenamefont {Baker},
  \citenamefont {Bellini}, \citenamefont {Ferreira}, \citenamefont {Lagos},
  \citenamefont {Noller},\ and\ \citenamefont {Sawicki}}]{Baker:2017hug}%
  \BibitemOpen
  \bibfield  {author} {\bibinfo {author} {\bibfnamefont {T.}~\bibnamefont
  {Baker}}, \bibinfo {author} {\bibfnamefont {E.}~\bibnamefont {Bellini}},
  \bibinfo {author} {\bibfnamefont {P.~G.}\ \bibnamefont {Ferreira}}, \bibinfo
  {author} {\bibfnamefont {M.}~\bibnamefont {Lagos}}, \bibinfo {author}
  {\bibfnamefont {J.}~\bibnamefont {Noller}}, \ and\ \bibinfo {author}
  {\bibfnamefont {I.}~\bibnamefont {Sawicki}},\ }\bibfield  {title} {\enquote
  {\bibinfo {title} {{Strong constraints on cosmological gravity from GW170817
  and GRB 170817A}},}\ }\href {\doibase 10.1103/PhysRevLett.119.251301}
  {\bibfield  {journal} {\bibinfo  {journal} {Phys. Rev. Lett.}\ }\textbf
  {\bibinfo {volume} {119}},\ \bibinfo {pages} {251301} (\bibinfo {year}
  {2017})},\ \Eprint {http://arxiv.org/abs/1710.06394} {arXiv:1710.06394
  [astro-ph.CO]} \BibitemShut {NoStop}%
\bibitem [{\citenamefont {Sakstein}\ and\ \citenamefont
  {Jain}(2017)}]{Sakstein:2017xjx}%
  \BibitemOpen
  \bibfield  {author} {\bibinfo {author} {\bibfnamefont {Jeremy}\ \bibnamefont
  {Sakstein}}\ and\ \bibinfo {author} {\bibfnamefont {Bhuvnesh}\ \bibnamefont
  {Jain}},\ }\bibfield  {title} {\enquote {\bibinfo {title} {{Implications of
  the Neutron Star Merger GW170817 for Cosmological Scalar-Tensor Theories}},}\
  }\href {\doibase 10.1103/PhysRevLett.119.251303} {\bibfield  {journal}
  {\bibinfo  {journal} {Phys. Rev. Lett.}\ }\textbf {\bibinfo {volume} {119}},\
  \bibinfo {pages} {251303} (\bibinfo {year} {2017})},\ \Eprint
  {http://arxiv.org/abs/1710.05893} {arXiv:1710.05893 [astro-ph.CO]}
  \BibitemShut {NoStop}%
\bibitem [{\citenamefont {Amendola}\ \emph
  {et~al.}(2018{\natexlab{a}})\citenamefont {Amendola}, \citenamefont
  {Bettoni}, \citenamefont {Domènech},\ and\ \citenamefont
  {Gomes}}]{Amendola:2018ltt}%
  \BibitemOpen
  \bibfield  {author} {\bibinfo {author} {\bibfnamefont {Luca}\ \bibnamefont
  {Amendola}}, \bibinfo {author} {\bibfnamefont {Dario}\ \bibnamefont
  {Bettoni}}, \bibinfo {author} {\bibfnamefont {Guillem}\ \bibnamefont
  {Domènech}}, \ and\ \bibinfo {author} {\bibfnamefont {Adalto~R.}\
  \bibnamefont {Gomes}},\ }\bibfield  {title} {\enquote {\bibinfo {title}
  {{Doppelgänger dark energy: modified gravity with non-universal couplings
  after GW170817}},}\ }\href {\doibase 10.1088/1475-7516/2018/06/029}
  {\bibfield  {journal} {\bibinfo  {journal} {JCAP}\ }\textbf {\bibinfo
  {volume} {1806}},\ \bibinfo {pages} {029} (\bibinfo {year}
  {2018}{\natexlab{a}})},\ \Eprint {http://arxiv.org/abs/1803.06368}
  {arXiv:1803.06368 [gr-qc]} \BibitemShut {NoStop}%
\bibitem [{\citenamefont {Amendola}\ \emph {et~al.}(2014)\citenamefont
  {Amendola}, \citenamefont {Ballesteros},\ and\ \citenamefont
  {Pettorino}}]{Amendola:2014wma}%
  \BibitemOpen
  \bibfield  {author} {\bibinfo {author} {\bibfnamefont {Luca}\ \bibnamefont
  {Amendola}}, \bibinfo {author} {\bibfnamefont {Guillermo}\ \bibnamefont
  {Ballesteros}}, \ and\ \bibinfo {author} {\bibfnamefont {Valeria}\
  \bibnamefont {Pettorino}},\ }\bibfield  {title} {\enquote {\bibinfo {title}
  {{Effects of modified gravity on B-mode polarization}},}\ }\href {\doibase
  10.1103/PhysRevD.90.043009} {\bibfield  {journal} {\bibinfo  {journal} {Phys.
  Rev.}\ }\textbf {\bibinfo {volume} {D90}},\ \bibinfo {pages} {043009}
  (\bibinfo {year} {2014})},\ \Eprint {http://arxiv.org/abs/1405.7004}
  {arXiv:1405.7004 [astro-ph.CO]} \BibitemShut {NoStop}%
\bibitem [{\citenamefont {Amendola}\ \emph
  {et~al.}(2018{\natexlab{b}})\citenamefont {Amendola}, \citenamefont {Kunz},
  \citenamefont {Saltas},\ and\ \citenamefont {Sawicki}}]{Amendola:2017orw}%
  \BibitemOpen
  \bibfield  {author} {\bibinfo {author} {\bibfnamefont {Luca}\ \bibnamefont
  {Amendola}}, \bibinfo {author} {\bibfnamefont {Martin}\ \bibnamefont {Kunz}},
  \bibinfo {author} {\bibfnamefont {Ippocratis~D.}\ \bibnamefont {Saltas}}, \
  and\ \bibinfo {author} {\bibfnamefont {Ignacy}\ \bibnamefont {Sawicki}},\
  }\bibfield  {title} {\enquote {\bibinfo {title} {{The fate of large-scale
  structure in modified gravity after GW170817 and GRB170817A}},}\ }\href
  {\doibase 10.1103/PhysRevLett.120.131101} {\bibfield  {journal} {\bibinfo
  {journal} {Phys. Rev. Lett.}\ }\textbf {\bibinfo {volume} {120}},\ \bibinfo
  {pages} {131101} (\bibinfo {year} {2018}{\natexlab{b}})},\ \Eprint
  {http://arxiv.org/abs/1711.04825} {arXiv:1711.04825 [astro-ph.CO]}
  \BibitemShut {NoStop}%
\bibitem [{\citenamefont {Nersisyan}\ \emph {et~al.}(2018)\citenamefont
  {Nersisyan}, \citenamefont {Lima},\ and\ \citenamefont
  {Amendola}}]{Nersisyan:2018auj}%
  \BibitemOpen
  \bibfield  {author} {\bibinfo {author} {\bibfnamefont {Henrik}\ \bibnamefont
  {Nersisyan}}, \bibinfo {author} {\bibfnamefont {Nelson~A.}\ \bibnamefont
  {Lima}}, \ and\ \bibinfo {author} {\bibfnamefont {Luca}\ \bibnamefont
  {Amendola}},\ }\bibfield  {title} {\enquote {\bibinfo {title} {{Gravitational
  wave speed: Implications for models without a mass scale}},}\ }\href@noop {}
  {\  (\bibinfo {year} {2018})},\ \Eprint {http://arxiv.org/abs/1801.06683}
  {arXiv:1801.06683 [astro-ph.CO]} \BibitemShut {NoStop}%
\bibitem [{\citenamefont {Amendola}\ \emph {et~al.}(2017)\citenamefont
  {Amendola}, \citenamefont {Sawicki}, \citenamefont {Kunz},\ and\
  \citenamefont {Saltas}}]{Amendola:2017ovw}%
  \BibitemOpen
  \bibfield  {author} {\bibinfo {author} {\bibfnamefont {Luca}\ \bibnamefont
  {Amendola}}, \bibinfo {author} {\bibfnamefont {Ignacy}\ \bibnamefont
  {Sawicki}}, \bibinfo {author} {\bibfnamefont {Martin}\ \bibnamefont {Kunz}},
  \ and\ \bibinfo {author} {\bibfnamefont {Ippocratis~D.}\ \bibnamefont
  {Saltas}},\ }\bibfield  {title} {\enquote {\bibinfo {title} {{Direct
  detection of gravitational waves can measure the time variation of the Planck
  mass}},}\ }\href@noop {} {\  (\bibinfo {year} {2017})},\ \Eprint
  {http://arxiv.org/abs/1712.08623} {arXiv:1712.08623 [astro-ph.CO]}
  \BibitemShut {NoStop}%
\bibitem [{\citenamefont {Maggiore}(2008)}]{maggiore2008gravitational}%
  \BibitemOpen
  \bibfield  {author} {\bibinfo {author} {\bibfnamefont {M.}~\bibnamefont
  {Maggiore}},\ }\href {https://books.google.de/books?id=AqVpQgAACAAJ} {\emph
  {\bibinfo {title} {Gravitational Waves: Volume 1: Theory and Experiments}}},\
  Gravitational Waves\ (\bibinfo  {publisher} {OUP Oxford},\ \bibinfo {year}
  {2008})\BibitemShut {NoStop}%
\bibitem [{\citenamefont {Nishizawa}(2018)}]{Nishizawa:2017nef}%
  \BibitemOpen
  \bibfield  {author} {\bibinfo {author} {\bibfnamefont {Atsushi}\ \bibnamefont
  {Nishizawa}},\ }\bibfield  {title} {\enquote {\bibinfo {title} {{Generalized
  framework for testing gravity with gravitational-wave propagation. I.
  Formulation}},}\ }\href {\doibase 10.1103/PhysRevD.97.104037} {\bibfield
  {journal} {\bibinfo  {journal} {Phys. Rev.}\ }\textbf {\bibinfo {volume}
  {D97}},\ \bibinfo {pages} {104037} (\bibinfo {year} {2018})},\ \Eprint
  {http://arxiv.org/abs/1710.04825} {arXiv:1710.04825 [gr-qc]} \BibitemShut
  {NoStop}%
\bibitem [{\citenamefont {Belgacem}\ \emph {et~al.}(2017)\citenamefont
  {Belgacem}, \citenamefont {Dirian}, \citenamefont {Foffa},\ and\
  \citenamefont {Maggiore}}]{Belgacem:2017ihm}%
  \BibitemOpen
  \bibfield  {author} {\bibinfo {author} {\bibfnamefont {Enis}\ \bibnamefont
  {Belgacem}}, \bibinfo {author} {\bibfnamefont {Yves}\ \bibnamefont {Dirian}},
  \bibinfo {author} {\bibfnamefont {Stefano}\ \bibnamefont {Foffa}}, \ and\
  \bibinfo {author} {\bibfnamefont {Michele}\ \bibnamefont {Maggiore}},\
  }\bibfield  {title} {\enquote {\bibinfo {title} {{The gravitational-wave
  luminosity distance in modified gravity theories}},}\ }\href@noop {} {\
  (\bibinfo {year} {2017})},\ \Eprint {http://arxiv.org/abs/1712.08108}
  {arXiv:1712.08108 [astro-ph.CO]} \BibitemShut {NoStop}%
\bibitem [{\citenamefont {Cyburt}\ \emph {et~al.}(2016)\citenamefont {Cyburt},
  \citenamefont {Fields}, \citenamefont {Olive},\ and\ \citenamefont
  {Yeh}}]{Cyburt:2015mya}%
  \BibitemOpen
  \bibfield  {author} {\bibinfo {author} {\bibfnamefont {Richard~H.}\
  \bibnamefont {Cyburt}}, \bibinfo {author} {\bibfnamefont {Brian~D.}\
  \bibnamefont {Fields}}, \bibinfo {author} {\bibfnamefont {Keith~A.}\
  \bibnamefont {Olive}}, \ and\ \bibinfo {author} {\bibfnamefont {Tsung-Han}\
  \bibnamefont {Yeh}},\ }\bibfield  {title} {\enquote {\bibinfo {title} {{Big
  Bang Nucleosynthesis: 2015}},}\ }\href {\doibase
  10.1103/RevModPhys.88.015004} {\bibfield  {journal} {\bibinfo  {journal}
  {Rev. Mod. Phys.}\ }\textbf {\bibinfo {volume} {88}},\ \bibinfo {pages}
  {015004} (\bibinfo {year} {2016})},\ \Eprint
  {http://arxiv.org/abs/1505.01076} {arXiv:1505.01076 [astro-ph.CO]}
  \BibitemShut {NoStop}%
\bibitem [{\citenamefont {{Riess}}\ \emph {et~al.}(2018)\citenamefont
  {{Riess}}, \citenamefont {{Casertano}}, \citenamefont {{Yuan}}, \citenamefont
  {{Macri}}, \citenamefont {{Anderson}}, \citenamefont {{MacKenty}},
  \citenamefont {{Bowers}}, \citenamefont {{Clubb}}, \citenamefont
  {{Filippenko}}, \citenamefont {{Jones}},\ and\ \citenamefont
  {{Tucker}}}]{2018ApJ...855..136R}%
  \BibitemOpen
  \bibfield  {author} {\bibinfo {author} {\bibfnamefont {A.~G.}\ \bibnamefont
  {{Riess}}}, \bibinfo {author} {\bibfnamefont {S.}~\bibnamefont
  {{Casertano}}}, \bibinfo {author} {\bibfnamefont {W.}~\bibnamefont {{Yuan}}},
  \bibinfo {author} {\bibfnamefont {L.}~\bibnamefont {{Macri}}}, \bibinfo
  {author} {\bibfnamefont {J.}~\bibnamefont {{Anderson}}}, \bibinfo {author}
  {\bibfnamefont {J.~W.}\ \bibnamefont {{MacKenty}}}, \bibinfo {author}
  {\bibfnamefont {J.~B.}\ \bibnamefont {{Bowers}}}, \bibinfo {author}
  {\bibfnamefont {K.~I.}\ \bibnamefont {{Clubb}}}, \bibinfo {author}
  {\bibfnamefont {A.~V.}\ \bibnamefont {{Filippenko}}}, \bibinfo {author}
  {\bibfnamefont {D.~O.}\ \bibnamefont {{Jones}}}, \ and\ \bibinfo {author}
  {\bibfnamefont {B.~E.}\ \bibnamefont {{Tucker}}},\ }\bibfield  {title}
  {\enquote {\bibinfo {title} {{New Parallaxes of Galactic Cepheids from
  Spatially Scanning the Hubble Space Telescope: Implications for the Hubble
  Constant}},}\ }\href {\doibase 10.3847/1538-4357/aaadb7} {\bibfield
  {journal} {\bibinfo  {journal} {\apj}\ }\textbf {\bibinfo {volume} {855}},\
  \bibinfo {eid} {136} (\bibinfo {year} {2018})},\ \Eprint
  {http://arxiv.org/abs/1801.01120} {arXiv:1801.01120 [astro-ph.SR]}
  \BibitemShut {NoStop}%
\bibitem [{\citenamefont {{Planck Collaboration}}\ \emph
  {et~al.}(2016)\citenamefont {{Planck Collaboration}}, \citenamefont {{Ade}},
  \citenamefont {{Aghanim}}, \citenamefont {{Arnaud}}, \citenamefont
  {{Ashdown}}, \citenamefont {{Aumont}}, \citenamefont {{Baccigalupi}},
  \citenamefont {{Banday}}, \citenamefont {{Barreiro}}, \citenamefont
  {{Bartlett}} \emph {et~al.}}]{2016A&A...594A..13P}%
  \BibitemOpen
  \bibfield  {author} {\bibinfo {author} {\bibnamefont {{Planck
  Collaboration}}}, \bibinfo {author} {\bibfnamefont {P.~A.~R.}\ \bibnamefont
  {{Ade}}}, \bibinfo {author} {\bibfnamefont {N.}~\bibnamefont {{Aghanim}}},
  \bibinfo {author} {\bibfnamefont {M.}~\bibnamefont {{Arnaud}}}, \bibinfo
  {author} {\bibfnamefont {M.}~\bibnamefont {{Ashdown}}}, \bibinfo {author}
  {\bibfnamefont {J.}~\bibnamefont {{Aumont}}}, \bibinfo {author}
  {\bibfnamefont {C.}~\bibnamefont {{Baccigalupi}}}, \bibinfo {author}
  {\bibfnamefont {A.~J.}\ \bibnamefont {{Banday}}}, \bibinfo {author}
  {\bibfnamefont {R.~B.}\ \bibnamefont {{Barreiro}}}, \bibinfo {author}
  {\bibfnamefont {J.~G.}\ \bibnamefont {{Bartlett}}},  \emph {et~al.},\
  }\bibfield  {title} {\enquote {\bibinfo {title} {{Planck 2015 results. XIII.
  Cosmological parameters}},}\ }\href {\doibase 10.1051/0004-6361/201525830}
  {\bibfield  {journal} {\bibinfo  {journal} {\aap}\ }\textbf {\bibinfo
  {volume} {594}},\ \bibinfo {eid} {A13} (\bibinfo {year} {2016})},\ \Eprint
  {http://arxiv.org/abs/1502.01589} {arXiv:1502.01589} \BibitemShut {NoStop}%
\bibitem [{\citenamefont {{Hildebrandt}}\ \emph {et~al.}(2017)\citenamefont
  {{Hildebrandt}}, \citenamefont {{Viola}}, \citenamefont {{Heymans}},
  \citenamefont {{Joudaki}}, \citenamefont {{Kuijken}}, \citenamefont
  {{Blake}}, \citenamefont {{Erben}}, \citenamefont {{Joachimi}}, \citenamefont
  {{Klaes}}, \citenamefont {{Miller}}, \citenamefont {{Morrison}},
  \citenamefont {{Nakajima}}, \citenamefont {{Verdoes Kleijn}}, \citenamefont
  {{Amon}}, \citenamefont {{Choi}}, \citenamefont {{Covone}}, \citenamefont
  {{de Jong}}, \citenamefont {{Dvornik}}, \citenamefont {{Fenech Conti}},
  \citenamefont {{Grado}}, \citenamefont {{Harnois-D{\'e}raps}}, \citenamefont
  {{Herbonnet}}, \citenamefont {{Hoekstra}}, \citenamefont {{K{\"o}hlinger}},
  \citenamefont {{McFarland}}, \citenamefont {{Mead}}, \citenamefont
  {{Merten}}, \citenamefont {{Napolitano}}, \citenamefont {{Peacock}},
  \citenamefont {{Radovich}}, \citenamefont {{Schneider}}, \citenamefont
  {{Simon}}, \citenamefont {{Valentijn}}, \citenamefont {{van den Busch}},
  \citenamefont {{van Uitert}},\ and\ \citenamefont {{Van
  Waerbeke}}}]{2017MNRAS.465.1454H}%
  \BibitemOpen
  \bibfield  {author} {\bibinfo {author} {\bibfnamefont {H.}~\bibnamefont
  {{Hildebrandt}}}, \bibinfo {author} {\bibfnamefont {M.}~\bibnamefont
  {{Viola}}}, \bibinfo {author} {\bibfnamefont {C.}~\bibnamefont {{Heymans}}},
  \bibinfo {author} {\bibfnamefont {S.}~\bibnamefont {{Joudaki}}}, \bibinfo
  {author} {\bibfnamefont {K.}~\bibnamefont {{Kuijken}}}, \bibinfo {author}
  {\bibfnamefont {C.}~\bibnamefont {{Blake}}}, \bibinfo {author} {\bibfnamefont
  {T.}~\bibnamefont {{Erben}}}, \bibinfo {author} {\bibfnamefont
  {B.}~\bibnamefont {{Joachimi}}}, \bibinfo {author} {\bibfnamefont
  {D.}~\bibnamefont {{Klaes}}}, \bibinfo {author} {\bibfnamefont
  {L.}~\bibnamefont {{Miller}}}, \bibinfo {author} {\bibfnamefont {C.~B.}\
  \bibnamefont {{Morrison}}}, \bibinfo {author} {\bibfnamefont
  {R.}~\bibnamefont {{Nakajima}}}, \bibinfo {author} {\bibfnamefont
  {G.}~\bibnamefont {{Verdoes Kleijn}}}, \bibinfo {author} {\bibfnamefont
  {A.}~\bibnamefont {{Amon}}}, \bibinfo {author} {\bibfnamefont
  {A.}~\bibnamefont {{Choi}}}, \bibinfo {author} {\bibfnamefont
  {G.}~\bibnamefont {{Covone}}}, \bibinfo {author} {\bibfnamefont {J.~T.~A.}\
  \bibnamefont {{de Jong}}}, \bibinfo {author} {\bibfnamefont {A.}~\bibnamefont
  {{Dvornik}}}, \bibinfo {author} {\bibfnamefont {I.}~\bibnamefont {{Fenech
  Conti}}}, \bibinfo {author} {\bibfnamefont {A.}~\bibnamefont {{Grado}}},
  \bibinfo {author} {\bibfnamefont {J.}~\bibnamefont {{Harnois-D{\'e}raps}}},
  \bibinfo {author} {\bibfnamefont {R.}~\bibnamefont {{Herbonnet}}}, \bibinfo
  {author} {\bibfnamefont {H.}~\bibnamefont {{Hoekstra}}}, \bibinfo {author}
  {\bibfnamefont {F.}~\bibnamefont {{K{\"o}hlinger}}}, \bibinfo {author}
  {\bibfnamefont {J.}~\bibnamefont {{McFarland}}}, \bibinfo {author}
  {\bibfnamefont {A.}~\bibnamefont {{Mead}}}, \bibinfo {author} {\bibfnamefont
  {J.}~\bibnamefont {{Merten}}}, \bibinfo {author} {\bibfnamefont
  {N.}~\bibnamefont {{Napolitano}}}, \bibinfo {author} {\bibfnamefont {J.~A.}\
  \bibnamefont {{Peacock}}}, \bibinfo {author} {\bibfnamefont {M.}~\bibnamefont
  {{Radovich}}}, \bibinfo {author} {\bibfnamefont {P.}~\bibnamefont
  {{Schneider}}}, \bibinfo {author} {\bibfnamefont {P.}~\bibnamefont
  {{Simon}}}, \bibinfo {author} {\bibfnamefont {E.~A.}\ \bibnamefont
  {{Valentijn}}}, \bibinfo {author} {\bibfnamefont {J.~L.}\ \bibnamefont {{van
  den Busch}}}, \bibinfo {author} {\bibfnamefont {E.}~\bibnamefont {{van
  Uitert}}}, \ and\ \bibinfo {author} {\bibfnamefont {L.}~\bibnamefont {{Van
  Waerbeke}}},\ }\bibfield  {title} {\enquote {\bibinfo {title} {{KiDS-450:
  cosmological parameter constraints from tomographic weak gravitational
  lensing}},}\ }\href {\doibase 10.1093/mnras/stw2805} {\bibfield  {journal}
  {\bibinfo  {journal} {\mnras}\ }\textbf {\bibinfo {volume} {465}},\ \bibinfo
  {pages} {1454--1498} (\bibinfo {year} {2017})},\ \Eprint
  {http://arxiv.org/abs/1606.05338} {arXiv:1606.05338} \BibitemShut {NoStop}%
\bibitem [{\citenamefont {Barros}\ \emph {et~al.}(2018)\citenamefont {Barros},
  \citenamefont {Amendola}, \citenamefont {Barreiro},\ and\ \citenamefont
  {Nunes}}]{Barros:2018efl}%
  \BibitemOpen
  \bibfield  {author} {\bibinfo {author} {\bibfnamefont {Bruno~J.}\
  \bibnamefont {Barros}}, \bibinfo {author} {\bibfnamefont {Luca}\ \bibnamefont
  {Amendola}}, \bibinfo {author} {\bibfnamefont {Tiago}\ \bibnamefont
  {Barreiro}}, \ and\ \bibinfo {author} {\bibfnamefont {Nelson~J.}\
  \bibnamefont {Nunes}},\ }\bibfield  {title} {\enquote {\bibinfo {title}
  {{Coupled quintessence with a $\Lambda$CDM background: removing the
  $\sigma_8$ tension}},}\ }\href@noop {} {\  (\bibinfo {year} {2018})},\
  \Eprint {http://arxiv.org/abs/1802.09216} {arXiv:1802.09216 [astro-ph.CO]}
  \BibitemShut {NoStop}%
\bibitem [{\citenamefont {Weinberg}\ \emph {et~al.}(2014)\citenamefont
  {Weinberg}, \citenamefont {Bullock}, \citenamefont {Governato}, \citenamefont
  {Kuzio~de Naray},\ and\ \citenamefont {Peter}}]{Weinberg:2013aya}%
  \BibitemOpen
  \bibfield  {author} {\bibinfo {author} {\bibfnamefont {David~H.}\
  \bibnamefont {Weinberg}}, \bibinfo {author} {\bibfnamefont {James~S.}\
  \bibnamefont {Bullock}}, \bibinfo {author} {\bibfnamefont {Fabio}\
  \bibnamefont {Governato}}, \bibinfo {author} {\bibfnamefont {Rachel}\
  \bibnamefont {Kuzio~de Naray}}, \ and\ \bibinfo {author} {\bibfnamefont
  {Annika H.~G.}\ \bibnamefont {Peter}},\ }\bibfield  {title} {\enquote
  {\bibinfo {title} {{Cold dark matter: controversies on small scales}},}\
  }\bibfield  {booktitle} {\emph {\bibinfo {booktitle} {{Sackler Colloquium:
  Dark Matter Universe: On the Threshhold of Discovery Irvine, USA, October
  18-20, 2012}}},\ }\href {\doibase 10.1073/pnas.1308716112} {\bibfield
  {journal} {\bibinfo  {journal} {Proc. Nat. Acad. Sci.}\ }\textbf {\bibinfo
  {volume} {112}},\ \bibinfo {pages} {12249--12255} (\bibinfo {year} {2014})},\
  \bibinfo {note} {[Proc. Nat. Acad. Sci.112,2249(2015)]},\ \Eprint
  {http://arxiv.org/abs/1306.0913} {arXiv:1306.0913 [astro-ph.CO]} \BibitemShut
  {NoStop}%
\bibitem [{\citenamefont {Abbott}\ \emph
  {et~al.}(2017{\natexlab{b}})\citenamefont {Abbott} \emph
  {et~al.}}]{Abbott:2017xzu}%
  \BibitemOpen
  \bibfield  {author} {\bibinfo {author} {\bibfnamefont {B.~P.}\ \bibnamefont
  {Abbott}} \emph {et~al.} (\bibinfo {collaboration} {LIGO Scientific,
  VINROUGE, Las Cumbres Observatory, DES, DLT40, Virgo, 1M2H, Dark Energy
  Camera GW-E, MASTER}),\ }\bibfield  {title} {\enquote {\bibinfo {title} {{A
  gravitational-wave standard siren measurement of the Hubble constant}},}\
  }\href {\doibase 10.1038/nature24471} {\bibfield  {journal} {\bibinfo
  {journal} {Nature}\ }\textbf {\bibinfo {volume} {551}},\ \bibinfo {pages}
  {85--88} (\bibinfo {year} {2017}{\natexlab{b}})},\ \Eprint
  {http://arxiv.org/abs/1710.05835} {arXiv:1710.05835 [astro-ph.CO]}
  \BibitemShut {NoStop}%
\bibitem [{\citenamefont {Kunz}(2009)}]{Kunz2009}%
  \BibitemOpen
  \bibfield  {author} {\bibinfo {author} {\bibfnamefont {Martin}\ \bibnamefont
  {Kunz}},\ }\bibfield  {title} {\enquote {\bibinfo {title} {{Degeneracy
  between the dark components resulting from the fact that gravity only
  measures the total energy-momentum tensor}},}\ }\href {\doibase
  10.1103/PhysRevD.80.123001} {\bibfield  {journal} {\bibinfo  {journal}
  {Physical Review D}\ }\textbf {\bibinfo {volume} {80}},\ \bibinfo {pages}
  {123001} (\bibinfo {year} {2009})}\BibitemShut {NoStop}%
\bibitem [{\citenamefont {{Eisenstein}}\ and\ \citenamefont
  {{Hu}}(1998)}]{1998ApJ...496..605E}%
  \BibitemOpen
  \bibfield  {author} {\bibinfo {author} {\bibfnamefont {D.~J.}\ \bibnamefont
  {{Eisenstein}}}\ and\ \bibinfo {author} {\bibfnamefont {W.}~\bibnamefont
  {{Hu}}},\ }\bibfield  {title} {\enquote {\bibinfo {title} {{Baryonic Features
  in the Matter Transfer Function}},}\ }\href {\doibase 10.1086/305424}
  {\bibfield  {journal} {\bibinfo  {journal} {\apj}\ }\textbf {\bibinfo
  {volume} {496}},\ \bibinfo {pages} {605--614} (\bibinfo {year} {1998})},\
  \Eprint {http://arxiv.org/abs/astro-ph/9709112} {astro-ph/9709112}
  \BibitemShut {NoStop}%
\bibitem [{\citenamefont {{Liske}}(2000)}]{2000MNRAS.319..557L}%
  \BibitemOpen
  \bibfield  {author} {\bibinfo {author} {\bibfnamefont {J.}~\bibnamefont
  {{Liske}}},\ }\bibfield  {title} {\enquote {\bibinfo {title} {{On the
  cosmological distance and redshift between any two objects}},}\ }\href
  {\doibase 10.1046/j.1365-8711.2000.03874.x} {\bibfield  {journal} {\bibinfo
  {journal} {\mnras}\ }\textbf {\bibinfo {volume} {319}},\ \bibinfo {pages}
  {557--561} (\bibinfo {year} {2000})},\ \Eprint
  {http://arxiv.org/abs/astro-ph/0007341} {astro-ph/0007341} \BibitemShut
  {NoStop}%
\bibitem [{\citenamefont {Baldi}\ \emph {et~al.}(2011)\citenamefont {Baldi},
  \citenamefont {Pettorino}, \citenamefont {Amendola},\ and\ \citenamefont
  {Wetterich}}]{Baldi:2011es}%
  \BibitemOpen
  \bibfield  {author} {\bibinfo {author} {\bibfnamefont {Marco}\ \bibnamefont
  {Baldi}}, \bibinfo {author} {\bibfnamefont {Valeria}\ \bibnamefont
  {Pettorino}}, \bibinfo {author} {\bibfnamefont {Luca}\ \bibnamefont
  {Amendola}}, \ and\ \bibinfo {author} {\bibfnamefont {Christof}\ \bibnamefont
  {Wetterich}},\ }\bibfield  {title} {\enquote {\bibinfo {title} {{Oscillating
  nonlinear large scale structure in growing neutrino quintessence}},}\ }\href
  {\doibase 10.1111/j.1365-2966.2011.19477.x} {\bibfield  {journal} {\bibinfo
  {journal} {Mon. Not. Roy. Astron. Soc.}\ }\textbf {\bibinfo {volume} {418}},\
  \bibinfo {pages} {214} (\bibinfo {year} {2011})},\ \Eprint
  {http://arxiv.org/abs/1106.2161} {arXiv:1106.2161 [astro-ph.CO]} \BibitemShut
  {NoStop}%
\bibitem [{\citenamefont {Kaiser}(1987)}]{kaiser_clustering_1987}%
  \BibitemOpen
  \bibfield  {author} {\bibinfo {author} {\bibfnamefont {Nick}\ \bibnamefont
  {Kaiser}},\ }\bibfield  {title} {\enquote {\bibinfo {title} {Clustering in
  real space and in redshift space},}\ }\href
  {http://adsabs.harvard.edu/abs/1987MNRAS.227....1K} {\bibfield  {journal}
  {\bibinfo  {journal} {Monthly Notices of the Royal Astronomical Society}\
  }\textbf {\bibinfo {volume} {227}},\ \bibinfo {pages} {1--21} (\bibinfo
  {year} {1987})}\BibitemShut {NoStop}%
\bibitem [{\citenamefont {{Seo}}\ and\ \citenamefont
  {{Eisenstein}}(2003)}]{2003ApJ...598..720S}%
  \BibitemOpen
  \bibfield  {author} {\bibinfo {author} {\bibfnamefont {H.-J.}\ \bibnamefont
  {{Seo}}}\ and\ \bibinfo {author} {\bibfnamefont {D.~J.}\ \bibnamefont
  {{Eisenstein}}},\ }\bibfield  {title} {\enquote {\bibinfo {title} {{Probing
  Dark Energy with Baryonic Acoustic Oscillations from Future Large Galaxy
  Redshift Surveys}},}\ }\href {\doibase 10.1086/379122} {\bibfield  {journal}
  {\bibinfo  {journal} {\apj}\ }\textbf {\bibinfo {volume} {598}},\ \bibinfo
  {pages} {720--740} (\bibinfo {year} {2003})},\ \Eprint
  {http://arxiv.org/abs/astro-ph/0307460} {astro-ph/0307460} \BibitemShut
  {NoStop}%
\bibitem [{\citenamefont {Rampf}\ \emph {et~al.}(2017)\citenamefont {Rampf},
  \citenamefont {Villa},\ and\ \citenamefont {Amendola}}]{Rampf:2017srp}%
  \BibitemOpen
  \bibfield  {author} {\bibinfo {author} {\bibfnamefont {Cornelius}\
  \bibnamefont {Rampf}}, \bibinfo {author} {\bibfnamefont {Eleonora}\
  \bibnamefont {Villa}}, \ and\ \bibinfo {author} {\bibfnamefont {Luca}\
  \bibnamefont {Amendola}},\ }\bibfield  {title} {\enquote {\bibinfo {title}
  {{Quasilinear observables in dark energy cosmologies}},}\ }\href {\doibase
  10.1103/PhysRevD.95.123516} {\bibfield  {journal} {\bibinfo  {journal} {Phys.
  Rev.}\ }\textbf {\bibinfo {volume} {D95}},\ \bibinfo {pages} {123516}
  (\bibinfo {year} {2017})},\ \Eprint {http://arxiv.org/abs/1703.09228}
  {arXiv:1703.09228 [astro-ph.CO]} \BibitemShut {NoStop}%
\bibitem [{\citenamefont {Motta}\ \emph {et~al.}(2013)\citenamefont {Motta}
  \emph {et~al.}}]{Motta2013}%
  \BibitemOpen
  \bibfield  {author} {\bibinfo {author} {\bibfnamefont {Mariele}\ \bibnamefont
  {Motta}} \emph {et~al.},\ }\bibfield  {title} {\enquote {\bibinfo {title}
  {{Probing dark energy through scale dependence}},}\ }\href {\doibase
  10.1103/PhysRevD.88.124035} {\bibfield  {journal} {\bibinfo  {journal}
  {Physical Review D}\ }\textbf {\bibinfo {volume} {88}},\ \bibinfo {pages}
  {124035} (\bibinfo {year} {2013})}\BibitemShut {NoStop}%
\bibitem [{\citenamefont {Amendola}\ \emph {et~al.}(2012)\citenamefont
  {Amendola} \emph {et~al.}}]{Amendola2012}%
  \BibitemOpen
  \bibfield  {author} {\bibinfo {author} {\bibfnamefont {Luca}\ \bibnamefont
  {Amendola}} \emph {et~al.},\ }\bibfield  {title} {\enquote {\bibinfo {title}
  {{Observables and unobservables in dark energy cosmologies}},}\ }\href
  {\doibase 10.1103/PhysRevD.87.023501} {\bibfield  {journal} {\bibinfo
  {journal} {Physics Review D}\ } (\bibinfo {year} {2012}),\
  10.1103/PhysRevD.87.023501},\ \Eprint {http://arxiv.org/abs/1210.0439}
  {arXiv:1210.0439} \BibitemShut {NoStop}%
\bibitem [{\citenamefont {{Lahav}}\ \emph {et~al.}(1991)\citenamefont
  {{Lahav}}, \citenamefont {{Lilje}}, \citenamefont {{Primack}},\ and\
  \citenamefont {{Rees}}}]{lahav_1991}%
  \BibitemOpen
  \bibfield  {author} {\bibinfo {author} {\bibfnamefont {O.}~\bibnamefont
  {{Lahav}}}, \bibinfo {author} {\bibfnamefont {P.~B.}\ \bibnamefont
  {{Lilje}}}, \bibinfo {author} {\bibfnamefont {J.~R.}\ \bibnamefont
  {{Primack}}}, \ and\ \bibinfo {author} {\bibfnamefont {M.~J.}\ \bibnamefont
  {{Rees}}},\ }\bibfield  {title} {\enquote {\bibinfo {title} {{Dynamical
  effects of the cosmological constant}},}\ }\href {\doibase
  10.1093/mnras/251.1.128} {\bibfield  {journal} {\bibinfo  {journal} {\mnras}\
  }\textbf {\bibinfo {volume} {251}},\ \bibinfo {pages} {128--136} (\bibinfo
  {year} {1991})}\BibitemShut {NoStop}%
\bibitem [{\citenamefont {Leonard}\ \emph {et~al.}(2015)\citenamefont
  {Leonard}, \citenamefont {Ferreira},\ and\ \citenamefont
  {Heymans}}]{Leonard2015}%
  \BibitemOpen
  \bibfield  {author} {\bibinfo {author} {\bibfnamefont {C.~Danielle}\
  \bibnamefont {Leonard}}, \bibinfo {author} {\bibfnamefont {Pedro~G.}\
  \bibnamefont {Ferreira}}, \ and\ \bibinfo {author} {\bibfnamefont
  {Catherine}\ \bibnamefont {Heymans}},\ }\bibfield  {title} {\enquote
  {\bibinfo {title} {{Testing gravity with {\$}E{\_}G{\$}: mapping theory onto
  observations}},}\ }\href {\doibase 10.1088/1475-7516/2015/12/051} {\bibfield
  {journal} {\bibinfo  {journal} {Journal of Cosmology and Astroparticle
  Physics}\ } (\bibinfo {year} {2015}),\ 10.1088/1475-7516/2015/12/051},\
  \Eprint {http://arxiv.org/abs/1510.04287} {arXiv:1510.04287} \BibitemShut
  {NoStop}%
\bibitem [{\citenamefont {Zhang}\ \emph {et~al.}(2007)\citenamefont {Zhang},
  \citenamefont {Liguori}, \citenamefont {Bean},\ and\ \citenamefont
  {Dodelson}}]{Zhang2007}%
  \BibitemOpen
  \bibfield  {author} {\bibinfo {author} {\bibfnamefont {Pengjie}\ \bibnamefont
  {Zhang}}, \bibinfo {author} {\bibfnamefont {Michele}\ \bibnamefont
  {Liguori}}, \bibinfo {author} {\bibfnamefont {Rachel}\ \bibnamefont {Bean}},
  \ and\ \bibinfo {author} {\bibfnamefont {Scott}\ \bibnamefont {Dodelson}},\
  }\bibfield  {title} {\enquote {\bibinfo {title} {{Probing Gravity at
  Cosmological Scales by Measurements which Test the Relationship between
  Gravitational Lensing and Matter Overdensity}},}\ }\href {\doibase
  10.1103/PhysRevLett.99.141302} {\bibfield  {journal} {\bibinfo  {journal}
  {Physical Review Letters}\ }\textbf {\bibinfo {volume} {99}},\ \bibinfo
  {pages} {141302} (\bibinfo {year} {2007})}\BibitemShut {NoStop}%
\bibitem [{\citenamefont {{Reyes}}\ \emph {et~al.}(2010)\citenamefont
  {{Reyes}}, \citenamefont {{Mandelbaum}}, \citenamefont {{Seljak}},
  \citenamefont {{Baldauf}}, \citenamefont {{Gunn}}, \citenamefont
  {{Lombriser}},\ and\ \citenamefont {{Smith}}}]{2010Natur.464..256R}%
  \BibitemOpen
  \bibfield  {author} {\bibinfo {author} {\bibfnamefont {R.}~\bibnamefont
  {{Reyes}}}, \bibinfo {author} {\bibfnamefont {R.}~\bibnamefont
  {{Mandelbaum}}}, \bibinfo {author} {\bibfnamefont {U.}~\bibnamefont
  {{Seljak}}}, \bibinfo {author} {\bibfnamefont {T.}~\bibnamefont {{Baldauf}}},
  \bibinfo {author} {\bibfnamefont {J.~E.}\ \bibnamefont {{Gunn}}}, \bibinfo
  {author} {\bibfnamefont {L.}~\bibnamefont {{Lombriser}}}, \ and\ \bibinfo
  {author} {\bibfnamefont {R.~E.}\ \bibnamefont {{Smith}}},\ }\bibfield
  {title} {\enquote {\bibinfo {title} {{Confirmation of general relativity on
  large scales from weak lensing and galaxy velocities}},}\ }\href {\doibase
  10.1038/nature08857} {\bibfield  {journal} {\bibinfo  {journal} {\nat}\
  }\textbf {\bibinfo {volume} {464}},\ \bibinfo {pages} {256--258} (\bibinfo
  {year} {2010})},\ \Eprint {http://arxiv.org/abs/1003.2185} {arXiv:1003.2185
  [astro-ph.CO]} \BibitemShut {NoStop}%
\bibitem [{\citenamefont {de~la Torre}\ \emph {et~al.}(2017)\citenamefont
  {de~la Torre} \emph {et~al.}}]{delaTorre:2016rxm}%
  \BibitemOpen
  \bibfield  {author} {\bibinfo {author} {\bibfnamefont {S.}~\bibnamefont
  {de~la Torre}} \emph {et~al.},\ }\bibfield  {title} {\enquote {\bibinfo
  {title} {{The VIMOS Public Extragalactic Redshift Survey (VIPERS). Gravity
  test from the combination of redshift-space distortions and galaxy-galaxy
  lensing at $0.5 < z < 1.2$}},}\ }\href {\doibase 10.1051/0004-6361/201630276}
  {\bibfield  {journal} {\bibinfo  {journal} {Astron. Astrophys.}\ }\textbf
  {\bibinfo {volume} {608}},\ \bibinfo {pages} {A44} (\bibinfo {year}
  {2017})},\ \Eprint {http://arxiv.org/abs/1612.05647} {arXiv:1612.05647
  [astro-ph.CO]} \BibitemShut {NoStop}%
\bibitem [{\citenamefont {Amon}\ \emph {et~al.}(2017)\citenamefont {Amon} \emph
  {et~al.}}]{Amon2017}%
  \BibitemOpen
  \bibfield  {author} {\bibinfo {author} {\bibfnamefont {A.}~\bibnamefont
  {Amon}} \emph {et~al.},\ }\bibfield  {title} {\enquote {\bibinfo {title}
  {{KiDS+2dFLenS+GAMA: Testing the cosmological model with the Eg
  statistic}},}\ }\href@noop {} {\  (\bibinfo {year} {2017})},\ \Eprint
  {http://arxiv.org/abs/1711.10999} {arXiv:1711.10999} \BibitemShut {NoStop}%
\bibitem [{\citenamefont {{Planck Collaboration}}\ \emph
  {et~al.}(2015)\citenamefont {{Planck Collaboration}}, \citenamefont {Ade},
  \citenamefont {Aghanim}, \citenamefont {Arnaud} \emph
  {et~al.}}]{PlanckCollaboration2015}%
  \BibitemOpen
  \bibfield  {author} {\bibinfo {author} {\bibfnamefont {P.~A.~R.}\
  \bibnamefont {{Planck Collaboration}}}, \bibinfo {author} {\bibfnamefont
  {P.~A.~R.}\ \bibnamefont {Ade}}, \bibinfo {author} {\bibfnamefont
  {N.}~\bibnamefont {Aghanim}}, \bibinfo {author} {\bibfnamefont
  {M.}~\bibnamefont {Arnaud}},  \emph {et~al.},\ }\bibfield  {title} {\enquote
  {\bibinfo {title} {{Planck 2015 results. XIII. Cosmological parameters}},}\
  }\href {\doibase 10.1051/0004-6361/201525830} {\bibfield  {journal} {\bibinfo
   {journal} {Astronomy {\&} Astrophysics}\ }\textbf {\bibinfo {volume}
  {594}},\ \bibinfo {pages} {A13} (\bibinfo {year} {2015})},\ \Eprint
  {http://arxiv.org/abs/1502.01589} {arXiv:1502.01589} \BibitemShut {NoStop}%
\bibitem [{\citenamefont {Trilleras}(2015)}]{AlejandroThesis}%
  \BibitemOpen
  \bibfield  {author} {\bibinfo {author} {\bibfnamefont {Alejandro~Guarnizo}\
  \bibnamefont {Trilleras}},\ }\emph {\bibinfo {title} {A Model-Independent
  Approach to Dark Energy Cosmologies: Current and Future Constraints}},\
  \href@noop {} {Ph.D. thesis},\ \bibinfo  {school}
  {Rupert-Karls-Universit\"{a}t Heidelberg} (\bibinfo {year}
  {2015})\BibitemShut {NoStop}%
\bibitem [{\citenamefont {Yu}\ \emph {et~al.}(2017)\citenamefont {Yu},
  \citenamefont {Ratra},\ and\ \citenamefont {Wang}}]{Yu2017}%
  \BibitemOpen
  \bibfield  {author} {\bibinfo {author} {\bibfnamefont {Hai}\ \bibnamefont
  {Yu}}, \bibinfo {author} {\bibfnamefont {Bharat}\ \bibnamefont {Ratra}}, \
  and\ \bibinfo {author} {\bibfnamefont {Fa-Yin}\ \bibnamefont {Wang}},\
  }\bibfield  {title} {\enquote {\bibinfo {title} {{Hubble Parameter and Baryon
  Acoustic Oscillation Measurement Constraints on the Hubble Constant, the
  Deviation from the Spatially-Flat Lambda cdm Model, The
  Deceleration-Acceleration Transition Redshift, and Spatial Curvature}},}\
  }\href@noop {} {\  (\bibinfo {year} {2017})},\ \Eprint
  {http://arxiv.org/abs/1711.03437} {arXiv:1711.03437} \BibitemShut {NoStop}%
\bibitem [{\citenamefont {Simon}\ \emph {et~al.}(2004)\citenamefont {Simon},
  \citenamefont {Verde},\ and\ \citenamefont {Jimenez}}]{Simon2004}%
  \BibitemOpen
  \bibfield  {author} {\bibinfo {author} {\bibfnamefont {Joan}\ \bibnamefont
  {Simon}}, \bibinfo {author} {\bibfnamefont {Licia}\ \bibnamefont {Verde}}, \
  and\ \bibinfo {author} {\bibfnamefont {Raul}\ \bibnamefont {Jimenez}},\
  }\bibfield  {title} {\enquote {\bibinfo {title} {{Constraints on the redshift
  dependence of the dark energy potential}},}\ }\href {\doibase
  10.1103/PhysRevD.71.123001} {\bibfield  {journal} {\bibinfo  {journal}
  {Physics Review D}\ }\textbf {\bibinfo {volume} {71}} (\bibinfo {year}
  {2004}),\ 10.1103/PhysRevD.71.123001},\ \Eprint
  {http://arxiv.org/abs/0412269} {arXiv:0412269 [astro-ph]} \BibitemShut
  {NoStop}%
\bibitem [{\citenamefont {Stern}\ \emph {et~al.}(2009)\citenamefont {Stern},
  \citenamefont {Jimenez}, \citenamefont {Verde} \emph {et~al.}}]{Stern2009}%
  \BibitemOpen
  \bibfield  {author} {\bibinfo {author} {\bibfnamefont {Daniel}\ \bibnamefont
  {Stern}}, \bibinfo {author} {\bibfnamefont {Raul}\ \bibnamefont {Jimenez}},
  \bibinfo {author} {\bibfnamefont {Licia}\ \bibnamefont {Verde}},  \emph
  {et~al.},\ }\bibfield  {title} {\enquote {\bibinfo {title} {{Cosmic
  Chronometers: Constraining the Equation of State of Dark Energy. II. A
  Spectroscopic Catalog of Red Galaxies in Galaxy Clusters}},}\ }\href
  {\doibase 10.1088/0067-0049/188/1/280} {\  (\bibinfo {year} {2009}),\
  10.1088/0067-0049/188/1/280},\ \Eprint {http://arxiv.org/abs/0907.3152}
  {arXiv:0907.3152} \BibitemShut {NoStop}%
\bibitem [{\citenamefont {Moresco}\ \emph {et~al.}(2012)\citenamefont {Moresco}
  \emph {et~al.}}]{Moresco:2012jh}%
  \BibitemOpen
  \bibfield  {author} {\bibinfo {author} {\bibfnamefont {M.}~\bibnamefont
  {Moresco}} \emph {et~al.},\ }\bibfield  {title} {\enquote {\bibinfo {title}
  {{Improved constraints on the expansion rate of the Universe up to z~1.1 from
  the spectroscopic evolution of cosmic chronometers}},}\ }\href {\doibase
  10.1088/1475-7516/2012/08/006} {\bibfield  {journal} {\bibinfo  {journal}
  {JCAP}\ }\textbf {\bibinfo {volume} {1208}},\ \bibinfo {pages} {006}
  (\bibinfo {year} {2012})},\ \Eprint {http://arxiv.org/abs/1201.3609}
  {arXiv:1201.3609 [astro-ph.CO]} \BibitemShut {NoStop}%
\bibitem [{\citenamefont {Moresco}(2015)}]{Moresco:2015cya}%
  \BibitemOpen
  \bibfield  {author} {\bibinfo {author} {\bibfnamefont {Michele}\ \bibnamefont
  {Moresco}},\ }\bibfield  {title} {\enquote {\bibinfo {title} {{Raising the
  bar: new constraints on the Hubble parameter with cosmic chronometers at $z
  \sim 2$}},}\ }\href {\doibase 10.1093/mnrasl/slv037} {\bibfield  {journal}
  {\bibinfo  {journal} {Mon. Not. Roy. Astron. Soc.}\ }\textbf {\bibinfo
  {volume} {450}},\ \bibinfo {pages} {L16--L20} (\bibinfo {year} {2015})},\
  \Eprint {http://arxiv.org/abs/1503.01116} {arXiv:1503.01116 [astro-ph.CO]}
  \BibitemShut {NoStop}%
\bibitem [{\citenamefont {Delubac}\ \emph {et~al.}(2015)\citenamefont {Delubac}
  \emph {et~al.}}]{Delubac:2014aqe}%
  \BibitemOpen
  \bibfield  {author} {\bibinfo {author} {\bibfnamefont {Timothée}\
  \bibnamefont {Delubac}} \emph {et~al.} (\bibinfo {collaboration} {BOSS}),\
  }\bibfield  {title} {\enquote {\bibinfo {title} {{Baryon acoustic
  oscillations in the Ly$\alpha$ forest of BOSS DR11 quasars}},}\ }\href
  {\doibase 10.1051/0004-6361/201423969} {\bibfield  {journal} {\bibinfo
  {journal} {Astron. Astrophys.}\ }\textbf {\bibinfo {volume} {574}},\ \bibinfo
  {pages} {A59} (\bibinfo {year} {2015})},\ \Eprint
  {http://arxiv.org/abs/1404.1801} {arXiv:1404.1801 [astro-ph.CO]} \BibitemShut
  {NoStop}%
\bibitem [{\citenamefont {Font-Ribera}\ \emph {et~al.}(2014)\citenamefont
  {Font-Ribera} \emph {et~al.}}]{Font-Ribera:2013wce}%
  \BibitemOpen
  \bibfield  {author} {\bibinfo {author} {\bibfnamefont {Andreu}\ \bibnamefont
  {Font-Ribera}} \emph {et~al.} (\bibinfo {collaboration} {BOSS}),\ }\bibfield
  {title} {\enquote {\bibinfo {title} {{Quasar-Lyman $\alpha$ Forest
  Cross-Correlation from BOSS DR11 : Baryon Acoustic Oscillations}},}\ }\href
  {\doibase 10.1088/1475-7516/2014/05/027} {\bibfield  {journal} {\bibinfo
  {journal} {JCAP}\ }\textbf {\bibinfo {volume} {1405}},\ \bibinfo {pages}
  {027} (\bibinfo {year} {2014})},\ \Eprint {http://arxiv.org/abs/1311.1767}
  {arXiv:1311.1767 [astro-ph.CO]} \BibitemShut {NoStop}%
\bibitem [{\citenamefont {{Moresco}}\ \emph {et~al.}(2016)\citenamefont
  {{Moresco}}, \citenamefont {{Pozzetti}}, \citenamefont {{Cimatti}},
  \citenamefont {{Jimenez}}, \citenamefont {{Maraston}}, \citenamefont
  {{Verde}}, \citenamefont {{Thomas}}, \citenamefont {{Citro}}, \citenamefont
  {{Tojeiro}},\ and\ \citenamefont {{Wilkinson}}}]{Moresco2016}%
  \BibitemOpen
  \bibfield  {author} {\bibinfo {author} {\bibfnamefont {M.}~\bibnamefont
  {{Moresco}}}, \bibinfo {author} {\bibfnamefont {L.}~\bibnamefont
  {{Pozzetti}}}, \bibinfo {author} {\bibfnamefont {A.}~\bibnamefont
  {{Cimatti}}}, \bibinfo {author} {\bibfnamefont {R.}~\bibnamefont
  {{Jimenez}}}, \bibinfo {author} {\bibfnamefont {C.}~\bibnamefont
  {{Maraston}}}, \bibinfo {author} {\bibfnamefont {L.}~\bibnamefont {{Verde}}},
  \bibinfo {author} {\bibfnamefont {D.}~\bibnamefont {{Thomas}}}, \bibinfo
  {author} {\bibfnamefont {A.}~\bibnamefont {{Citro}}}, \bibinfo {author}
  {\bibfnamefont {R.}~\bibnamefont {{Tojeiro}}}, \ and\ \bibinfo {author}
  {\bibfnamefont {D.}~\bibnamefont {{Wilkinson}}},\ }\bibfield  {title}
  {\enquote {\bibinfo {title} {{A 6\% measurement of the Hubble parameter at
  z\~{}0.45: direct evidence of the epoch of cosmic re-acceleration}},}\ }\href
  {\doibase 10.1088/1475-7516/2016/05/014} {\bibfield  {journal} {\bibinfo
  {journal} {Journal of Cosmology and Astroparticle Physics}\ }\textbf
  {\bibinfo {volume} {5}},\ \bibinfo {eid} {014} (\bibinfo {year} {2016})},\
  \Eprint {http://arxiv.org/abs/1601.01701} {arXiv:1601.01701} \BibitemShut
  {NoStop}%
\bibitem [{\citenamefont {Zhang}\ \emph {et~al.}(2012)\citenamefont {Zhang},
  \citenamefont {Zhang}, \citenamefont {Yuan}, \citenamefont {Liu},
  \citenamefont {Zhang},\ and\ \citenamefont {Sun}}]{Zhang2012}%
  \BibitemOpen
  \bibfield  {author} {\bibinfo {author} {\bibfnamefont {Cong}\ \bibnamefont
  {Zhang}}, \bibinfo {author} {\bibfnamefont {Han}\ \bibnamefont {Zhang}},
  \bibinfo {author} {\bibfnamefont {Shuo}\ \bibnamefont {Yuan}}, \bibinfo
  {author} {\bibfnamefont {Siqi}\ \bibnamefont {Liu}}, \bibinfo {author}
  {\bibfnamefont {Tong-Jie}\ \bibnamefont {Zhang}}, \ and\ \bibinfo {author}
  {\bibfnamefont {Yan-Chun}\ \bibnamefont {Sun}},\ }\bibfield  {title}
  {\enquote {\bibinfo {title} {{Four New Observational {\$}H(z){\$} Data From
  Luminous Red Galaxies of Sloan Digital Sky Survey Data Release Seven}},}\
  }\href {\doibase 10.1088/1674--4527/14/10/002} {\  (\bibinfo {year} {2012}),\
  10.1088/1674--4527/14/10/002},\ \Eprint {http://arxiv.org/abs/1207.4541}
  {arXiv:1207.4541} \BibitemShut {NoStop}%
\bibitem [{\citenamefont {Alam}\ \emph {et~al.}(2016)\citenamefont {Alam},
  \citenamefont {Ata}, \citenamefont {Bailey} \emph {et~al.}}]{Alam2016}%
  \BibitemOpen
  \bibfield  {author} {\bibinfo {author} {\bibfnamefont {Shadab}\ \bibnamefont
  {Alam}}, \bibinfo {author} {\bibfnamefont {Metin}\ \bibnamefont {Ata}},
  \bibinfo {author} {\bibfnamefont {Stephen}\ \bibnamefont {Bailey}},  \emph
  {et~al.},\ }\bibfield  {title} {\enquote {\bibinfo {title} {{The clustering
  of galaxies in the completed SDSS-III Baryon Oscillation Spectroscopic
  Survey: cosmological analysis of the DR12 galaxy sample}},}\ }\href {\doibase
  10.1093/mnras/stx721} {\bibfield  {journal} {\bibinfo  {journal} {Monthly
  Notices of the Royal Astronomical Society}\ } (\bibinfo {year} {2016}),\
  10.1093/mnras/stx721},\ \Eprint {http://arxiv.org/abs/1607.03155}
  {arXiv:1607.03155} \BibitemShut {NoStop}%
\bibitem [{\citenamefont {Blake}\ \emph {et~al.}(2012)\citenamefont {Blake},
  \citenamefont {Brough}, \citenamefont {Colless} \emph {et~al.}}]{Blake2012}%
  \BibitemOpen
  \bibfield  {author} {\bibinfo {author} {\bibfnamefont {C.}~\bibnamefont
  {Blake}}, \bibinfo {author} {\bibfnamefont {S.}~\bibnamefont {Brough}},
  \bibinfo {author} {\bibfnamefont {M.}~\bibnamefont {Colless}},  \emph
  {et~al.},\ }\bibfield  {title} {\enquote {\bibinfo {title} {{The WiggleZ Dark
  Energy Survey: joint measurements of the expansion and growth history at
  {\textless}i{\textgreater}z{\textless}/i{\textgreater} {\textless} 1}},}\
  }\href {\doibase 10.1111/j.1365-2966.2012.21473.x} {\bibfield  {journal}
  {\bibinfo  {journal} {Monthly Notices of the Royal Astronomical Society}\
  }\textbf {\bibinfo {volume} {425}},\ \bibinfo {pages} {405--414} (\bibinfo
  {year} {2012})}\BibitemShut {NoStop}%
\bibitem [{\citenamefont {Riess}\ \emph {et~al.}(2017)\citenamefont {Riess},
  \citenamefont {Rodney}, \citenamefont {Scolnic} \emph {et~al.}}]{Riess2017}%
  \BibitemOpen
  \bibfield  {author} {\bibinfo {author} {\bibfnamefont {Adam~G.}\ \bibnamefont
  {Riess}}, \bibinfo {author} {\bibfnamefont {Steven~A.}\ \bibnamefont
  {Rodney}}, \bibinfo {author} {\bibfnamefont {Daniel~M.}\ \bibnamefont
  {Scolnic}},  \emph {et~al.},\ }\bibfield  {title} {\enquote {\bibinfo {title}
  {{Type Ia Supernova Distances at z {\textgreater} 1.5 from the Hubble Space
  Telescope Multi-Cycle Treasury Programs: The Early Expansion Rate}},}\
  }\href@noop {} {\  (\bibinfo {year} {2017})},\ \Eprint
  {http://arxiv.org/abs/1710.00844} {arXiv:1710.00844} \BibitemShut {NoStop}%
\bibitem [{\citenamefont {Blake}\ \emph {et~al.}(2016)\citenamefont {Blake},
  \citenamefont {Joudaki}, \citenamefont {Heymans} \emph {et~al.}}]{Blake2016}%
  \BibitemOpen
  \bibfield  {author} {\bibinfo {author} {\bibfnamefont {C.}~\bibnamefont
  {Blake}}, \bibinfo {author} {\bibfnamefont {S.}~\bibnamefont {Joudaki}},
  \bibinfo {author} {\bibfnamefont {C.}~\bibnamefont {Heymans}},  \emph
  {et~al.},\ }\bibfield  {title} {\enquote {\bibinfo {title} {{RCSLenS: Testing
  gravitational physics through the cross-correlation of weak lensing and
  large-scale structure}},}\ }\href {\doibase 10.1093/mnras/stv2875} {\bibfield
   {journal} {\bibinfo  {journal} {Monthly Notices of the Royal Astronomical
  Society}\ } (\bibinfo {year} {2016}),\ 10.1093/mnras/stv2875},\ \Eprint
  {http://arxiv.org/abs/1507.03086} {arXiv:1507.03086} \BibitemShut {NoStop}%
\bibitem [{\citenamefont {de~la Torre}\ \emph {et~al.}(2016)\citenamefont
  {de~la Torre}, \citenamefont {Jullo}, \citenamefont {Giocoli} \emph
  {et~al.}}]{DelaTorre2016}%
  \BibitemOpen
  \bibfield  {author} {\bibinfo {author} {\bibfnamefont {S.}~\bibnamefont
  {de~la Torre}}, \bibinfo {author} {\bibfnamefont {E.}~\bibnamefont {Jullo}},
  \bibinfo {author} {\bibfnamefont {C.}~\bibnamefont {Giocoli}},  \emph
  {et~al.},\ }\bibfield  {title} {\enquote {\bibinfo {title} {{The VIMOS Public
  Extragalactic Redshift Survey (VIPERS). Gravity test from the combination of
  redshift-space distortions and galaxy-galaxy lensing at {\$}0.5 {\textless} z
  {\textless} 1.2{\$}}},}\ }\href@noop {} {\bibfield  {journal} {\bibinfo
  {journal} {eprint arXiv:1612.05647}\ } (\bibinfo {year} {2016})},\ \Eprint
  {http://arxiv.org/abs/1612.05647} {arXiv:1612.05647} \BibitemShut {NoStop}%
\bibitem [{\citenamefont {Beutler}\ \emph {et~al.}(2012)\citenamefont
  {Beutler}, \citenamefont {Blake} \emph {et~al.}}]{Beutler2012}%
  \BibitemOpen
  \bibfield  {author} {\bibinfo {author} {\bibfnamefont {F.}~\bibnamefont
  {Beutler}}, \bibinfo {author} {\bibfnamefont {M.}~\bibnamefont {Blake},
  \bibfnamefont {C.and~Colless}},  \emph {et~al.},\ }\bibfield  {title}
  {\enquote {\bibinfo {title} {{The 6dF Galaxy Survey: z=0 measurements of the
  growth rate and $\sigma$8}},}\ }\href {\doibase
  10.1111/j.1365-2966.2012.21136.x} {\bibfield  {journal} {\bibinfo  {journal}
  {Monthly Notices of the Royal Astronomical Society}\ }\textbf {\bibinfo
  {volume} {423}},\ \bibinfo {pages} {3430--3444} (\bibinfo {year}
  {2012})}\BibitemShut {NoStop}%
\bibitem [{\citenamefont {Okumura}\ \emph {et~al.}(2015)\citenamefont
  {Okumura}, \citenamefont {Hikage}, \citenamefont {Totani} \emph
  {et~al.}}]{Okumura2015}%
  \BibitemOpen
  \bibfield  {author} {\bibinfo {author} {\bibfnamefont {Teppei}\ \bibnamefont
  {Okumura}}, \bibinfo {author} {\bibfnamefont {Chiaki}\ \bibnamefont
  {Hikage}}, \bibinfo {author} {\bibfnamefont {Tomonori}\ \bibnamefont
  {Totani}},  \emph {et~al.},\ }\bibfield  {title} {\enquote {\bibinfo {title}
  {{The Subaru FMOS galaxy redshift survey (FastSound). IV. New constraint on
  gravity theory from redshift space distortions at {\$}z = 1.4{\$}}},}\ }\href
  {\doibase 10.1093/pasj/psw029} {\bibfield  {journal} {\bibinfo  {journal}
  {Publications of the Astronomical Society of Japan, Volume 68, Issue 3, id.38
  24 pp.}\ }\textbf {\bibinfo {volume} {68}} (\bibinfo {year} {2015}),\
  10.1093/pasj/psw029},\ \Eprint {http://arxiv.org/abs/1511.08083}
  {arXiv:1511.08083} \BibitemShut {NoStop}%
\bibitem [{\citenamefont {Song}\ and\ \citenamefont
  {Percival}(2008)}]{Song2008}%
  \BibitemOpen
  \bibfield  {author} {\bibinfo {author} {\bibfnamefont {Yong-Seon}\
  \bibnamefont {Song}}\ and\ \bibinfo {author} {\bibfnamefont {Will~J.}\
  \bibnamefont {Percival}},\ }\bibfield  {title} {\enquote {\bibinfo {title}
  {{Reconstructing the history of structure formation using redshift
  distortions}},}\ }\href {\doibase 10.1088/1475-7516/2009/10/004} {\bibfield
  {journal} {\bibinfo  {journal} {Journal of Cosmology and Astroparticle
  Physics}\ } (\bibinfo {year} {2008}),\ 10.1088/1475-7516/2009/10/004},\
  \Eprint {http://arxiv.org/abs/0807.0810} {arXiv:0807.0810} \BibitemShut
  {NoStop}%
\bibitem [{\citenamefont {Hawken}\ \emph {et~al.}(2016)\citenamefont {Hawken},
  \citenamefont {Granett}, \citenamefont {Iovino} \emph {et~al.}}]{Hawken2016}%
  \BibitemOpen
  \bibfield  {author} {\bibinfo {author} {\bibfnamefont {A.~J.}\ \bibnamefont
  {Hawken}}, \bibinfo {author} {\bibfnamefont {B.~R.}\ \bibnamefont {Granett}},
  \bibinfo {author} {\bibfnamefont {A.}~\bibnamefont {Iovino}},  \emph
  {et~al.},\ }\bibfield  {title} {\enquote {\bibinfo {title} {{The VIMOS Public
  Extragalactic Redshift Survey: Measuring the growth rate of structure around
  cosmic voids}},}\ }\href {\doibase 10.1051/0004-6361/201629678} {\bibfield
  {journal} {\bibinfo  {journal} {Astronomy {\&} Astrophysics}\ }\textbf
  {\bibinfo {volume} {607}} (\bibinfo {year} {2016}),\
  10.1051/0004-6361/201629678},\ \Eprint {http://arxiv.org/abs/1611.07046}
  {arXiv:1611.07046} \BibitemShut {NoStop}%
\bibitem [{\citenamefont {de~la Torre}\ \emph {et~al.}(2013)\citenamefont
  {de~la Torre}, \citenamefont {Guzzo}, \citenamefont {Peacock} \emph
  {et~al.}}]{DelaTorre2013}%
  \BibitemOpen
  \bibfield  {author} {\bibinfo {author} {\bibfnamefont {S.}~\bibnamefont
  {de~la Torre}}, \bibinfo {author} {\bibfnamefont {L.}~\bibnamefont {Guzzo}},
  \bibinfo {author} {\bibfnamefont {J.~A.}\ \bibnamefont {Peacock}},  \emph
  {et~al.},\ }\bibfield  {title} {\enquote {\bibinfo {title} {{The VIMOS Public
  Extragalactic Redshift Survey (VIPERS). Galaxy clustering and redshift-space
  distortions at z=0.8 in the first data release}},}\ }\href {\doibase
  10.1051/0004-6361/201321463} {\bibfield  {journal} {\bibinfo  {journal}
  {Astronomy {\&} Astrophysics, Volume 557, id.A54, 19 pp.}\ }\textbf {\bibinfo
  {volume} {557}} (\bibinfo {year} {2013}),\ 10.1051/0004-6361/201321463},\
  \Eprint {http://arxiv.org/abs/1303.2622} {arXiv:1303.2622} \BibitemShut
  {NoStop}%
\bibitem [{\citenamefont {Mohammad}\ \emph {et~al.}(2017)\citenamefont
  {Mohammad}, \citenamefont {Granett}, \citenamefont {Guzzo} \emph
  {et~al.}}]{Mohammad2017}%
  \BibitemOpen
  \bibfield  {author} {\bibinfo {author} {\bibfnamefont {F.~G.}\ \bibnamefont
  {Mohammad}}, \bibinfo {author} {\bibfnamefont {B.~R.}\ \bibnamefont
  {Granett}}, \bibinfo {author} {\bibfnamefont {L.}~\bibnamefont {Guzzo}},
  \emph {et~al.},\ }\bibfield  {title} {\enquote {\bibinfo {title} {{The VIMOS
  Public Extragalactic Redshift Survey (VIPERS): An unbiased estimate of the
  growth rate of structure at z = 0.85 using the clustering of luminous blue
  galaxies}},}\ }\href@noop {} {\  (\bibinfo {year} {2017})},\ \Eprint
  {http://arxiv.org/abs/1708.00026} {arXiv:1708.00026} \BibitemShut {NoStop}%
\bibitem [{\citenamefont {Howlett}\ \emph {et~al.}(2015)\citenamefont
  {Howlett}, \citenamefont {Ross}, \citenamefont {Samushia}, \citenamefont
  {Percival},\ and\ \citenamefont {Manera}}]{Howlett2015}%
  \BibitemOpen
  \bibfield  {author} {\bibinfo {author} {\bibfnamefont {Cullan}\ \bibnamefont
  {Howlett}}, \bibinfo {author} {\bibfnamefont {Ashley~J.}\ \bibnamefont
  {Ross}}, \bibinfo {author} {\bibfnamefont {Lado}\ \bibnamefont {Samushia}},
  \bibinfo {author} {\bibfnamefont {Will~J.}\ \bibnamefont {Percival}}, \ and\
  \bibinfo {author} {\bibfnamefont {Marc}\ \bibnamefont {Manera}},\ }\bibfield
  {title} {\enquote {\bibinfo {title} {{The clustering of the SDSS main galaxy
  sample – II. Mock galaxy catalogues and a measurement of the growth of
  structure from redshift space distortions at z = 0.15}},}\ }\href {\doibase
  10.1093/mnras/stu2693} {\bibfield  {journal} {\bibinfo  {journal} {Monthly
  Notices of the Royal Astronomical Society}\ }\textbf {\bibinfo {volume}
  {449}},\ \bibinfo {pages} {848--866} (\bibinfo {year} {2015})}\BibitemShut
  {NoStop}%
\bibitem [{\citenamefont {Samushia}\ \emph {et~al.}(2012)\citenamefont
  {Samushia}, \citenamefont {Percival},\ and\ \citenamefont
  {Raccanelli}}]{Samushia2012}%
  \BibitemOpen
  \bibfield  {author} {\bibinfo {author} {\bibfnamefont {L.}~\bibnamefont
  {Samushia}}, \bibinfo {author} {\bibfnamefont {W.~J.}\ \bibnamefont
  {Percival}}, \ and\ \bibinfo {author} {\bibfnamefont {A.}~\bibnamefont
  {Raccanelli}},\ }\bibfield  {title} {\enquote {\bibinfo {title}
  {{Interpreting large-scale redshift-space distortion measurements}},}\ }\href
  {\doibase 10.1111/j.1365-2966.2011.20169.x} {\bibfield  {journal} {\bibinfo
  {journal} {Monthly Notices of the Royal Astronomical Society}\ }\textbf
  {\bibinfo {volume} {420}},\ \bibinfo {pages} {2102--2119} (\bibinfo {year}
  {2012})}\BibitemShut {NoStop}%
\bibitem [{\citenamefont {Tojeiro}\ \emph {et~al.}(2012)\citenamefont {Tojeiro}
  \emph {et~al.}}]{Tojeiro:2012rp}%
  \BibitemOpen
  \bibfield  {author} {\bibinfo {author} {\bibfnamefont {Rita}\ \bibnamefont
  {Tojeiro}} \emph {et~al.},\ }\bibfield  {title} {\enquote {\bibinfo {title}
  {{The clustering of galaxies in the SDSS-III Baryon Oscillation Spectroscopic
  Survey: measuring structure growth using passive galaxies}},}\ }\href
  {\doibase 10.1111/j.1365-2966.2012.21404.x} {\bibfield  {journal} {\bibinfo
  {journal} {Mon. Not. Roy. Astron. Soc.}\ }\textbf {\bibinfo {volume} {424}},\
  \bibinfo {pages} {2339} (\bibinfo {year} {2012})},\ \Eprint
  {http://arxiv.org/abs/1203.6565} {arXiv:1203.6565 [astro-ph.CO]} \BibitemShut
  {NoStop}%
\bibitem [{\citenamefont {{Chuang}}\ and\ \citenamefont
  {{Wang}}(2013)}]{Chuang2012}%
  \BibitemOpen
  \bibfield  {author} {\bibinfo {author} {\bibfnamefont {C.-H.}\ \bibnamefont
  {{Chuang}}}\ and\ \bibinfo {author} {\bibfnamefont {Y.}~\bibnamefont
  {{Wang}}},\ }\bibfield  {title} {\enquote {\bibinfo {title} {{Modelling the
  anisotropic two-point galaxy correlation function on small scales and
  single-probe measurements of H(z), D$_{A}$(z) and f(z){$\sigma$}$_{8}$(z)
  from the Sloan Digital Sky Survey DR7 luminous red galaxies}},}\ }\href
  {\doibase 10.1093/mnras/stt1290} {\bibfield  {journal} {\bibinfo  {journal}
  {Monthly Notices of the Royal Astronomical Society}\ }\textbf {\bibinfo
  {volume} {435}},\ \bibinfo {pages} {255--262} (\bibinfo {year} {2013})},\
  \Eprint {http://arxiv.org/abs/1209.0210} {arXiv:1209.0210} \BibitemShut
  {NoStop}%
\bibitem [{\citenamefont {Gil-Marín}\ \emph {et~al.}(2016)\citenamefont
  {Gil-Marín} \emph {et~al.}}]{Gil-Marin:2015sqa}%
  \BibitemOpen
  \bibfield  {author} {\bibinfo {author} {\bibfnamefont {Héctor}\ \bibnamefont
  {Gil-Marín}} \emph {et~al.},\ }\bibfield  {title} {\enquote {\bibinfo
  {title} {{The clustering of galaxies in the SDSS-III Baryon Oscillation
  Spectroscopic Survey: RSD measurement from the LOS-dependent power spectrum
  of DR12 BOSS galaxies}},}\ }\href {\doibase 10.1093/mnras/stw1096} {\bibfield
   {journal} {\bibinfo  {journal} {Mon. Not. Roy. Astron. Soc.}\ }\textbf
  {\bibinfo {volume} {460}},\ \bibinfo {pages} {4188--4209} (\bibinfo {year}
  {2016})},\ \Eprint {http://arxiv.org/abs/1509.06386} {arXiv:1509.06386
  [astro-ph.CO]} \BibitemShut {NoStop}%
\bibitem [{\citenamefont {Gil-Mar{\'{i}}n}\ \emph {et~al.}(2016)\citenamefont
  {Gil-Mar{\'{i}}n}, \citenamefont {Percival}, \citenamefont {Cuesta} \emph
  {et~al.}}]{Gil-Marin2016}%
  \BibitemOpen
  \bibfield  {author} {\bibinfo {author} {\bibfnamefont {H.}~\bibnamefont
  {Gil-Mar{\'{i}}n}}, \bibinfo {author} {\bibfnamefont {W.}~\bibnamefont
  {Percival}}, \bibinfo {author} {\bibfnamefont {A.}~\bibnamefont {Cuesta}},
  \emph {et~al.},\ }\bibfield  {title} {\enquote {\bibinfo {title} {{The
  clustering of galaxies in the SDSS-III Baryon Oscillation Spectroscopic
  Survey: BAO measurement from the LOS-dependent power spectrum of DR12 BOSS
  galaxies}},}\ }\href {\doibase 10.1093/mnras/stw1264} {\bibfield  {journal}
  {\bibinfo  {journal} {Monthly Notices of the Royal Astronomical Society}\
  }\textbf {\bibinfo {volume} {460}},\ \bibinfo {pages} {4210--4219} (\bibinfo
  {year} {2016})}\BibitemShut {NoStop}%
\bibitem [{\citenamefont {Chuang}\ \emph {et~al.}(2016)\citenamefont {Chuang},
  \citenamefont {Prada}, \citenamefont {Pellejero-Ibanez} \emph
  {et~al.}}]{Chuang2016}%
  \BibitemOpen
  \bibfield  {author} {\bibinfo {author} {\bibfnamefont {C.}~\bibnamefont
  {Chuang}}, \bibinfo {author} {\bibfnamefont {F.}~\bibnamefont {Prada}},
  \bibinfo {author} {\bibfnamefont {M.}~\bibnamefont {Pellejero-Ibanez}},
  \emph {et~al.},\ }\bibfield  {title} {\enquote {\bibinfo {title} {{The
  clustering of galaxies in the SDSS-III Baryon Oscillation Spectroscopic
  Survey: single-probe measurements from CMASS anisotropic galaxy
  clustering}},}\ }\href@noop {} {\bibfield  {journal} {\bibinfo  {journal}
  {Monthly Notices of the Royal Astronomical Society}\ }\textbf {\bibinfo
  {volume} {461}} (\bibinfo {year} {2016})}\BibitemShut {NoStop}%
\bibitem [{\citenamefont {Cabr{\'{e}}}\ and\ \citenamefont
  {Gazta{\~{n}}aga}(2009)}]{Cabre2009}%
  \BibitemOpen
  \bibfield  {author} {\bibinfo {author} {\bibfnamefont {Anna}\ \bibnamefont
  {Cabr{\'{e}}}}\ and\ \bibinfo {author} {\bibfnamefont {Enrique}\ \bibnamefont
  {Gazta{\~{n}}aga}},\ }\bibfield  {title} {\enquote {\bibinfo {title}
  {{Clustering of luminous red galaxies - I. Large-scale redshift-space
  distortions}},}\ }\href {\doibase 10.1111/j.1365-2966.2008.14281.x}
  {\bibfield  {journal} {\bibinfo  {journal} {Monthly Notices of the Royal
  Astronomical Society}\ }\textbf {\bibinfo {volume} {393}},\ \bibinfo {pages}
  {1183--1208} (\bibinfo {year} {2009})}\BibitemShut {NoStop}%
\bibitem [{\citenamefont {Guzzo}\ \emph {et~al.}(2008)\citenamefont {Guzzo},
  \citenamefont {Pierleoni}, \citenamefont {Meneux} \emph
  {et~al.}}]{Guzzo2008}%
  \BibitemOpen
  \bibfield  {author} {\bibinfo {author} {\bibfnamefont {L.}~\bibnamefont
  {Guzzo}}, \bibinfo {author} {\bibfnamefont {M.}~\bibnamefont {Pierleoni}},
  \bibinfo {author} {\bibfnamefont {B.}~\bibnamefont {Meneux}},  \emph
  {et~al.},\ }\bibfield  {title} {\enquote {\bibinfo {title} {{A test of the
  nature of cosmic acceleration using galaxy redshift distortions}},}\ }\href
  {\doibase 10.1038/nature06555} {\bibfield  {journal} {\bibinfo  {journal}
  {Nature}\ }\textbf {\bibinfo {volume} {451}},\ \bibinfo {pages} {541--544}
  (\bibinfo {year} {2008})}\BibitemShut {NoStop}%
\bibitem [{\citenamefont {Rasmussen}\ and\ \citenamefont
  {Williams}(2006)}]{GPbook}%
  \BibitemOpen
  \bibfield  {author} {\bibinfo {author} {\bibfnamefont {C.}~\bibnamefont
  {Rasmussen}}\ and\ \bibinfo {author} {\bibfnamefont {C.}~\bibnamefont
  {Williams}},\ }\href@noop {} {\emph {\bibinfo {title} {Gaussian Processes for
  Machine Learning}}}\ (\bibinfo  {publisher} {MIT Press},\ \bibinfo {year}
  {2006})\BibitemShut {NoStop}%
\bibitem [{\citenamefont {Seikel}\ \emph {et~al.}(2012)\citenamefont {Seikel},
  \citenamefont {Clarkson},\ and\ \citenamefont {Smith}}]{Seikel2012}%
  \BibitemOpen
  \bibfield  {author} {\bibinfo {author} {\bibfnamefont {Marina}\ \bibnamefont
  {Seikel}}, \bibinfo {author} {\bibfnamefont {Chris}\ \bibnamefont
  {Clarkson}}, \ and\ \bibinfo {author} {\bibfnamefont {Mathew}\ \bibnamefont
  {Smith}},\ }\bibfield  {title} {\enquote {\bibinfo {title} {{Reconstruction
  of dark energy and expansion dynamics using Gaussian processes}},}\ }\href
  {\doibase 10.1088/1475-7516/2012/06/036} {\bibfield  {journal} {\bibinfo
  {journal} {Journal of Cosmology and Astroparticle Physics}\ } (\bibinfo
  {year} {2012}),\ 10.1088/1475-7516/2012/06/036},\ \Eprint
  {http://arxiv.org/abs/1204.2832} {arXiv:1204.2832} \BibitemShut {NoStop}%
\bibitem [{\citenamefont {Laureijs}\ \emph {et~al.}(2011)\citenamefont
  {Laureijs} \emph {et~al.}}]{Laureijs:2011gra}%
  \BibitemOpen
  \bibfield  {author} {\bibinfo {author} {\bibfnamefont {R.}~\bibnamefont
  {Laureijs}} \emph {et~al.} (\bibinfo {collaboration} {EUCLID}),\ }\bibfield
  {title} {\enquote {\bibinfo {title} {{Euclid Definition Study Report}},}\
  }\href@noop {} {\  (\bibinfo {year} {2011})},\ \Eprint
  {http://arxiv.org/abs/1110.3193} {arXiv:1110.3193 [astro-ph.CO]} \BibitemShut
  {NoStop}%
\bibitem [{\citenamefont {Amendola}\ \emph
  {et~al.}(2013{\natexlab{b}})\citenamefont {Amendola} \emph
  {et~al.}}]{Amendola:2012ys}%
  \BibitemOpen
  \bibfield  {author} {\bibinfo {author} {\bibfnamefont {Luca}\ \bibnamefont
  {Amendola}} \emph {et~al.} (\bibinfo {collaboration} {Euclid Theory Working
  Group}),\ }\bibfield  {title} {\enquote {\bibinfo {title} {{Cosmology and
  fundamental physics with the Euclid satellite}},}\ }\href {\doibase
  10.12942/lrr-2013-6} {\bibfield  {journal} {\bibinfo  {journal} {Living Rev.
  Rel.}\ }\textbf {\bibinfo {volume} {16}},\ \bibinfo {pages} {6} (\bibinfo
  {year} {2013}{\natexlab{b}})},\ \Eprint {http://arxiv.org/abs/1206.1225}
  {arXiv:1206.1225 [astro-ph.CO]} \BibitemShut {NoStop}%
\bibitem [{\citenamefont {{Rawlings}}(2011)}]{SKAmission}%
  \BibitemOpen
  \bibfield  {author} {\bibinfo {author} {\bibfnamefont {S.}~\bibnamefont
  {{Rawlings}}},\ }\bibfield  {title} {\enquote {\bibinfo {title} {{Cosmology
  with the Square Kilometre Array}},}\ }\href@noop {} {\bibfield  {journal}
  {\bibinfo  {journal} {ArXiv e-prints}\ } (\bibinfo {year} {2011})},\ \Eprint
  {http://arxiv.org/abs/1105.6333} {arXiv:1105.6333 [astro-ph.CO]} \BibitemShut
  {NoStop}%
\bibitem [{\citenamefont {{LSST Science Collaboration}}\ \emph
  {et~al.}(2009)\citenamefont {{LSST Science Collaboration}}, \citenamefont
  {{Abell}}, \citenamefont {{Allison}}, \citenamefont {{Anderson}},
  \citenamefont {{Andrew}}, \citenamefont {{Angel}}, \citenamefont {{Armus}},
  \citenamefont {{Arnett}}, \citenamefont {{Asztalos}}, \citenamefont
  {{Axelrod}},\ and\ \citenamefont {et~al.}}]{2009arXiv0912.0201L}%
  \BibitemOpen
  \bibfield  {author} {\bibinfo {author} {\bibnamefont {{LSST Science
  Collaboration}}}, \bibinfo {author} {\bibfnamefont {P.~A.}\ \bibnamefont
  {{Abell}}}, \bibinfo {author} {\bibfnamefont {J.}~\bibnamefont {{Allison}}},
  \bibinfo {author} {\bibfnamefont {S.~F.}\ \bibnamefont {{Anderson}}},
  \bibinfo {author} {\bibfnamefont {J.~R.}\ \bibnamefont {{Andrew}}}, \bibinfo
  {author} {\bibfnamefont {J.~R.~P.}\ \bibnamefont {{Angel}}}, \bibinfo
  {author} {\bibfnamefont {L.}~\bibnamefont {{Armus}}}, \bibinfo {author}
  {\bibfnamefont {D.}~\bibnamefont {{Arnett}}}, \bibinfo {author}
  {\bibfnamefont {S.~J.}\ \bibnamefont {{Asztalos}}}, \bibinfo {author}
  {\bibfnamefont {T.~S.}\ \bibnamefont {{Axelrod}}}, \ and\ \bibinfo {author}
  {\bibnamefont {et~al.}},\ }\bibfield  {title} {\enquote {\bibinfo {title}
  {{LSST Science Book, Version 2.0}},}\ }\href@noop {} {\bibfield  {journal}
  {\bibinfo  {journal} {ArXiv e-prints}\ } (\bibinfo {year} {2009})},\ \Eprint
  {http://arxiv.org/abs/0912.0201} {arXiv:0912.0201 [astro-ph.IM]} \BibitemShut
  {NoStop}%
\bibitem [{\citenamefont {{Levi}}\ \emph {et~al.}(2013)\citenamefont {{Levi}},
  \citenamefont {{Bebek}}, \citenamefont {{Beers}}, \citenamefont {{Blum}},
  \citenamefont {{Cahn}}, \citenamefont {{Eisenstein}}, \citenamefont
  {{Flaugher}}, \citenamefont {{Honscheid}}, \citenamefont {{Kron}},
  \citenamefont {{Lahav}}, \citenamefont {{McDonald}}, \citenamefont {{Roe}},
  \citenamefont {{Schlegel}},\ and\ \citenamefont {{representing the DESI
  collaboration}}}]{2013arXiv1308.0847L}%
  \BibitemOpen
  \bibfield  {author} {\bibinfo {author} {\bibfnamefont {M.}~\bibnamefont
  {{Levi}}}, \bibinfo {author} {\bibfnamefont {C.}~\bibnamefont {{Bebek}}},
  \bibinfo {author} {\bibfnamefont {T.}~\bibnamefont {{Beers}}}, \bibinfo
  {author} {\bibfnamefont {R.}~\bibnamefont {{Blum}}}, \bibinfo {author}
  {\bibfnamefont {R.}~\bibnamefont {{Cahn}}}, \bibinfo {author} {\bibfnamefont
  {D.}~\bibnamefont {{Eisenstein}}}, \bibinfo {author} {\bibfnamefont
  {B.}~\bibnamefont {{Flaugher}}}, \bibinfo {author} {\bibfnamefont
  {K.}~\bibnamefont {{Honscheid}}}, \bibinfo {author} {\bibfnamefont
  {R.}~\bibnamefont {{Kron}}}, \bibinfo {author} {\bibfnamefont
  {O.}~\bibnamefont {{Lahav}}}, \bibinfo {author} {\bibfnamefont
  {P.}~\bibnamefont {{McDonald}}}, \bibinfo {author} {\bibfnamefont
  {N.}~\bibnamefont {{Roe}}}, \bibinfo {author} {\bibfnamefont
  {D.}~\bibnamefont {{Schlegel}}}, \ and\ \bibinfo {author} {\bibnamefont
  {{representing the DESI collaboration}}},\ }\bibfield  {title} {\enquote
  {\bibinfo {title} {{The DESI Experiment, a whitepaper for Snowmass 2013}},}\
  }\href@noop {} {\bibfield  {journal} {\bibinfo  {journal} {ArXiv e-prints}\ }
  (\bibinfo {year} {2013})},\ \Eprint {http://arxiv.org/abs/1308.0847}
  {arXiv:1308.0847 [astro-ph.CO]} \BibitemShut {NoStop}%
\bibitem [{\citenamefont {Zumalacarregui}\ \emph {et~al.}(2017)\citenamefont
  {Zumalacarregui}, \citenamefont {Bellini}, \citenamefont {Sawicki},
  \citenamefont {Lesgourgues},\ and\ \citenamefont
  {Ferreira}}]{Zumalacarregui:2016pph}%
  \BibitemOpen
  \bibfield  {author} {\bibinfo {author} {\bibfnamefont {Miguel}\ \bibnamefont
  {Zumalacarregui}}, \bibinfo {author} {\bibfnamefont {Emilio}\ \bibnamefont
  {Bellini}}, \bibinfo {author} {\bibfnamefont {Ignacy}\ \bibnamefont
  {Sawicki}}, \bibinfo {author} {\bibfnamefont {Julien}\ \bibnamefont
  {Lesgourgues}}, \ and\ \bibinfo {author} {\bibfnamefont {Pedro~G.}\
  \bibnamefont {Ferreira}},\ }\bibfield  {title} {\enquote {\bibinfo {title}
  {{hiclass: Horndeski in the Cosmic Linear Anisotropy Solving System}},}\
  }\href {\doibase 10.1088/1475-7516/2017/08/019} {\bibfield  {journal}
  {\bibinfo  {journal} {JCAP}\ }\textbf {\bibinfo {volume} {1708}},\ \bibinfo
  {pages} {019} (\bibinfo {year} {2017})},\ \Eprint
  {http://arxiv.org/abs/1605.06102} {arXiv:1605.06102 [astro-ph.CO]}
  \BibitemShut {NoStop}%
\end{thebibliography}%

\end{document}